# On the Fragility of Third-Party Punishment: The Context Effect of a Dominated Risky Investment Option[a]

Changkuk Im[b] and Jinkwon Lee[c]

**Abstract**

Experimental studies regularly show that third-party punishment (TPP) substantially exists in various settings. This study further investigates the robustness of TPP under an environment where context effects are involved. In our experiment, we offer a third party an additional but unattractive risky investment option. We find that, when the dominated investment option irrelevant to prosocial behavior is available, the demand for punishment decreases, whereas the demand for investment increases. These findings support our hypothesis that the seemingly unrelated and dominated investment option may work as a compromise and suggest the fragility of TPP in this setting.

---

[a] Declarations of interest: None. This research did not receive any specific grant from funding agencies in the public, commercial, or not-for-profit sectors. The authors obtained IRB approval from Sogang University (SGUIRB-A-1710-34).

[b] Changkuk Im. Department of Economics, The Ohio State University, Columbus, 43210, U.S.A. Email: im.95@osu.edu.

[c] Corresponding author, Jinkwon Lee. School of Economics, Sogang University, Seoul, 04107, South Korea. Tel. +82-2-705-8511. Email: jlee22@sogang.ac.kr. ORCID: 0000-0002-5695-1556.

## 1. Introduction

Third-party punishment (TPP)[1] is an important way of enforcing people to abide by social norms and achieve social stability [1, 2, 3, 4]. Indeed, a number of experimental studies report the substantial existence of TPP in numerous circumstances. For instance, the willingness of TPP is found in various populations [5, 6, 7, 8] and institutions including competition, voting, and a bottom-up system [9, 10, 11]. Some studies still find a sizable amount of TPP even when multiple third parties interact during the punishment decision [12, 13, 14], and when a third party has alternative choices, such as rewarding or second-party punishment (SPP) [15, 16]. The existence of TPP found in a controlled laboratory setting is not an artifact of its experimental design [17]. Instead, two psychological motivations for TPP are known. One is to vent negative emotions, such as anger after witnessing a violation of social norms [17, 18, 19], and the other is to reduce the degree of inequalities between one's own payoff and others' payoffs [12, 20, 21].

Yet, to our knowledge, it has not been investigated whether alternatives that are unattractive and irrelevant to prosocial behavior could even affect a third party's punitive behavior. In fact, recent studies that find context effects[2] are substantial in some field decision-making environments [22, 23] open a possibility that such alternatives could deter the willingness of TPP if context effects are involved in a third party's decision-making process. Thus, exploring when TPP can be fragile by context effects is an interesting question and would enhance our knowledge of human punitive behavior. In this paper, we experimentally test the robustness of TPP under an environment where context effects can possibly occur. Specifically, in our experiment, we offer a third party a seemingly unrelated and dominated alternative in addition to a punishment option and examine whether it deters the willingness of TPP.

In our three-player dictator game, similar to the experimental design of Fehr and Fischbacher [24], a third party has an opportunity to punish a dictator with a material cost given the information of the dictator's allocation. In addition to the punishment option, we provide a simple risky investment option for a third party whose expected net return is equal to or less than zero. From the perspective of material payoffs, the lottery is not an attractive choice to risk-neutral and risk-averse third parties. From the perspective of psychological payoffs, the investment option, compared with the punishment option, is not the best instrument for venting negative emotions or reducing the extent of payoff inequalities. Hence, at first glance, the investment option that we additionally offer seems unlikely to affect risk-averse and risk-neutral third parties' behavior because it is a dominated option in terms of both material and psychological payoffs.

---

[1] Third-party punishment (TPP) refers to an individual's behavior when one imposes costly punishment to a norm violator even if the punisher's payoff is not affected by the violator [24]. Second-party punishment (SPP) refers to punishment toward the violator exercised by an individual who has got harms by a norm violator [64].

[2] Examples of context effects are a compromise effect and a decoy effect. A compromise effect refers to a phenomenon that an individual perceives an option more attractive when it is placed at an intermediate in all attributes rather than an extreme position [65]. A decoy effect refers to a phenomenon when the choice frequency of a target option increases if an option dominated by the target option across all attributes is also available [66]. Note that they have been widely examined in marketing, psychological, and economic studies [28, 45, 46, 47] .

However, combining neuroeconomic studies on risky decision-making, punitive behaviors, and context effects opens a possibility of the dominated investment option to work as a compromise or decoy and affect a third party's punishment decision. According to those studies, some parts of the brain, such as the prefrontal cortex, the anterior insula, the amygdala, the anterior/posterior cingulate cortex, and nucleus accumbens, are commonly activated for choices involving risks [25, 26, 27, 28, 29], punishment decisions including SPP and TPP [30, 31, 32, 33, 34, 35], and context-dependent choices [36, 37, 38, 39, 40].[3] It is notable that while the activation of anterior insula is very common for all the tasks (implying a negative emotion process and/or intuitive decision-making involved in the tasks), the activation of nucleus accumbens related to reward processing is firmly found in a risky choice versus a safe choice [25, 29], but its strong activation was found only for SPP and much weaker for TPP [34]. In summary, combining all the results suggests a possible common neuro-mechanism in risky, punishment, and context-dependent choices: that is, the anterior insula and amygdala would gauge emotional states while expressing it with a cost is cognitively controlled by the prefrontal cortex and nucleus accumbens, and the conflict may be mediated by the anterior cingulate cortex [30, 31]. Although conjecturing a cognitive functioning simply through the activated area of the brain is difficult [41, 42], the co-incidental activations in similar areas by different environmental stimuli can still be interacted and confound the role of the brain [43, 44]. This suggests that a context effect can play a crucial role when both punishment and investment alternatives are available in our experiment.

---

[3] The prefrontal cortex is known to be related to integrating separate cognitive operations and decision-making [67, 68]. The anterior insula and the amygdala are known to be closely connected [69, 70], and both are related to visceral salience detection, intuitive decision-making, and negative emotion process [43, 71, 72, 73, 74, 75, 76] . The anterior cingulate cortex is known to involve in a conflict resolution between cognitive and emotional processes [31]. The posterior cingulate cortex is known to function as a mediator between memory and emotional processes among others [77]. The nucleus accumbens is related to the cognitive processing of reward and incentive salience among others [29].

Our framework assumes that material and psychological payoffs are two distinguished attributes involved in a third party's decision-making process. Note that the psychological-payoff attribute can be interpreted as two points of view: the "emotion-venting" viewpoint or the "inequity-averse" viewpoint. Depending on which point of view we focus on in our analysis, the investment option can be perceived differently. Focusing on the emotion-venting viewpoint, we can perceive the dominated investment option as an *intermediate* choice in both attributes and become more attractive by a compromise effect. If focusing on the inequity-averse viewpoint, the investment option could work as a *decoy* for the safe option which is an option to choose to do nothing.[4] Both the compromise and decoy effects predict that the demand for TPP would be smaller when both the punishment and the investment options are available relative to the demand for TPP when the investment option is not allowed. However, they provide different hypotheses for other behavior. If we observe the decreased demand for TPP in our experiment, we can distinguish which context effect has influenced a third party's punitive behavior by our experimental design.

The main finding of our experiment is that both the choice frequency and the expenditure of punishment significantly decrease when a third party is allowed to choose also the dominated investment option. We also find that the choice frequency and the expenditure of investment tend to increase when all options are available. The increased demand for investment does not support the hypothesis that the investment option is a decoy for the safe option. Besides, it is inconsistent with the argument that the reduced demand for TPP results from an experimenter demand effect caused by the difference in the number of available alternatives between treatments. Instead, our findings imply the existence of a substitution effect between punishment and investment and weakly support the hypothesis that the investment option works as a compromise.

Findings from our experiment contribute to the existing literature in manifold aspects. First,

[4] The detailed discussions are presented in Section 3.

we show how an individual's punitive behavior and context effects can be linked based on the recent findings from neuroeconomics. This suggests that context effects commonly studied in the areas of marketing and consumer decisions [23, 45, 46, 47] can also be well applied to the field of prosocial behavior. Moreover, to our knowledge, as this paper is the first study to examine the relationship between TPP and context effects, it extends the previous studies testing the robustness of TPP under numerous environments. Answering the question of whether TPP is robust under diverse circumstances is important to improve our understanding of human prosocial behavior. For instance, findings that dictator allocations are sensitive to experimental designs in dictator games suggest that rules of fair allocation in the game differ by context, are not broadly shared across individuals, and are often in conflict with self-interest.[5] This study provides evidence that TPP can be fragile under an environment where context effects are involved.

Second, our paper is particularly related to previous studies testing the robustness of TPP by enlarging a choice set of a third party. Nikiforakis and Mitchell [15] offer a third party an option to reward others in addition to the punishment option in a dictator game with TPP. The paper finds that the rewarding option works as a substitute for punishment because not rewarding when available can be perceived as an indirect way to punish a dictator who unequally allocated the endowment. Carpenter and Matthews [16] examine individuals' punishment behavior within a group (SPP) and between groups (TPP) simultaneously. In their experiment, the extent of TPP is small but still meaningful to sustain a high cooperation level along with SPP. Even if the extent of TPP decreases when other choices are available in these studies, they imply that TPP and alternatives related to prosocial behavior interact together and could achieve the stability of a society. Although we expand a choice set of a third party, the novelty of our experiment is that the additional option we provide is seemingly irrelevant to prosocial behavior and unattractive to most third parties in terms of material and psychological payoffs. Furthermore, unlike the previous studies, the substitution relationship between punishment and the alternative found in our experiment implies social instability because the investment option has no role in inducing people to follow social norms.

---

[5] In psychological terms, the dictator game having such properties is called a "weak situation" [78].

Third, extending the earlier discussion, our study is connected to studies concerning the low extent of TPP in the field. Guala [48] argued that there is scarce real-world evidence that voluntary material punishment, including SPP and TPP, exists. A well-known reason for the low extent of TPP is that it is sensitive to costs. For instance, the presence of a risk to be counter-punished by a violator significantly discourages a third party's willingness to punish the violator [49, 50, 51]. Goeschl and Jarke [52] incorporate monitoring costs for observing other players' choices in a prisoner's dilemma game and find that the amounts of both SPP and TPP decrease as the cost rises.[6] Our paper suggests another possible reason for the scarce evidence of TPP in the real world. That is, given that multiple types of alternatives exist in our daily life, a certain type of context effects can be associated with a third party's decision and eventually deters the motivation of TPP.

The remainder of this paper is organized as follows. Section 2 describes our experimental design. Section 3 presents two sets of hypotheses based on compromise and decoy effects. Section 4 analyzes the experimental results. Section 5 concludes the paper.

## 2. Experimental design and procedure

The experiment was conducted at SEE Lab (Sogang Experimental Economics Laboratory) at Sogang University, South Korea, from December 2017 to March 2018. It was programmed using the z-Tree software package [53]. A total of 318 undergraduate and graduate students majoring in various academic fields participated in the experiment. The total number of sessions was 16, and each session consisted of one of five treatments. Each subject participated only in one session.

Each treatment consists of two independent tasks: Task 1 and Task 2. All subjects faced the same Task 1 regardless of treatments. After all the subjects made decisions for Task 1, they faced Task 2. The various treatments in the experiment depended only on Task 2. The instructions for Task 2 were displayed after Task 1 was completed, and the contents of the instructions were different depending on treatments and roles in Task 2. The total payoff was accumulated for both tasks. However, the realization of the Task 1 outcome was conducted after all experiments were completed to mitigate a possible wealth effect.

[6] A few studies report a low extent of TPP in some environments. Nevertheless, it does not always mean the collapse of social norms as Nikiforakis and Mitchell [15] and Carpenter and Matthews [16] showed. For another example, Kamei [12] found that a third party's willingness to punish decreases when there are multiple third parties in a group. However, the aggregate punishment level of a group was large enough to attain high cooperation level.

The average duration of each treatment was 40 minutes. The conversion rate was 80 KRW (roughly, 0.08 USD) per token. The show-up fee was 3,000 KRW, and the average earning, including the show-up fee, was about 12,000 KRW.

### *2.1. Task 1: Risk preference elicitation*

In Task 1, we elicited subjects' risk preferences following Holt and Laury [54]. As shown in Table 1, Task 1 consists of individual decision problems of ten-paired lottery choices . The payoffs in Lottery L are less variable than the payoffs in Lottery R. Thus, the switching point from Lottery L to R represents the extent of risk aversion. For instance, switching at Question 5 represents a risk-neutral preference for any expected utility framework,[7] and a higher switching point represents a higher risk aversion.

### *2.2. Task 2: Modified dictator game with third-party punishment*

At the beginning of Task 2, subjects were randomly assigned to Player A, B, or C,[8] and three distinct types of subjects were randomly matched as a group. Player A (dictator), B (recipient), and C (third party) received endowments of 100, 0, and 50 tokens, respectively. Depending on Player C's decision problem, we implemented five different treatments denoted as punishment treatment (P treatment), punishment and investment with zero expected net return treatment (P&I0 treatment), investment with zero expected net return treatment (I0 treatment), punishment and investment with negative expected net return treatment (P&Ineg treatment), and investment with negative expected net return treatment (Ineg treatment).

---

[7] Technically, switching at Question 5 allows either risk-averse, risk-neutral, or risk-loving preferences. For instance, assuming a utility function $u = \frac{w^{1-r}}{1-r}$ $(r \neq 1)$, $u = \ln w$ $(r = 1)$, where $w$ is wealth and $r$ is a constant relative risk aversion (CRRA) coefficient, the switching point at Question 5 represents the CRRA coefficient interval $(-0.15, 0.15)$. However, since the extent of risk-loving and risk-averse is small, we define an individual who switches at Question 5 as risk-neutral. Moreover, further analysis in Section 4.3 mitigates this issue.

[8] All subjects participated in Task 1 regardless of their roles in Task 2.

Table 1. Ten pair-wise lottery choices in Task 1

| Choice question | Lottery L | Lottery R | E(L) −E(R) |
|---|---|---|---|
| 1 | (3,750, 1/10; 3,550, 9/10) | (8,000, 1/10; 100, 9/10) | 2,680 |
| 2 | (3,750, 2/10; 3,550, 8/10) | (8,000, 2/10; 100, 8/10) | 1,910 |
| 3 | (3,750, 3/10; 3,550, 7/10) | (8,000, 3/10; 100, 7/10) | 1,140 |
| 4 | (3,750, 4/10; 3,550, 6/10) | (8,000, 4/10; 100, 6/10) | 370 |
| 5 | (3,750, 5/10; 3,550, 5/10) | (8,000, 5/10; 100, 5/10) | −400 |
| 6 | (3,750, 6/10; 3,550, 4/10) | (8,000, 6/10; 100, 4/10) | −1,170 |
| 7 | (3,750, 7/10; 3,550, 3/10) | (8,000, 7/10; 100, 3/10) | −1,940 |
| 8 | (3,750, 8/10; 3,550, 2/10) | (8,000, 8/10; 100, 2/10) | −2,710 |
| 9 | (3,750, 9/10; 3,550, 1/10) | (8,000, 9/10; 100, 1/10) | −3,480 |
| 10 | (3,750, 10/10; 3,550, 0/10) | (8,000, 10/10; 100, 0/10) | −4,250 |

Note: This table was equally displayed on all subjects' computer monitors, except for the difference of two lotteries' expected payoffs. The unit of monetary value is KRW (Korean won). The switching point at Question 5 represents a risk-neutral preference. A higher switching point represents a risk-averse preference, and a lower switching point represents a risk-loving preference.

P treatment is a generic version of the dictator game with TPP as in Fehr and Fischbacher [24]. Player A could transfer 0, 10, 20, 30, 40, or 50 tokens to Player B. Then, Player C had an opportunity to impose deduction points on Player A, which is referred to as a "punishment option." One deduction point costs 1 token and reduces the opponent's tokens by 3. Thus, if Player C imposes X number of deduction points on Player A, then Player C's payoff is reduced by X tokens and Player A's payoff is reduced by 3 times X tokens. After purchasing deduction points, Player C's remaining tokens are kept in one's private account referred to as a "safe option." Hence, Player C's decision problem is to allocate the endowment to the punishment option and the safe option. Player C has a distinct choice set for each treatment, whereas Player A's set of choices is identical across all treatments.

In P&I0 treatment, we offered Player C an additional alternative, called an "investment option," which is similar to Gneezy and Potters [55]. The investment option is a simple lottery that costs one token per point and returns double with half probability or nothing with half probability. Thus, if Player C spends Y tokens on investment points, the payoff is decreased or increased by Y tokens with equal probability. In this treatment, Player C's decision problem is to allocate the endowment to the punishment option, the investment option, and the safe option.

In I0 treatment, we offered only the investment option and the safe option to Player C. Hence, Player C's decision problem in this treatment is to allocate the endowment to the investment option and the safe option.[9]

We additionally implemented P&Ineg and Ineg treatments where a lottery yields a negative expected net return. In particular, the structures of P&Ineg and Ineg treatments are identical to those of P&I0 and I0 treatments, respectively, except for the return rate of the investment option, which is reduced from 2 to 1.5. These two treatments were implemented to check the robustness of the results for the case where a possible indifference between the investment option and the safe option is excluded. Further, the supplementary treatments can remove the possibility of the safe option's non-dominance over investment option in terms of material payoffs. This possibility can occur because of either the risk elicitation method that we used in Task 1 or the discrepancy of risk attitudes from different risky tasks used in Tasks 1 and 2. Details are explained in Section 3. Table 2 summarizes available alternatives in each treatment.

Table 2. Choice set for each treatment

| Treatment | Available options | | |
|---|---|---|---|
| | Punishment | Investment | Expected net return of investment |
| P | Yes | No | — |
| P&I0 | Yes | Yes | Zero |
| I0 | No | Yes | Zero |
| P&Ineg | Yes | Yes | Negative |
| Ineg | No | Yes | Negative |

Note: The safe option is always available across all treatments.

In all treatments, we used a strategy method to observe Player C's choices. This means that Player C had to decide the number of deduction and investment points for all six cases of Player A's possible transfers, that is, 0, 10, 20, 30, 40, and 50 tokens.[10] To avoid unintended experimenter demand effects, we used neutral words in the instructions.[11] Furthermore, we clarified in the instruction that Player C can spend all, none, or partial amount of the endowment on deduction and/or investment points. We asked control questions to prevent decision-making without fully understanding the experiment. Specifically, subjects could move on to the actual decision-making screen only if they answered all the questions correctly. Those who wrote wrong answers had to read the instruction again and revise the answers correctly. Instructions and examples of control questions are provided in Appendix A.

[9] Meanwhile, Player B reported expectations on how many tokens Player A would transfer and how many tokens Player C would use for punishing Player A and/or purchase for investment points to an experimenter depending on the treatment. These guesses are not incentivized and did not affect anyone's material payoffs at all.

[10] Jordan et al. [17] find that the existence of TPP in a controlled laboratory setting is not an artifact of using the strategy method.

After Task 2, all the subjects were required to fill out a brief questionnaire, including questions on personality traits and socio-demographic information. After the questionnaire section, they privately received the final payoffs in cash in an enclosed envelope and left the laboratory.

## 3. Conceptual framework and hypotheses

In this section, we discuss the implication of the investment option as a compromise and a decoy. An important assumption here is that an individual has preferences over material and psychological payoffs. Thus, we regard material and psychological payoffs as two distinct attributes. In particular, an individual evaluates an alternative in terms of the material-payoff attribute and the psychological-payoff attribute separately and then combines them as a whole for the final decision.

We assume that an individual's risk attitude is involved when one evaluates risky alternatives based on the material-payoff attribute. Therefore, risk-neutral and risk-averse third parties would rank the safe option the best, the investment option the second-best, and the costly punishment option the worst in terms of the material-payoff attribute.

According to the literature, we can consider the psychological-payoff attribute from two different points of view. Given that TPP is driven by negative emotions, such as anger [17, 18, 19], and related to the activation of the brain regions linked to negative emotion processes, such as the anterior insula and amygdala [30, 34, 35], we can view the psychological-payoff attribute in terms of the "emotion-venting" aspect. Since inequity aversion is known as another motivation for TPP [12, 20, 21], we can also view the psychological-payoff attribute in terms of the "inequity-averse" aspect. By focusing on the emotion-venting viewpoint, the investment option can be perceived as a compromise. It can be perceived as a decoy if we focus on the inequity-averse viewpoint. We provide more explanations in the following.[12]

[11] For example, we indicated each subject as "Player A," "Player B," or "Player C" instead of calling "dictator," "recipient," or "third party." Likewise, we wrote "deduction points" when explaining the role of Player C and never used words, such as "punish" or "sanction."

First, consider the emotion-venting viewpoint. Here, we assume that an individual ranks an alternative higher if it is a good means for expressing one's emotion. Then the punishment option is the best as it is a good instrument to directly express one's negative emotions and oppositions after observing a dictator's unfair allocation [15, 56]. By contrast, the safe option would be the worst because keeping one's endowment in the private account means doing nothing. Given that a risky alternative could indirectly improve the emotional state with impulsive risk-taking behavior [57], we argue that the investment option is the second-best choice. Table 3 summarizes the rankings of all alternatives for each attribute. Considering each attribute separately, we deem the investment option unattractive to risk-neutral and risk-averse third parties because it is always dominated by either the punishment option or the safe option in each attribute. However, it is an intermediate choice among three alternatives. Therefore, combining both attributes, a compromise effect could play a role and induce those third parties to invest in the lottery rather than to punish a norm violator when all the three alternatives are available.

Table 3. Rankings of alternatives for each attribute

| Attribute | Rank | | |
|---|---|---|---|
| | 1st | 2nd | 3rd |
| Material-payoff | Safe | Investment | Punishment |
| Psychological-payoff: Emotion-venting | Punishment | Investment | Safe |
| Inequity-averse | Punishment | Safe | Investment |

Note: Punishment, Investment, and Safe indicate the punishment option, the investment option, and the safe option. Rank is an order of alternatives evaluated by risk-neutral and risk-averse third parties.

One may reasonably argue that, for some risk-neutral and risk-averse individuals, the investment option with zero expected net return is not dominated by the safe option in terms of the material-payoff attribute. One reason is from the risk elicitation method in Task 1. The way we define "risk-neutral" in Task 1 allows some extent of "risk-loving" attitudes. Thus, some "risk-loving" third parties, yet defined as "risk-neutral" in Task 1, would evaluate the investment option better than the safe option in terms of the material-payoff attribute in Task 2. Another reason is from the distinct risk elicitation methods in Tasks 1 and 2. Crosetto and Filippin [58] find that the degree of risk aversion elicited by Holt and Laury [54] tends to be higher than that by Gneezy and Potters [55]. We follow the design of Holt and Laury [54] in Task 1 and provide a lottery that is similar to Gneezy and Potters [55] in Task 2. Thus, an individual who is sorted as "risk-neutral" or even "risk-averse" in Task 1 may act like a "risk lover" in Task 2. Hence, those third parties would evaluate the investment option better than the safe option in terms of the material-payoff attribute. We also conduct P&Ineg and Ineg treatments offering a lottery with a negative expected net return for the investment option to deal with this issue. Those supplementary treatments allow us to test whether the results from P, P&I0, and I0 treatments are robust or not.

[12] Here, we informally describe intuitions of compromise and decoy effects in our experiment. In Appendix B.1, we formally demonstrate how the compromise effect works in our setting using a normalized contextual concavity model studied by Kivetz et al. [79]. In Appendix B.2, we use Saito [80] model to formally verify how the investment option can become a decoy for the safe option in our experiment.

Based on the compromise effect argument, we construct the following two hypotheses:

**Hypothesis 1.** *For the risk-neutral and the risk-averse, the choice frequency and the expenditure of punishment in P&I0 [P&Ineg] treatment are lesser than those in P treatment.*

**Hypothesis 2.** *For the risk-neutral and the risk-averse, the choice frequency and the expenditure of investment in P&I0 [P&Ineg] treatment are greater than those in I0 [Ineg] treatment.*

Second, we consider the psychological-payoff attribute in the inequity-averse point of view instead of the emotion-venting point of view. Briefly speaking, the punishment option would be ranked the best in this case because it effectively reduces the degree of payoff inequalities among players. In contrast, the investment option would be ranked the worst because it can make the player become a highest-payoff player or the lowest-payoff player in a group. This extreme position harms the investor in terms of the inequity aversion aspect. Hence, as Table 3 summarizes, the investment option is always worse than the safe option under all attributes for both risk-neutral and risk-averse third parties. This implies that the investment option becomes a decoy for the safe option. The decoy effect suggests that the safe option would be more attractive when all three options are available than when the investment option is not allowed to choose. Due to the increased demand for the safe option, the willingness to punish would decrease when all options are available. The same logic applies to the investment option with a negative expected net return. Based on the decoy effect argument, we construct the following hypotheses:

**Hypothesis 3.** *For the risk-neutral and the risk-averse, the choice frequency and the expenditure of punishment in P&I0 [P&Ineg] treatment are lesser than those in P treatment.*

**Hypothesis 4.** *For the risk-neutral and the risk-averse, the choice frequency and the expenditure of the safe option in P&I0 [P&Ineg] treatment are larger than those in P treatment.*

Note that Hypotheses 1 and 3 are similar regarding TPP, meaning both emotion-venting and inequity-averse viewpoints agree that the demand for TPP would decrease if the investment option is additionally available. While the punishment behavior is our main interest, we can discern which context effect is likely to influence a third party's willingness of TPP by testing Hypotheses 2 and 4.

## 4. Results

In this section, we test our hypotheses using experimental data. Table 4 displays the total number of third parties and their risk attitudes that were elicited in Task 1. We omit observations whose risk preference is inconsistent[13] and focus on risk-neutral and risk-averse Player Cs throughout the analysis.[14]

[13] We define an individual to have inconsistent risk preferences if one makes a switch from Lottery B to A or chooses Lottery A in the last choice problem in Task 1.

[14] The results presented in Section 4 are not qualitatively affected even if we consider the entire subjects regardless of their choices in Task 1 (see Appendix C.1) or subjects whose risk preferences are consistent (see Appendix C.2). In addition, analyses on the transfer behavior of risk-consistent dictators are presented in Appendix D.

Table 4. Sorting third parties based on risk attitudes

| | | Treatment | | | | | Total |
|---|---|---|---|---|---|---|---|
| | | P | P&I0 | I0 | P&Ineg | Ineg | |
| Number of third-parties | | 27 | 28 | 23 | 13 | 15 | 106 |
| Risk attitude: | Risk-neutral | 7 | 9 | 7 | 3 | 4 | 30 |
| | Risk-averse | 16 | 13 | 12 | 7 | 7 | 55 |
| | Risk-loving | 4 | 4 | 3 | 3 | 4 | 18 |
| | Risk-inconsistent | 0 | 2 | 1 | 0 | 0 | 3 |

Note: Risk attitudes were elicited in Task 1. None of the differences in the distributions of risk attitudes between any two treatments are statistically significant based on the Kolmogorov–Smirnov test (exact p>0.542 for all pair-wise cases).

### *4.1. Fragility of TPP*

Figure 1a reports that, in P treatment, more than 60% (n=23) of risk-neutral and risk-averse Player Cs imposed a positive number of deduction points on Player A who transferred nothing to Player B. The percentage is constant until the transfer level of 20 is reached, and it starts to decline and approaches zero at the fair allocation. The pattern also appears in P&I0 treatment; roughly 40% (n=22) of risk-neutral and risk-averse Player Cs punished Player A at the transfer level of zero, and the percentage approaches zero at the transfer level of 50. Figure 1b reports that, on average, Player C assigned about 15 deduction points in P treatment and 5 deduction points in P&I0 treatment to Player A who transferred nothing. The expenditures monotonically decrease and converge to zero as the transfer level increases in both treatments.

Despite the similar punishing pattern between the two treatments, the percentage points and expenditures considerably differ at unfair transfer levels. We find that the percentage of punishers is significantly lower in P&I0 treatment than the percentage in P treatment at the transfer levels of 10, 20, and 30 at the significance level of 5% (p=[0.076, 0.038, 0.017, 0.013, 0.071, 1.000], two-sided Fisher's exact test)[15]. The expenditure of deduction points in P&I0 treatment is also significantly smaller than the expenditure in P treatment for each unfair transfer level (p=[0.026, 0.009, 0.003, 0.003, 0.031, 0.328], two-sided Wilcoxon rank-sum test). Hence, our data show that a third party's willingness to punish a dictator who allocated the endowment unequally significantly decreases when the dominated investment option is available.

[15] p is a vector collecting computed p-values at transfer levels of 0, 10, 20, 30, 40, and 50.

For a thorough examination of the diminished demand for TPP, we also conduct a regression analysis. First, we regress *Punisher* on *Transfer*, *Treatment,* and other controlling variables[16] using the linear probability model (LPM) and the Probit model. *Punisher* is defined as 0 if Player C assigned no deduction point and 1 if Player C imposed any positive number of deduction points on Player A. *Transfer* contains 0 to 50 with 10 increments. *Treatment* is 0 if the treatment is P treatment and 1 if P&I0 treatment. Columns 1 and 2 in Table 5 show that the coefficients of *Treatment* are negative and significantly different from zero. This implies that the probability of being a punisher is lower in P&I0 treatment than in P treatment. In addition, we define *Punishment* as the token amounts Player C spent for imposing deducting points on Player A. We regress *Punishment* on *Transfer*, *Treatment,* and other controlling variables using the ordinary least squares (OLS) and Tobit models. Columns 3 and 4 in Table 5 show that the coefficients of *Treatment* are negative and significantly different from zero, which implies that the amount of punishment is smaller in P&I0 treatment than in P treatment.[17]

[16] Controlling variables include *Switching point*, socio-demographic, and Big Five personality traits. *Switching point* is the question number where an individual switches the choice from Lottery L to R in Task 1. Socio-demographic variables include *Gender*, *Income*, and *Economics major*.

[17] For further analysis, we include an *Interaction* term, a product of *Transfer* and *Treatment*, as an independent variable in the models used in Tables 5 and 7. We find that the *Interaction* term does not qualitatively affect the regression results reported in Section 4.1 (see Appendix C.3).

Figure 1. Punishment and investment in P, I0, and P&I0 treatments (with risk-neutral and risk-averse)

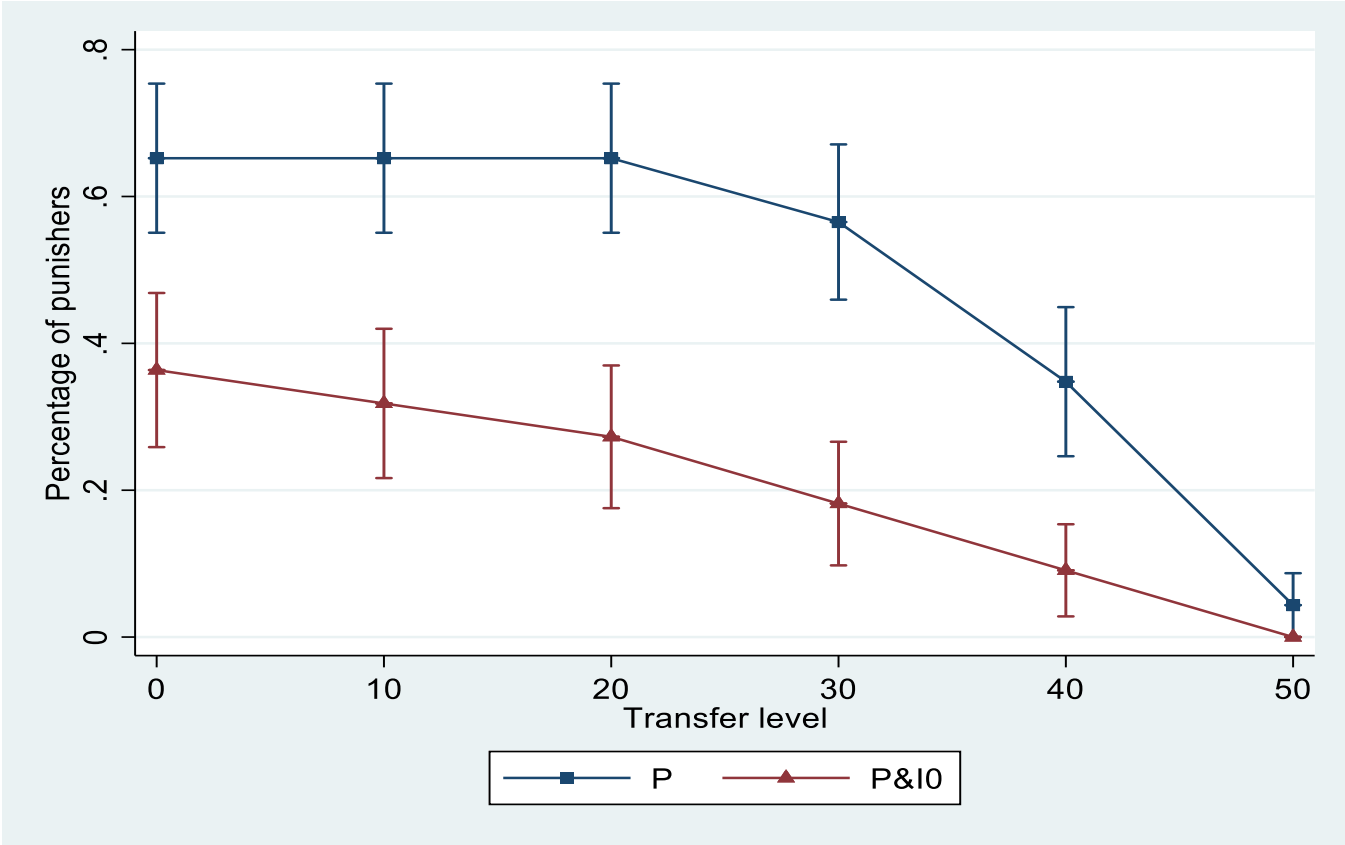

**a.** Percentage of punishers

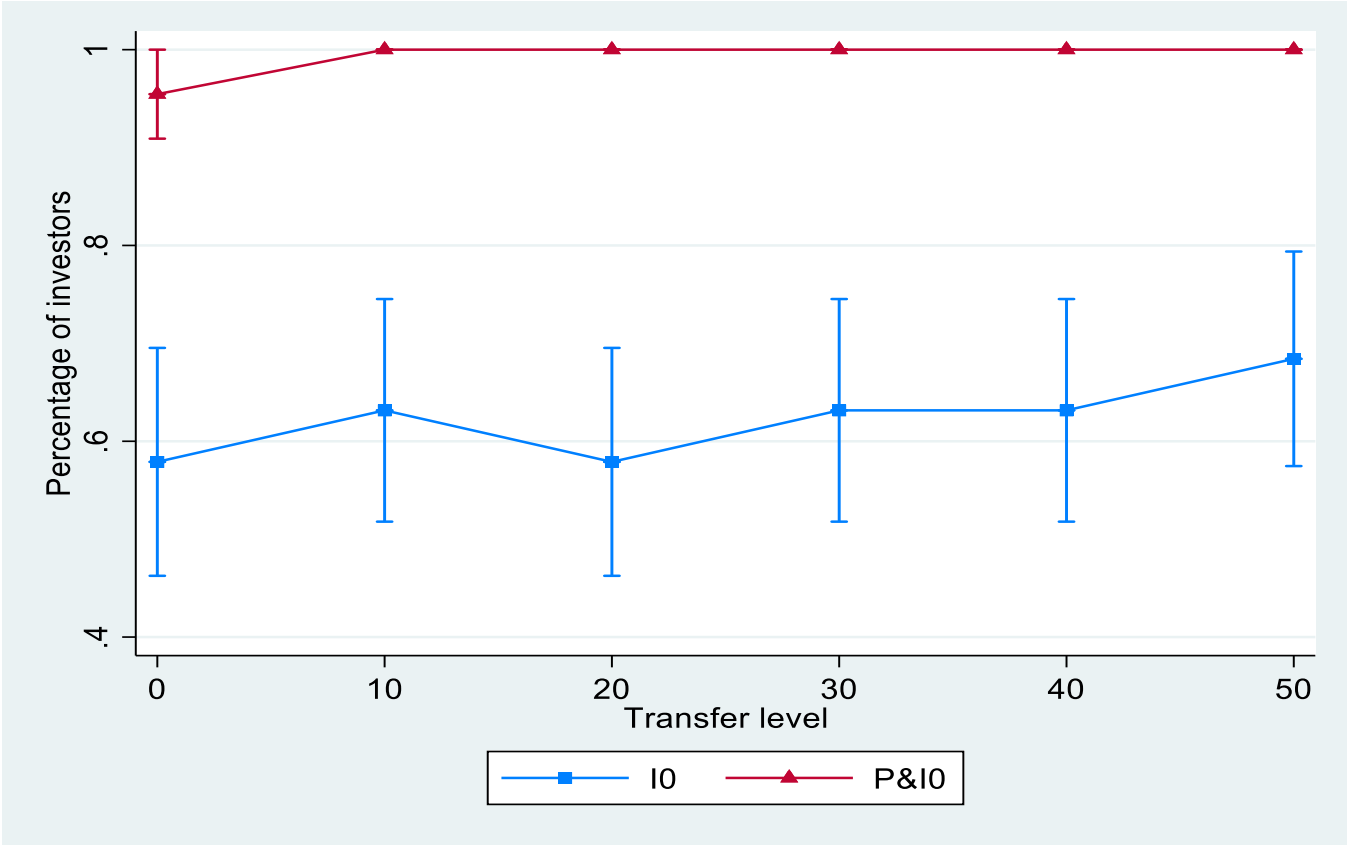

**c.** Percentage of investors

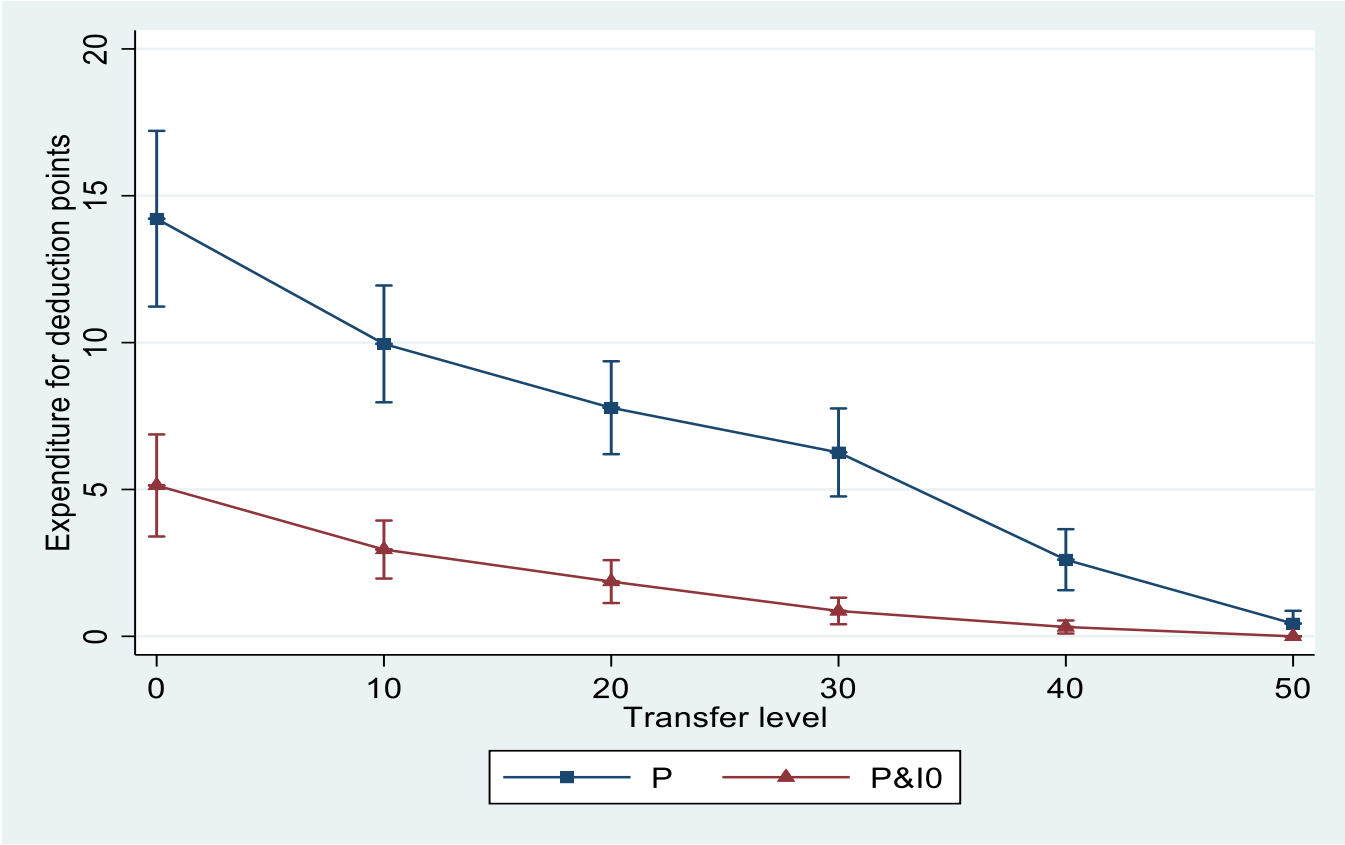

**b.** Average expenditure for deduction points

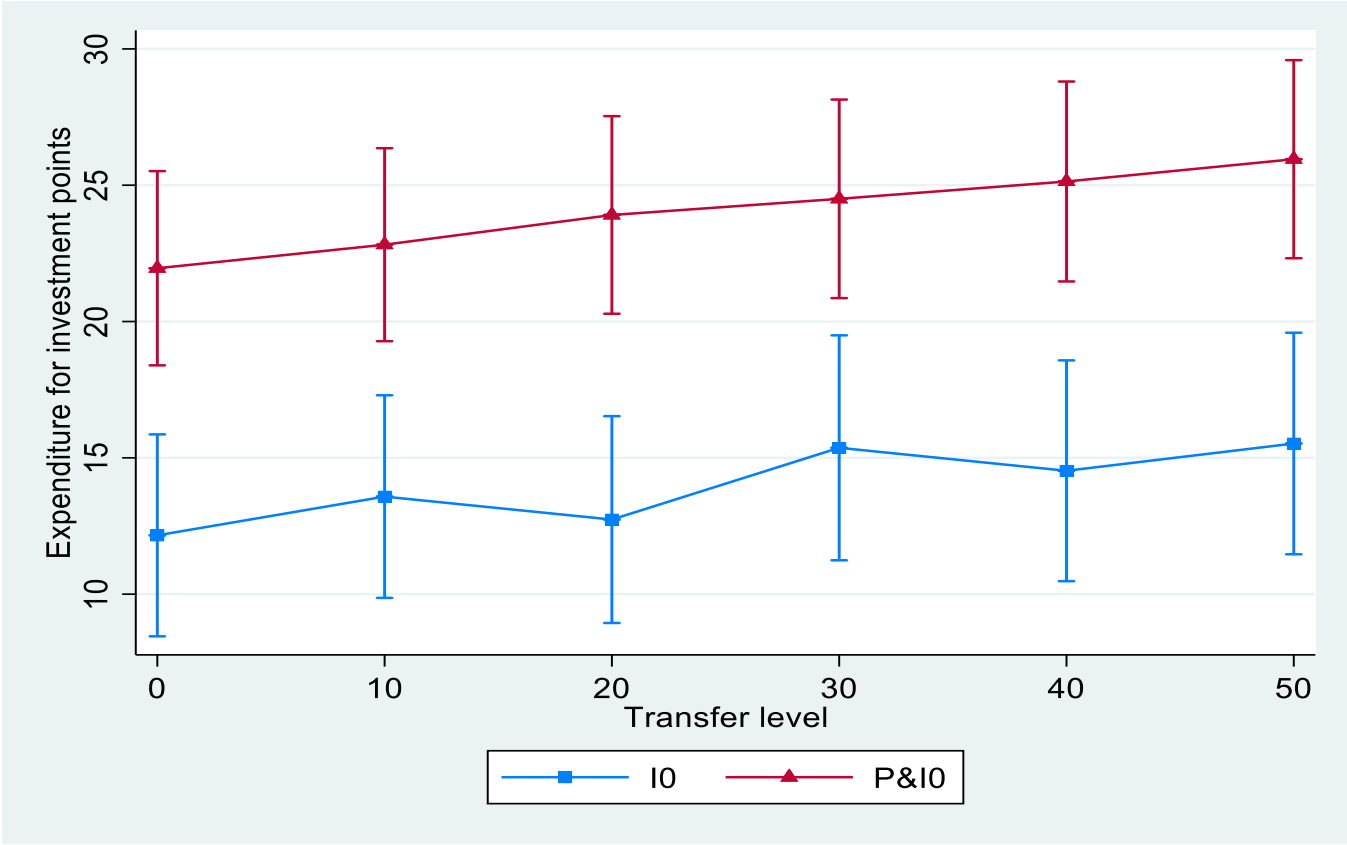

**d.** Average expenditure for investment points

Note: A punisher [investor] is defined as Player C who spent at least one token to deduction [investment] points. Expenditure for deduction [investment] points is the number of tokens that Player C spent for deduction [investment] points among 50 tokens. Transfer level is the number of tokens that Player A transferred to Player B. A vertical bar indicates plus and minus one standard error.

Table 5. Regression on punishment (with risk-neutral and risk-averse)

| Treatment: | P, P&I0 | | | | | | | |
|---|---|---|---|---|---|---|---|---|
| Level of data: | Transfer level | | Transfer level | | Individual level | | Individual level | |
| Model: | LPM | Probit | OLS | Tobit | OLS | Tobit | OLS | Tobit |
| Dependent variable: | Punisher | | Punishment | | Mean punishment | | Median punishment | |
| | (1) | (2) | (3) | (4) | (5) | (6) | (7) | (8) |
| Transfer | −0.010*** | −0.038*** | −0.183*** | −0.509*** | — | — | — | — |
| | (0.002) | (0.005) | (0.034) | (0.056) | — | — | — | — |
| Treatment: P&I0 | −0.283*** | −1.210*** | −4.567*** | −14.594*** | −4.567** | −8.646*** | −5.284*** | −11.640*** |
| | (0.094) | (0.358) | (1.507) | (3.986) | (1.774) | (2.998) | (1.826) | (3.452) |
| Switching point | −0.052 | −0.234 | −0.117 | −1.765 | −0.117 | −0.943 | −0.487 | −2.018 |
| | (0.045) | (0.155) | (0.672) | (1.528) | (0.768) | (1.165) | (0.790) | (1.312) |
| Constant | 0.175 | −1.532 | 2.023 | −12.680 | −2.561 | −14.444 | 2.404 | −8.965 |
| | (0.588) | (2.099) | (10.733) | (22.382) | (9.590) | (14.964) | (9.869) | (16.155) |
| Socio-demographic | Yes | Yes | Yes | Yes | Yes | Yes | Yes | Yes |
| Personality traits | Yes | Yes | Yes | Yes | Yes | Yes | Yes | Yes |
| Observations | 270 | 270 | 270 | 270 | 45 | 45 | 45 | 45 |
| Left-censored | — | — | — | 176 | — | 22 | — | 24 |
| Right-censored | — | — | — | 1 | — | 0 | — | 0 |
| Number of individuals | 45 | 45 | 45 | 45 | 45 | 45 | 45 | 45 |
| (Pseudo) R-squared | 0.326 | 0.317 | 0.334 | 0.132 | 0.327 | 0.099 | 0.355 | 0.118 |

Note: The dependent variable in Tobit models is truncated at 0 (left-censored) and 50 (right-censored). Socio-demographic variables include gender, income, and Economics major. In columns 1–4, clustered standard errors at the individual level are in parentheses. In columns 5–8, standard errors are in parentheses. ***, **, and * denote significance at 1%, 5%, and 10%, respectively.

Table 6. Punishment and investment using individual-level data

| | Punishment | | | | Investment | | | |
|---|---|---|---|---|---|---|---|---|
| | Risk-neutral and Risk-averse | | | | Risk-neutral and Risk-averse | | | |
| | Mean | p-value | Median | p-value | Mean | p-value | Median | p-value |
| P [I0] | 6.88 | | 7.02 | | 13.98 | | 13.63 | |
| | (6.89) | 0.012 | (7.26) | 0.002 | (16.44) | 0.024 | (16.68) | 0.018 |
| P&I0 | 1.86 | | 1.36 | | 24.05 | | 24.20 | |
| | (2.89) | | (2.59) | | (16.72) | | (17.02) | |

Note: Mean and Median refer to averages using mean and median individual-level data, respectively. Standard deviations are in parentheses. p-values are from the two-sided Wilcoxon rank-sum test.

Moreover, we examine Player C's punitive behavior using subjects' individual-level data. We define a subject's *mean punishment* [*mean investment*] as one's mean expenditure for deduction [investment] points across all transfer levels. Similarly, a subject's *median punishment* [*median investment*] is defined as one's median expenditure for deduction [investment] points across all transfer levels. Analyses of these individual-level data are summarized in Table 6. The averages of mean [median] punishment are 6.88 [7.02] in P treatment and 1.86 [1.36] in P&I0 treatment, and the difference is statistically significant (p=0.012 [p=0.002], two-sided Wilcoxon rank-sum test). Columns 5–8 in Table 5 report the results of the OLS and the Tobit regressions using the individual-level data. All coefficients of *Treatment* are negative and significantly different from zero. Thus, results using the individual-level data also support the reduced demand for TPP.

As discussed in Section 3, some risk-neutral Player Cs may not perceive the investment option to be strictly dominated by the safe option in terms of the material-payoff attribute. Hence, whether we still observe similar punishing behavior by considering the risk-averse only is worth checking. An analysis reported in Appendix C.4 confirms that the demand for TPP decreases even if we only consider risk-averse third parties. We summarize these findings as follows:

**Result 1.** *Both the frequency of punisher and the expenditure of deduction points decrease when the dominated investment option whose expected net return is zero is available in addition to the punishment option and the safe option. Thus, Hypothesis 1and 3 are supported.*

One may argue that an experimenter demand effect of the different number of available alternatives induces the diminished demand for TPP. Specifically, to avoid doing nothing, the third party in P&I0 treatment has two options: the punishment option and the investment option.

Here, doing nothing means allocating the entire tokens to the safe option. Meanwhile, the third party in P treatment has only the punishment option to avoid doing nothing. Then the experimenter demand effect suggests the possibility that the narrower choice set offered in P treatment could induce Player C to spend more tokens for deduction points than in P&I0 [P&Ineg] treatment because one might feel awkward about doing nothing during the experiment. If this is a convincing explanation for Result 1, the investment demand should also be lower in P&I0 [P&Ineg] treatment than in I0 [Ineg] treatment because the choice set in I0 [Ineg] treatment is more restricted. In contrast, Hypothesis 2 predicts that the investment demand should be larger in P&I0 [P&Ineg] treatment than in I0 [Ineg] treatment. The following subsection provides evidence that the investment behavior is indeed consistent with Hypothesis 2, suggesting that the diminished demand for TPP is not an artifact of the experimenter demand effect.

*4.2. Increased demand for investment*

Figure 1c reports that, for all transfer levels, roughly 60% of risk-neutral and risk-averse Player Cs purchased a positive number of investment points in I0 treatment, whereas almost all Player Cs purchased a positive number of investment points in P&I0 treatment. The difference in the proportions between the treatments is statistically significant at all transfer levels at the significance level of 5% (p=[0.006, 0.002, 0.001, 0.002, 0.002, 0.006], two-sided Fisher's exact test). Figure 1d reports that Player C purchased about 14 investment points, on average, for all transfer levels in I0 treatment. In P&I0 treatment, Player C spent about 22 points, on average, at the transfer level zero and monotonically increased the expenditure as the transfer level rises. For each transfer level, the difference in the expenditures between the treatments is statistically significant (p=[0.017, 0.029, 0.012, 0.041, 0.012, 0.020], two-sided Wilcoxon rank-sum test). Thus, our data show a higher percentage of investors and more expenditures for investment points when all options are available than when only the investment and safe options are available to Player C.[18]

[18] Interesting observations here are the strictly positive percentage of investors and investment expenditure in I0 and Ineg treatments. Recent studies of preferences for randomization [59, 60, 61, 62, 63] may partially rationalize this investment behavior. For example, if a third party is willing to randomize or mix over safe and risky alternatives, then one might prefer to allocate some positive amount of tokens to the dominated investment option regardless of one's risk attitude. Yet, those preferences do not predict the increased investment demand when the punishment option is available.

A regression analysis partially confirms the increased demand for investment. We define *Investor* as 0 if Player C did not invest in the lottery at all, and 1 if Player C purchased any positive number of investment points. *Investment* is defined as the token amount Player C spent for investment points. Independent variables contain *Transfer, Treatment*, and other controlling variables. Here, *Treatment* is defined as 0 if I0 treatment and 1 if P&I0 treatment. The regression results of *Investor* using the LPM and the Probit model are presented in columns 1 and 2 in Table 7. The coefficients of *Treatment* are positive and significantly different from zero, which means that the probability of investing in the lottery is higher in P&I0 treatment than in I0 treatment. The regression results of *Investment* using the OLS and Tobit models are presented in columns 3 and 4 in Table 7. The coefficients of *Treatment* are positive in both models, and they are significantly different from zero if the censored data are controlled through the Tobit model. This implies that the expenditure of investment in P&I0 treatment is higher than in I0 treatment.[19]

We also examine Player C's investment behavior using the individual-level data. According to Table 6, the averages of mean [median] investment are 13.98 [13.63] and 24.05 [24.20] in I0 and P&I0 treatments, respectively, and they are significantly different (p=0.024 [p=0.018], two-sided Wilcoxon rank-sum test). The OLS and Tobit regression results using the individual-level data are presented in columns 5–8 in Table 7. The coefficients of *Treatment* are all positive, and they are significantly different from zero if the censored median investment individual-level data are controlled through the Tobit model. A further analysis presented in Appendix C.4 weakly confirms the increased demand for investment even if we exclude the risk-neutral and consider the risk-averse only. The findings are summarized as follows:

**Result 2.** *Both the frequency of investors and the expenditure of investment points increase, or at least do not decrease, when all the options are available. Thus, Hypothesis 2 is supported.*

[19] Similar to Section 4.1, *Interaction* term does not qualitatively affect the results (see Appendix C.3).

Table 7. Regression on investment (with risk-neutral and risk-averse)

| Treatment: | I0, P&I0 | | | | | | | |
|---|---|---|---|---|---|---|---|---|
| Level of data: | Transfer level | | Transfer level | | Individual level | | Individual level | |
| Model: | LPM | Probit | OLS | Tobit | OLS | Tobit | OLS | Tobit |
| Dependent variable: | Investor | | Investment | | Mean investment | | Median investment | |
| | (1) | (2) | (3) | (4) | (5) | (6) | (7) | (8) |
| Transfer | 0.001 | 0.010* | 0.072*** | 0.117*** | — | — | — | — |
| | (0.001) | (0.006) | (0.026) | (0.043) | — | — | — | — |
| Treatment: P&I0 | 0.343*** | 2.054*** | 7.392* | 14.578** | 7.392 | 11.234* | 7.742 | 14.924** |
| | (0.091) | (0.594) | (3.896) | (5.819) | (5.012) | (5.709) | (5.045) | (6.349) |
| Switching point | 0.024 | 0.145 | −0.908 | −0.507 | −0.908 | −0.864 | −0.860 | 0.316 |
| | (0.038) | (0.214) | (1.637) | (2.568) | (1.844) | (2.079) | (1.856) | (2.291) |
| Constant | 0.070 | −1.039 | −37.870** | −74.472** | −36.075* | −52.256** | −38.602* | −78.507** |
| | (0.427) | (1.590) | (18.053) | (29.031) | (21.068) | (24.345) | (21.210) | (29.372) |
| Socio-demographic | Yes | Yes | Yes | Yes | Yes | Yes | Yes | Yes |
| Personality traits | Yes | Yes | Yes | Yes | Yes | Yes | Yes | Yes |
| Observations | 246 | 246 | 246 | 246 | 41 | 41 | 41 | 41 |
| Left-censored | — | — | — | 44 | — | 5 | — | 7 |
| Right-censored | — | — | — | 44 | — | 6 | — | 6 |
| Number of individuals | 41 | 41 | 41 | 41 | 41 | 41 | 41 | 41 |
| (Pseudo) R-square | 0.405 | 0.504 | 0.512 | 0.114 | 0.533 | 0.112 | 0.544 | 0.126 |

Note: The dependent variable in Tobit models is truncated at 0 (left-censored) and 50 (right-censored). Socio-demographic variables include gender, income, and Economics major. In columns 1–4, clustered standard errors at individual level are in parentheses. In columns 5–8, standard errors are in parentheses. ***, **, and * denote significance at 1%, 5%, and 10%, respectively.

Results 1 and 2 clearly show that our findings are not an artifact of the experimenter demand effect induced by the different number of alternatives. Instead, a substitution effect between punishment and investment seems to exist, thereby supporting Hypothesis 2 that the investment option works as a compromise.

To examine a substitution effect between punishment and investment in more detail, we regress *Investment* on *Transfer* and other controlling variables in P&I0 and I0 treatments, respectively. Given that *Punishment* is negatively correlated with *Transfer* (see Table 5), *Investment* and *Transfer* in P&I0 treatment would be positively correlated if the substitution effect exists. This positive correlation between *Investment* and *Transfer* would disappear in I0 treatment because the punishment option that connects those two variables is no longer available. Indeed, as presented in columns 1–4 in Table 8, we find a significantly positive correlation in P&I0 treatment but an insignificant correlation in I0 treatment.

Another way to verify the substitution effect is to compare a non-punisher's investment expenditure in P&I0 treatment with the investment expenditure in I0 treatment. Note that being a non-punisher in P&I0 treatment has two possible reasons. One is that a non-punisher is actually a third party who never chooses to punish in any circumstance. Here, the non-punisher's decision problem in P&I0 treatment should be identical to the decision problem in I0 treatment because the availability of punishment has no impact. The other reason is the substitution effect; that is, a third party reduces punishment (and becomes a non-punisher) to allocate more tokens to the investment option when it is available. Consequently, if the substitution effect exists, a non-punisher's investment expenditure would be relatively higher in P&I0 treatment than in I0 treatment; otherwise, the investment expenditures between the treatments would have no difference. Our data show that a non-punisher's investment expenditure in P&I0 treatment is significantly higher than the expenditure in I0 treatment at all transfer levels (p=[0.007, 0.028, 0.020, 0.034, 0.014, 0.020], two-sided Wilcoxon rank-sum test). We also regress *Investment* conditional on a non-punisher on *Treatment* and other variables. Columns 5 and 6 in Table 8 show that the coefficients of *Treatment* are positive, and they are significantly different from zero if the censored data are controlled through the Tobit model. This means that the investment expenditure of a non-punisher is higher in P&I0 treatment than the expenditure in I0 treatment.

Table 8. The substitution effect (with risk-neutral and risk-averse)

| Treatment: | P&I0 | | I0 | | P&I0, I0 | |
|---|---|---|---|---|---|---|
| Level of data: | Transfer level | | Transfer level | | Transfer level | |
| Model: | OLS | Tobit | OLS | Tobit | OLS | Tobit |
| Dependent variable: | Investment | | Investment | | Investment (conditional on a non-punisher) | |
| | (1) | (2) | (3) | (4) | (5) | (6) |
| Transfer | 0.079** | 0.110** | 0.064 | 0.132 | 0.060* | 0.102** |
| | (0.032) | (0.047) | (0.045) | (0.085) | (0.032) | (0.048) |
| Treatment: P&I0 | — | — | — | — | 7.313* | 14.904** |
| | — | — | — | — | (3.741) | (6.237) |
| Switching point | 2.554 | 4.753 | −0.344 | −2.861 | −1.500 | −1.774 |
| | (2.240) | (3.297) | (2.562) | (5.030) | (1.832) | (3.066) |
| Constant | −63.310 | −97.166 | −27.938 | −77.174** | −38.029** | −76.836** |
| | (46.596) | (62.429) | (19.408) | (35.033) | (18.375) | (30.025) |
| Socio-demographic | Yes | Yes | Yes | Yes | Yes | Yes |
| Personality traits | Yes | Yes | Yes | Yes | Yes | Yes |
| Observations | 132 | 132 | 114 | 114 | 219 | 219 |
| Left-censored | — | 1 | — | 43 | — | 43 |
| Right-censored | — | 29 | — | 15 | — | 44 |
| Number of individuals | 22 | 22 | 19 | 19 | 41 | 41 |
| (Pseudo) R-squared | 0.497 | 0.103 | 0.537 | 0.124 | 0.556 | 0.127 |

Note: The dependent variable in Tobit models is truncated at 0 (left-censored) and 50 (right-censored). Socio-demographic variables include gender, income, and Economics major. Clustered standard errors at individual level are in parentheses. ***, **, and * denote significance at 1%, 5%, and 10%, respectively.

The existence of a substitution effect between punishment and investment seems consistent with the argument that the investment option works as a compromise instead of a decoy. In fact, from our data, we find no evidence supporting Hypothesis 4 predicting the demand for the safe option increases when the investment option is available. In P treatment, 96% or all of the risk-neutral and risk-averse third parties spent at least one token for the safe option at the transfer level of 0 or the remaining transfer levels, respectively. In P&I0 treatment, 73% of those third parties spent at least one token for the safe option at all transfer levels. The difference in proportions between the treatments for each transfer level is even statistically significant (p=[0.047, 0.009, 0.009, 0.009, 0.009, 0.009], two-sided Fisher's exact test). For the safe option expenditure, risk-neutral and risk-averse third parties spent 35 to 47 tokens on average across all transfer levels in P treatment, whereas those third parties spent 22 to 24 tokens on average across all transfer levels in P&I0 treatment. The difference in expenditures between the treatments at each transfer level is also statistically significant (p=[0.008, 0.001, 0.001, 0.001, 0.001, 0.001], two-sided Wilcoxon rank-sum test). The individual-level data also support the aforementioned findings. Define a subject's *mean [median] safe* as one's mean [median] expenditure for the safe option across all transfer levels. The averages of mean [median] safe are 43.12 [42.98] in P treatment and 24.10 [24.39] in P&I0 treatment. The difference in mean [median] safe expenditures is statistically significant (p<0.001 [p<0.001], two-sided Wilcoxon rank-sum test). Therefore, in our experiment, we find that the safe option losses its attraction if the investment option is additionally available, which is inconsistent with Hypothesis 4. The following summarizes the findings:

**Result 3.** *A substitution effect exists between punishment and investment, supporting the investment option works as a compromise consistent with the emotion-venting motivation rather than a decoy consistent with the inequity-averse motivation. Thus, Hypothesis 2 is supported, whereas Hypothesis 4 is not.*

*4.3. Robustness check using the supplementary treatments*

As discussed in Section 3, some Player Cs who are sorted as "risk-neutral" or even "risk-averse" in Task 1 might act as if they are "risk lover" in Task 2. Then the investment option with zero expected net return is no longer dominated by the safe option to those third parties. To deal with this issue, we conducted supplementary treatments, namely, P&Ineg and Ineg treatments, where the lottery yields a negative expected net return. Note that the investment option offered in these treatments should be dominated by the safe option in terms of the material-payoff attribute even for moderately risk-loving third parties.

Figures 2a and 2b show that risk-neutral and risk-averse Player Cs are less likely to punish Player A in P&Ineg treatment than in P treatment. The percentage of punishers is significantly lower in P&Ineg treatment than in P treatment at transfer levels of 0 to 30 at the significance level of 5% (p=[0.026, 0.026, 0.026, 0.002, 0.071, 1.000], two-sided Fisher's exact test). The expenditure for deduction points is significantly smaller in P&Ineg treatment than in P treatment for each unfair transfer level (p=[0.026, 0.017, 0.020, 0.004, 0.037, 0.510], two-sided Wilcoxon rank-sum test). Table 9 presents the averages of punishment using the individual-level data. The averages of mean [median] punishment are 6.88 [7.02] and 1.25 [0.25] in P and P&Ineg treatments, respectively, and the difference is statistically significant (p=0.019 [p=0.003], two-sided Wilcoxon rank-sum test). Related regression results presented in Appendix C.5 are consistent with the diminished demand for punishment. Thus, Result 1 is supported even if the expected net return of the investment option is negative.

Next, Figure 2c shows that the proportion of investors is higher in P&Ineg treatment than in Ineg treatment for each transfer level, and the difference is significant at the transfer level of 10 (p=[0.260, 0.024, 0.135, 0.063, 0.063, 0.063], two-sided Fisher's exact test). Figure 2d show that the expenditure of investment is also higher in P&Ineg treatment than in Ineg treatment for each transfer level, and the difference is significant at the transfer level of 10 (p=[0.303, 0.028, 0.135, 0.118, 0.148, 0.147], two-sided Wilcoxon rank-sum test). Table 9 shows that the averages of mean [median] investment are 6.12 [4.86] and 15.08 [15.00] in Ineg and P&Ineg treatments, respectively, where the differences are insignificant (p=0.131 [p=0.080], two-sided Wilcoxon rank-sum tests). Despite several insignificant statistical results, our data confirm that risk-neutral and risk-averse Player Cs are directionally more willing to invest in P&Ineg treatment than in Ineg treatment. Related regression results reported in Appendix C.5 support these findings. Lastly, Appendix C.6 shows a substitution effect between punishment and investment under the supplementary treatments, and Appendix C.7 finds no evidence that the investment option with a negative expected net return works as a decoy for the safe option. Thus, Results 2 and 3 are weakly supported by the supplementary treatments.

Table 9. Punishment and investment using individual-level data

| | Punishment<br>Risk-neutral and Risk-aversion | | | | Investment<br>Risk-neutral and Risk-aversion | | | |
|---|---|---|---|---|---|---|---|---|
| | Mean | p-value | Median | p-value | Mean | p-value | Median | p-value |
| P [Ineg] | 6.88 | 0.019 | 7.02 | 0.003 | 6.12 | 0.131 | 4.86 | 0.080 |
| | (6.89) | | (7.26) | | (9.56) | | (9.16) | |
| P&Ineg | 1.25 | | 0.25 | | 15.08 | | 15.00 | |
| | (2.12) | | (0.79) | | (19.20) | | (19.44) | |

Note: Mean and Median refer to an average using mean and median individual-level data, respectively. Standard deviations are in parentheses. p-values are from the two-sided Wilcoxon rank-sum test.

Figure 2. Punishment and investment in P, Ineg and P&Ineg treatments (with risk-neutral and risk-averse)

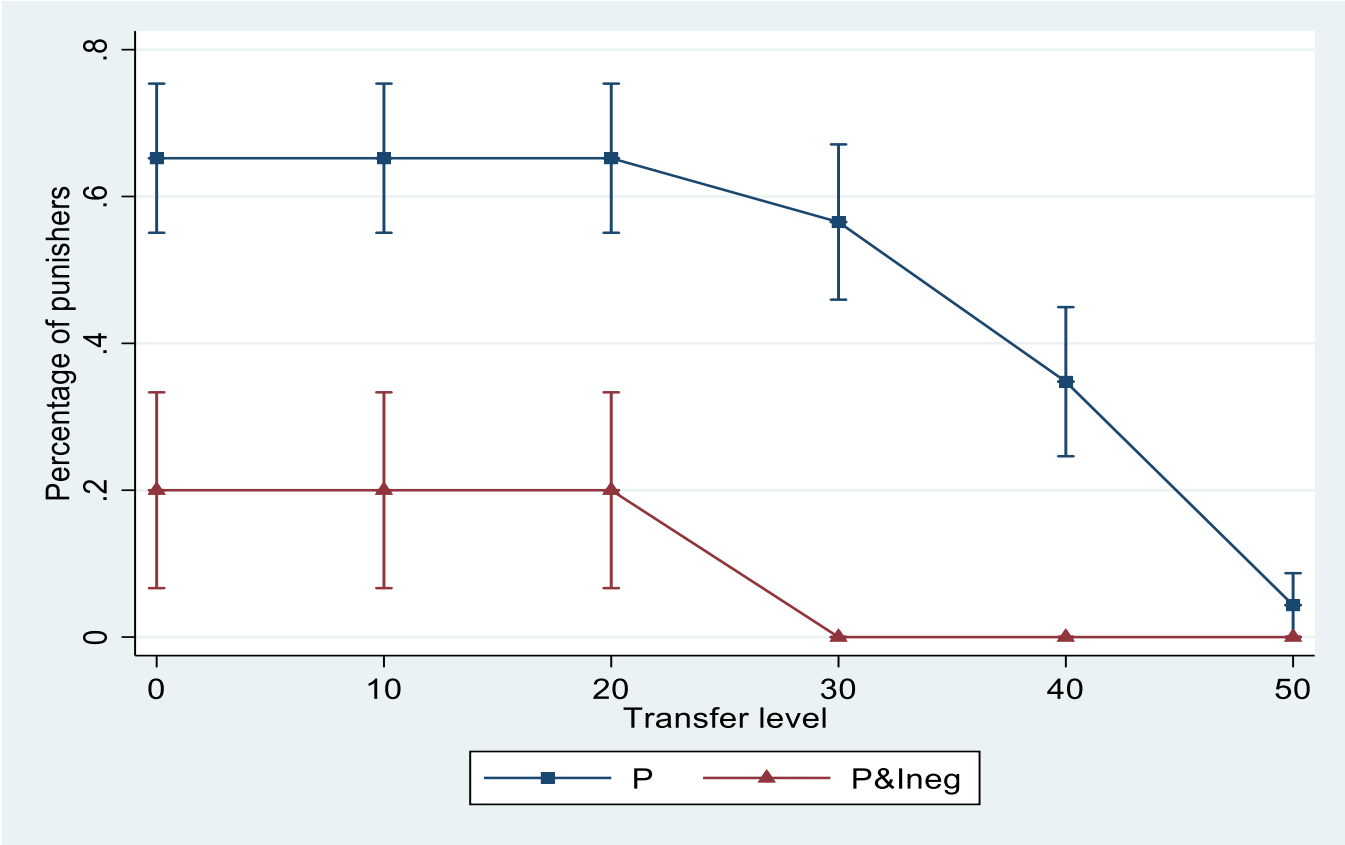

**a.** Percentage of punishers

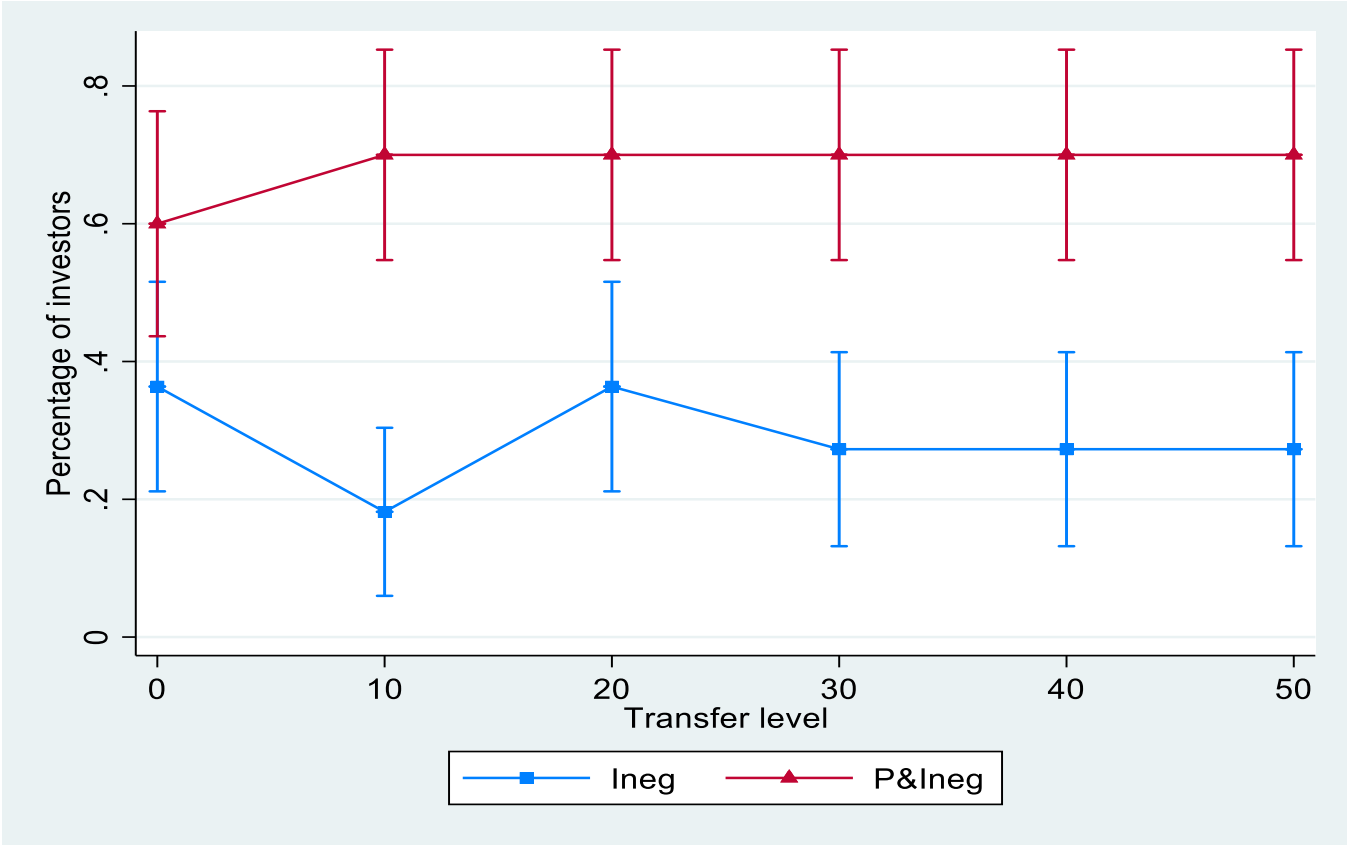

**c.** Percentage of investors

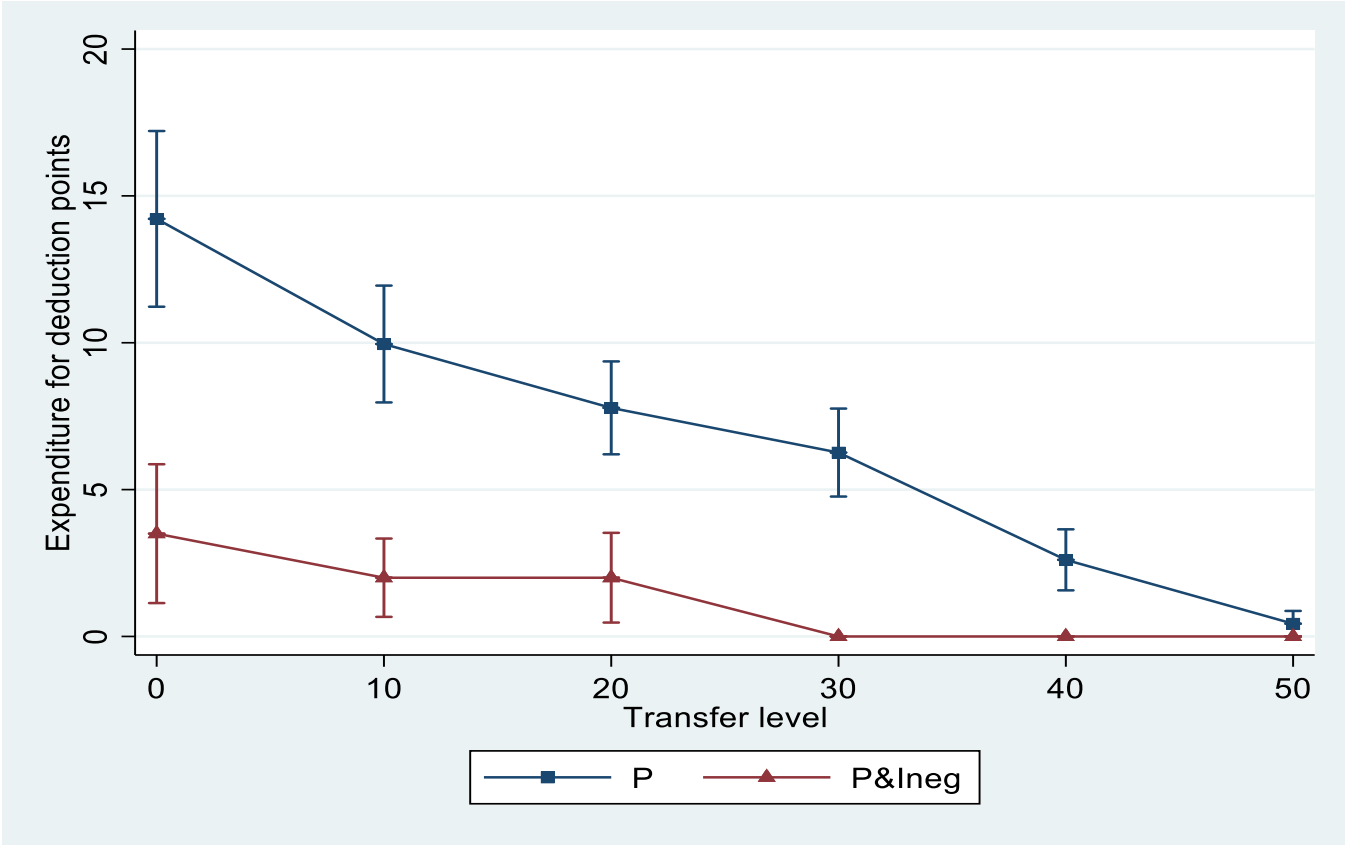

**b.** Average expenditure for deduction points

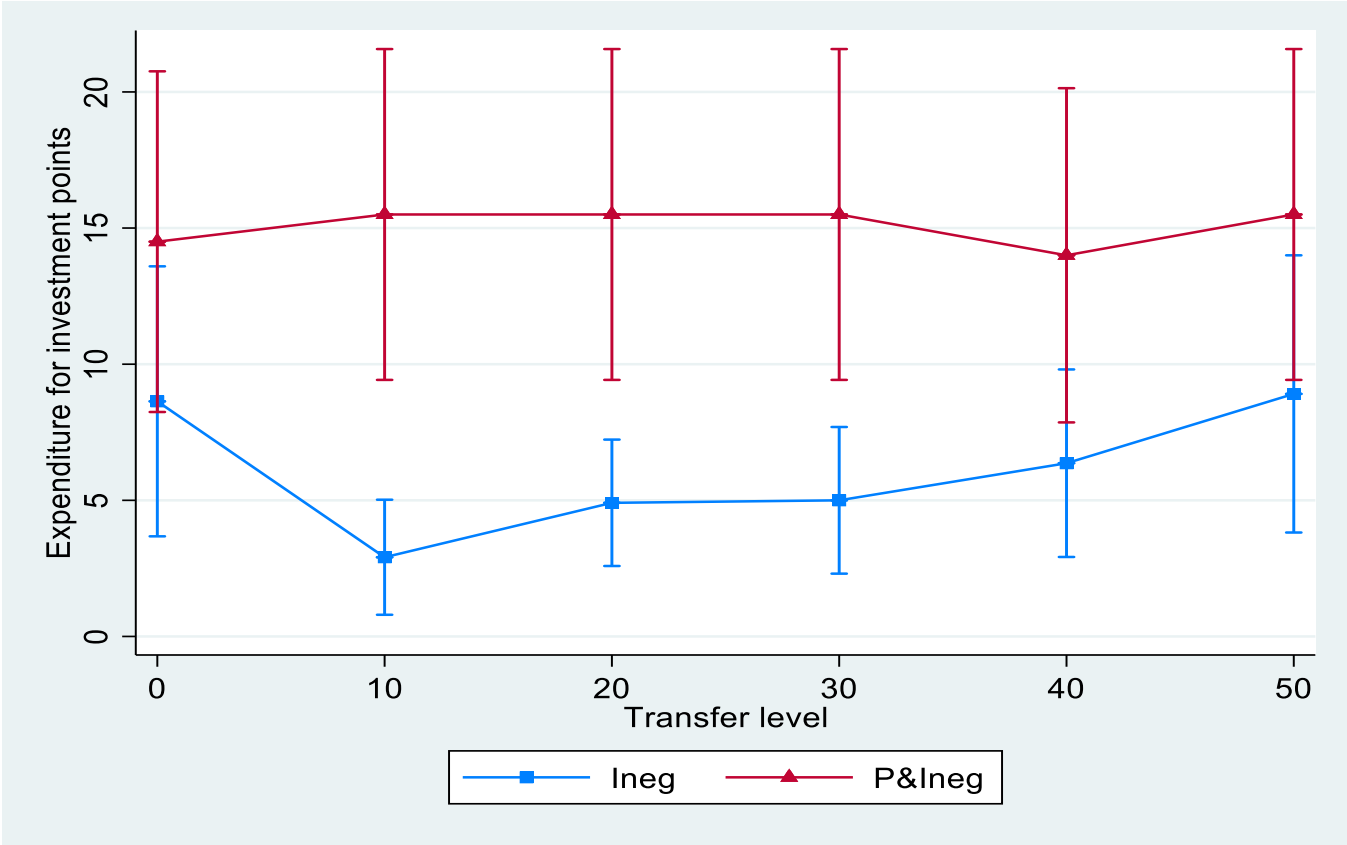

**d.** Average expenditure for investment points

Note: A punisher [investor] is defined as Player C who spent at least one token to deduction [investment] points. Expenditure for deduction [investment] points is the number of tokens that Player C spent for deduction [investment] points among 50 tokens. Transfer level is the number of tokens that Player A transferred to Player B. A vertical bar indicates plus and minus one standard error.

## 5. Concluding remarks

This study experimentally tests whether TPP is robust to context effects and finds that the availability of the seemingly unrelated and dominated investment option reduces both the choice frequency and the amount of TPP. Moreover, the increased (at least not decreased) demand for investment indicates a substitution effect between the punishment and investment options. These findings support that the investment option may work as a compromise consistent with an emotion-venting motivation (rather than as a decoy consistent with an inequity-averse motivation), and confirm that TPP is fragile to such a context effect.

Our study finds a case where TPP can be fragile by a context effect and thus suggests several future research topics related to prosocial behavior and context effects. For example, it would be interesting to examine what kinds of alternatives could also generate context effects and affect the willingness of TPP. In particular, context effects even imply that different types of alternatives could enhance the demand for TPP. Another interesting topic would be exploring whether other prosocial behaviors, such as SPP or rewarding, are robust to context effects. For instance, if rewarding is sensitive to a decision environment as in our study, it could provide a rationale for or against the so-called "Good Samaritan Law," which enforces bystanders to help an injured person by legally sanctioning an individual who ignores and does not assist the person.

In addition, in our experiment, we find that a sizable fraction of risk-neutral and risk-averse third parties choose to invest in a lottery that yields zero or negative expected net return even when the context effect plays no role, that is, when the punishment option is not available. A possible explanation for this behavior is that people may have preferences for randomization [59, 60, 61, 62, 63]. For instance, apart from one's risk attitude, a third party may prefer to mix between safe and risky investment options. Then, even risk-averse third parties may allocate some positive token amount to the dominated risky option. The investment behavior in our experiment may be one example of why people sometimes choose an unfavorable alternative, and investigating its fundamental motivation would be an interesting further study.

**Acknowledgments**

The authors are very grateful for the comments and suggestions by the associate editor and two anonymous referees which greatly improved the quality of this study.

**Supplementary Online Appendices for "On the Fragility of Third-Party Punishment: The Context Effect of a Dominated Risky Investment Option"**

## Appendix A. Instructions

These instructions are English translation. The instructions written in Korean were used in the experiment. The Korean instructions are available upon request.

### A.1. General Instructions

The following instructions provide basic information for the experiment. Please read the instruction carefully and make your decision prudently for each task. You will participate in two tasks and join two distinct decision-making processes for each task. For convenience, we refer to the first task as Task 1 and the second task as Task 2. Specific instructions for Tasks 1 and 2 will be displayed on the monitor before starting each task. Your final payoff depends on your decisions determined in each task. If you have any questions about the instructions, please raise your hand quietly. An assistant will answer your question. Mobile phones and communication with other participants are not allowed during the experiment.

***Payoff rules***

Payoffs for each task will be displayed on the monitor after the entire experiment is completed. The unit of the payoff is "KRW (South Korean won)" for Task 1 and "token" for Task 2. Your final payoff for the experiment will be "[3000 KRW (show-up fee)] + [payoff in task 1] + [payoff in task 2] × 80 KRW." At the end of the experiment, a questionnaire session will follow. The payment will be provided after the session.

Specific instructions for Task 1 will be presented on the next screen. All subjects will participate identically in Task 1. After finish reading the instructions, click the OK button and wait for a moment.

### A.2. Specific Instructions for Task 1

The following are instructions for Task 1. In Task 1, ten questions are displayed as follows. Each question offers Lottery L and Lottery R. Select one lottery that you prefer for each question and click the OK button. Write L if you prefer Lottery L, and write R if you prefer Lottery R in the blank (You should write in capital letters).

For instance, after all participants complete Tasks 1 and 2, your payoff for Task 1 will be realized by the decision determined in Task 1. In other words, one question in Task 1 will be selected by the computerized random procedure. Suppose that Question 1 is selected and you have chosen Lottery R for the given question. Then, the computer executes the lottery. Since Lottery R provides "8,000 KRW with probability 1/10 and 100 KRW with probability 9/10," the computer randomly draws a ball in a basket containing ten balls composed of 1 "white" ball and 9 "red" balls. You will receive 8,000 KRW if a white ball is drawn and 100 KRW if a red ball is drawn. This will be your payoff for Task 1.

### A.3. Specific Instructions for Task 2

The following are instructions for Task 2. In Task 2, three distinct types of Player have different roles (Players A, B, and C). Before Task 2 starts, the computer will randomly assign one Player type (Player A, B, or C) to each participant. Then, Players A, B, and C are randomly matched as a group. In Task 2, you make a decision using tokens. Your payoff in Task 2 is determined by decisions made by you and other participants in the same group. Your final payoff is informed after all participants complete making their decisions.

#### A.3.1. Player A

You are Player A. Your group has one Player B and one Player C. At the beginning of Task 2, you receive 100 tokens as an endowment. Also, Players B and C receive 0 token and 50 tokens, respectively.

***Stage 1***

In Stage 1, only Player A makes a decision. You, as Player A, decide how many tokens you would transfer to Player B from your own endowment. You can choose one of {0, 10, 20, 30, 40, 50} tokens for the transfer. Suppose that you transfer X token amount to Player B. Then, your endowment changes to 100−X tokens, and Player B's endowment changes to X tokens in Stage 1. You can enter the transferred amount in the blank space appearing on the monitor.

***Stage 2 [P treatment]***

In Stage 2, Player C can observe the token amount you transferred to Player B in Stage 1. In addition, Player C decides how many "deduction points" to impose on you. An explanation of deduction points is provided below.

Suppose that Player C imposes Y number of deduction points on you. Then, Player C's endowment is reduced by Y tokens, and your endowment is reduced by 3 times the Y tokens. Hence, one deduction point costs 1 token and reduces your token by 3. Moreover, Player C can freely assign tokens from 0 to 50 in integer unit for imposing deduction points.

After all the participants have completed their own decision-making, you will be able to observe the token amount you transferred to Player B and the number of deduction points Player C imposed on you.

Therefore, the final payoffs of each Player in Task 2 are

- Player A (You): (100 – [Token amount you transferred to Player B] – 3×[Deduction points]),
- Player B: ([Token amount you transferred to Player B]), and
- Player C: (50 – [Cost of imposed deduction points]).

***Stage 2 [P&I0 treatment]***

In Stage 2, Player C can observe the token amount you transferred to Player B in Stage 1. In addition, Player C decides how many "investment points" to purchase and how many "deduction points" to impose on you. Explanations of investment and deduction points are provided below.

Suppose that Player C imposes Y number of deduction points on you. Then, Player C's endowment is reduced by Y tokens and your endowment is reduced by 3 times the Y tokens. Hence, one deduction point costs 1 token and reduces your tokens by 3.

Suppose that Player C purchases Z number of investment points. Then, Player C receives 2 times the Z tokens at half probability, or zero token at half probability. Hence, one investment point costs 1 token, and the investment profit is randomly determined.

Note that the investment point does not affect your payoff. Moreover, Player C can freely assign tokens from 0 to 50 in integer unit for purchasing investment points and imposing deduction points.

For instance, Player C can both purchase investment points and impose deduction points on you. Another option for Player C is to purchase investment points and impose no deduction point on you or purchase no investment points and impose deduction points on you. Lastly, Player C can purchase no investment point and impose no deduction point on you. However, for all cases, the total number of investment and deduction points should be between 0 and 50.

After all participants have completed their own decision-making, you will be able to observe the token amount you transferred to Player B and the number of deduction points Player C imposed on you.

Therefore, the final payoffs of each Player in Task 2 are

- Player A (You): 100 – [Token amount you transferred to Player B] – 3×[Deduction points],
- Player B: [Token amount you transferred to Player B], and
- Player C: 50 – [Cost of imposed deduction points] – [Cost of purchased investment points] + [Investment profit of Player C].

***Stage 2 [I0 treatment]***

In Stage 2, Player C can observe the token amount you transferred to Player B in Stage 1. In addition, Player C decides how many "investment points" to purchase. An explanation of investment points is provided below.

Suppose that Player C purchases Z number of investment points. Then, Player C receives 2 times the Z tokens at half probability, or zero token at half probability. Hence, one investment point costs 1 token, and the investment profit is randomly determined. Note that the investment point does not affect your payoff. Moreover, Player C can freely assign tokens from 0 to 50 in integer unit for purchasing investment points.

Therefore, the final payoffs of each Player in Task 2 are

- Player A (You): (100 – [Token amount you transferred to Player B]),
- Player B: ([Token amount you transferred to Player B]), and
- Player C: (50 – [Cost of purchased investment points] + [Investment profit of Player C]).

**A.3.2. Player B**

You are Player B. Your group has one Player A and one Player C. At the beginning of Task 2, you receive 0 token as an endowment. Also, Player A receives 100 tokens and Player C receives 50 tokens.

***Stage 1***

In Stage 1, only Player A makes a decision. Player A decides how many tokens to transfer to you from one's endowment. Player A can choose one of {0, 10, 20, 30, 40, 50} tokens for the transfer. Suppose that Player A transfers X token amount to you. Then, Player A's endowment changes to 100–X tokens, and your endowment changes to X tokens in Stage 1.

First, we explain Player C's role. Then, we explain your role in Task 2.

***Stage 2 [P treatment]***

In Stage 2, Player C can observe the token amount Player A transferred to you in Stage 1. In addition, Player C decides how many "deduction points" to impose on Player A. An explanation of deduction points is provided below.

Suppose that Player C imposes Y number of deduction points on Player A. Then Player C's endowment is reduced by Y tokens and Player A's endowment is reduced by 3 times Y tokens. Hence, one deduction point costs 1 token and reduces Player A's tokens by 3. Moreover, Player C can freely assign tokens from 0 to 50 integer unit for imposing deduction points.

Now, we explain your role in Task 2. Only Player A decides in Stage 1; hence, you immediately move on to Stage 2 for your decision-making. In Stage 2, you do not know Player A and C's decisions.

You write the "expected amount" of tokens Player A would transfer to you, that is, 0, 10, 20, 30, 40, or 50. In addition, you write the "expected amount" of deduction points Player C would impose on Player A. You should write the expected deduction points for each possible transfer level ({0, 10, 20, 30, 40, 50}). The maximum and minimum values for the expected deduction points are 0 and 50, respectively.

Your expected values are irrelevant to your payoff. However, we would appreciate it if you would honestly write your expectations during the task.

After all the participants have completed their own decision-making, you will be able to observe the token amount Player A transferred to you and the number of deduction points decided by Player C.

Therefore, the final payoffs of each Player in Task 2 are

- Player A: (100 – [Token amount Player A transferred to you] – 3×[Deduction points]),
- Player B (You): ([Token amount Player A transferred to you]), and
- Player C: (50 – [Cost of imposed deduction points]).

***Stage 2 [P&I0 treatment]***

In Stage 2, Player C can observe the token amount Player A transferred to you in Stage 1. In addition, Player C decides how many "investment points" to purchase and how many "deduction points" to impose on Player A. Explanations of investment and deduction points are provided below.

Suppose that Player C imposes Y number of deduction points on Player A. Then Player C's endowment is reduced by Y tokens, and Player A's endowment is reduced by 3 times the Y tokens. Hence, one deduction point costs 1 token and reduces Player A's tokens by 3.

Suppose that Player C purchases Z number of investment points. Then, Player C receives 2 times the Z tokens at half probability or zero token at half probability. Hence, one investment point costs 1 token, and the investment profit is randomly determined.

Note that the investment point does not affect Player A's payoff. Moreover, Player C can freely assign tokens from 0 to 50 in integer unit for purchasing investment points and imposing deduction points.

For instance, Player C can purchase investment points and impose deduction points on Player A. Another option for Player C is to purchase investment points and impose no deduction point on Player A or purchase no investment points and impose deduction points on Player A. Lastly, Player C can purchase no investment point and impose no deduction point on Player A. However, for all cases, the total number of investment and deduction points should be between 0 and 50.

Now, we explain your role in Task 2. Since only Player A decides in Stage 1, you immediately move on to Stage 2 for your decision-making stage. In Stage 2, you do not know Player A and C's decisions.

You write the "expected amount" of tokens Player A would transfer to you, that is, 0, 10, 20, 30, 40, or 50. In addition, you write the "expected amount" of deduction points Player C would impose on Player A. You should write the expected deduction points for each possible transfer level {0, 10, 20, 30, 40, 50}. The maximum and minimum values for the expected deduction points are 0 and 50, respectively. Lastly, you write the "expected amount" of investment points Player C would purchase. You should write the expected investment points for each possible transfer level {0, 10, 20, 30, 40, 50}. The maximum and minimum values for the expected investment points are 0 and 50, respectively. Hence, for all cases, the total number of expected investment and deduction points should be between 0 and 50.

Your expected values are irrelevant to your payoff. However, we would appreciate it if you would honestly write your expectations during the task.

After all participants have completed their own decision-making, you will be able to observe the token amount Player A transferred to you and the number of deduction and investment points decided by Player C.

Therefore, the final payoffs of each Player in Task 2 are

- Player A: (100 – [Token amount Player A transferred to you] – 3×[Deduction points]),
- Player B (You): ([Token amount Player A transferred to you]), and
- Player C: (50 – [Cost of imposed deduction points] – [Cost of purchased investment points] + [Investment profit of Player C]).

***Stage 2 [I0 treatment]***

In Stage 2, Player C can observe the token amount Player A transferred to you in Stage 1. In addition, Player C decides how many “investment points” to purchase. An explanation of investment points is provided below.

Suppose that Player C purchases Z number of investment points. Then, Player C receives 2 times the Z tokens at half probability or zero token at half probability. Hence, one investment point costs 1 token, and the investment profit is randomly determined.

Note that the investment point does not affect Player A’s payoff. Moreover, Player C can freely assign tokens from 0 to 50 in integer unit for purchasing investment points.

Now, we explain your role in Task 2. Only Player A decides in Stage 1; hence, you immediately move on to Stage 2 for your decision-making. In Stage 2, you do not know Player A and C’s decisions.

You write the “expected amount” of tokens Player A would transfer to you, that is, 0, 10, 20, 30, 40, or 50. In addition, you write the “expected amount” of investment points Player C would purchase. You should write the expected investment points for each possible transfer level ({0, 10, 20, 30, 40, 50}). The maximum and minimum values for the expected investment points are 0 and 50, respectively.

Your expected values are irrelevant to your payoff. However, we would appreciate it if you would honestly write your expectations during the task.

After all the participants have completed their own decision-making, you will be able to observe the token amount Player A transferred to you and the number of investment points decided by Player C.

Therefore, the final payoffs of each Player in Task 2 are

- Player A: (100 – [Token amount Player A transferred to you]),
- Player B (You): ([Token amount Player A transferred to you]), and
- Player C: (50 – [Cost of purchased investment points] + [Investment profit of Player C]).

### A.3.3. Player C

You are Player C. Your group has one Player A and one Player B. At the beginning of Task 2, you receive 50 tokens as an endowment. Also, Players A and B receive 100 tokens and 0 token, respectively.

***Stage 1***

In Stage 1, only Player A makes a decision. Player A decides how many tokens to transfer to Player B from his or her endowment. Player A can choose one of {0, 10, 20, 30, 40, 50} tokens for the transfer. Suppose that Player A transfers X token amount to Player B. Then, Player A's endowment changes to 100−X tokens, and Player B's endowment changes to X tokens in Stage 1.

Now, we explain your role in Stage 2. Since only Player A decides in Stage 1, you immediately move on to Stage 2 for your decision-making.

***Stage 2 [P treatment]***

You, as Player C, decide how many "deduction points" to impose on Player A. An explanation of deduction points is provided below.

Suppose that you impose Y number of deduction points on Player A. Then your endowment is reduced by Y tokens, and Player A's endowment is reduced by 3 times the Y tokens. Hence, one deduction point costs 1 token and reduces Player A's tokens by 3.

Note that you must decide the token amount for imposing deduction points before you observe Player A's actual transfer. Thus, you must write your decisions concerning each possible transfer level that Player A can choose. In other words, you can freely assign tokens from 0 to 50 in integer unit for imposing deduction points concerning Player A's each possible transfer level, that is, 0, 10, 20, 30, 40, and 50. Hence, during your decision-making stage, a table of six possible cases ({0, 10, 20, 30, 40, 50}) will appear on the monitor. For each case, you can freely assign your tokens for deduction points. The token amount that remains after subtracting the costs of imposing deduction points from your initial endowment (50 tokens) becomes reserves in your private account.

After all the participants have completed their own decision-making, you will be able to observe Player A's actual transfer. Then, the corresponding deduction points you have determined during the decision-making stage will be realized. We call this the realized deduction points.

Therefore, the final payoffs of each Player in Task 2 are

- Player A: (100 – [Token amount Player A transferred to Player B] – 3×[Realized deduction points]),
- Player B: ([Token amount Player A transferred to Player B]), and
- Player C (You): ([Realized private account]), that is, (50 – [Cost of realized deduction points]).

***Stage 2 [P&I0 treatment]***

You, as Player C, decide how many "investment points" to purchase and how many "deduction points" to impose on Player A. Explanations of investment and deduction points are provided below.

Suppose that you impose Y number of deduction points on Player A. Then your endowment is reduced by Y tokens, and Player A's endowment is reduced by 3 times the Y tokens. Hence, one deduction point costs 1 token and reduces Player A's tokens by 3.

Suppose that you purchase Z number of investment points. Then, you receive 2 times the Z tokens at half probability, or zero token at half probability. Hence, one investment point costs 1 token. Note that investment points do not affect Player A's endowment. Your investment profit is determined by a computerized random procedure. Random procedure draws a "white ball" at half probability or a "red ball" at half probability. If you draw a white ball, you receive 2 times the Z tokens. If you draw a red ball, you receive nothing.

Note that you must decide the token amount for purchasing investment points and imposing deduction points before you observe Player A's actual transfer. Thus, you must write your decisions concerning each possible transfer level that Player A can choose. In other words, you can freely assign tokens from 0 to 50 in integer unit for purchasing investment points and imposing deduction points concerning each possible transfer level of Player A, that is, 0, 10, 20, 30, 40, and 50. Hence, during your decision-making stage, a table of six possible cases ({0, 10, 20, 30, 40, 50}) will appear on the monitor. For each case, you can freely assign your tokens for investment and deduction points.

For instance, you can both purchase investment points and impose deduction points on Player A. As another option, you can purchase investment points and impose no deduction point on Player A or purchase no investment points and impose deduction points on Player A. The last option would be purchasing no investment point and imposing no deduction point on Player A. However, for all cases, the total number of investment and deduction points should be between 0 and 50. The token amount that remains after subtracting the costs of imposing deduction points and purchasing investment points from your initial endowment (50 tokens) becomes reserves in your private account.

After all the participants have completed their own decision-making, you will be able to observe Player A's actual transfer. Then, the corresponding investment and deduction points you have already determined during the decision-making stage will be realized. We call these as realized investment points and realized deduction points, respectively. In addition, the outcome of the lottery is informed.

Therefore, the final payoffs of each Player in Task 2 are

- Player A: (100 – [Token amount Player A transferred to Player B] – 3×[Realized deduction points]),
- Player B: ([Token amount Player A transferred to Player B]), and
- Player C (You): ([Realized private account] + [Realized investment profit]), that is,
  if you draw a "white ball," then (50 – [Cost of realized deduction points] – [Cost of realized investment points] + [Realized investment profit]),
  if you draw a "red ball," then (50 – [Cost of realized deduction points] – [Cost of realized investment points]).

***Stage 2 [I0 treatment]***

You, as Player C, decide how many "investment points" to purchase. An explanation of investment points is provided below.

Suppose that you purchase Z number of investment points. Then, you receive 2 times the Z tokens at half probability, or zero token at half probability. Hence, one investment point costs 1 token. Note that investment points do not affect Player A's endowment. Your investment profit is determined by a computerized random procedure. Random procedure draws a "white ball" at half probability or a "red ball" at half probability. If you draw a white ball, you receive 2 times the Z tokens. If you draw a red ball, you receive nothing.

Note that you must decide the token amount for purchasing investment points before you observe Player A's actual transfer. Thus, you must write your decisions concerning each possible transfer level that Player A can choose. In other words, you can freely assign tokens from 0 to 50 in integer unit for purchasing investment points concerning each possible transfer level of Player A, that is, 0, 10, 20, 30, 40, and 50. Hence, during your decision-making stage, a table of six possible cases ({0, 10, 20, 30, 40, 50}) will appear on the monitor. You can purchase identical or different number of investment points for all six cases. The token amount that remains after subtracting the costs purchasing investment points from your initial endowment (50 tokens) becomes reserves in your private account.

After all the participants have completed their own decision-making, you will be able to observe Player A's actual transfer. Then, the corresponding investment points you have already determined during the decision-making stage will be realized. We call this as realized investment points. In addition, the outcome of the lottery is informed.

Therefore, the final payoffs of each Player in Task 2 are

- Player A: (100 – [Token amount Player A transferred to Player B]),
- Player B: ([Token amount Player A transferred to Player B]), and
- Player C (You): ([Realized private account] + [Realized investment profit]), that is,
  if you draw a "white ball," then (50 – [Cost of realized investment points] + [Realized investment profit]),
  if you draw a "red ball," then (50 – [Cost of realized investment points]).

**A.4. Examples of Control Questions [Player C in P&I0 treatment]**

On the next screen, a simple quiz will be provided to help your understanding of Task 2. After reading the questions, write the correct answer. You can move on to the decision-making stage only if you write the correct answers to all the questions. If you incorrectly answer any question, you will not be able to move on to the next stage. In that case, please read the instructions again carefully and fix the incorrect answer(s). The result of the quiz is irrelevant to your payoff in the experiment.

1. Player A decided to transfer 10 tokens to Player B. Player C purchased 0 investment point and

imposed 0 deduction point on player A when transferring 10 tokens to Player B.

- If you draw a “white ball,” what would be Player A, B, and C’s final tokens in Task 2, respectively?
- If you draw a “red ball,” what would be Player A, B, and C’s final tokens in Task 2, respectively?

2. Player A decided to transfer 10 tokens to Player B. Player C purchased 14 investment points and imposed 0 deduction point on Player A when transferring 10 tokens to Player B.
*(skip)*

3. Player A decided to transfer 10 tokens to Player B. Player C purchased 0 investment point and imposed 18 deduction points on Player A when transferring 10 tokens to Player B.
*(skip)*

4. Player A decided to transfer 10 tokens to Player B. Player C purchased 14 investment points and imposed 18 deduction points on Player A when transferring 10 tokens to Player B.
*(skip)*

## A.5. Examples of Screenshots

The following are examples of screenshots for Player C in P&I0 treatment. The screenshots for the other players and the other treatments are similar besides the specific instructions for Task 2, the contents of quizzes, and the decisions to be made by Player C.

### A.5.1. General Instructions

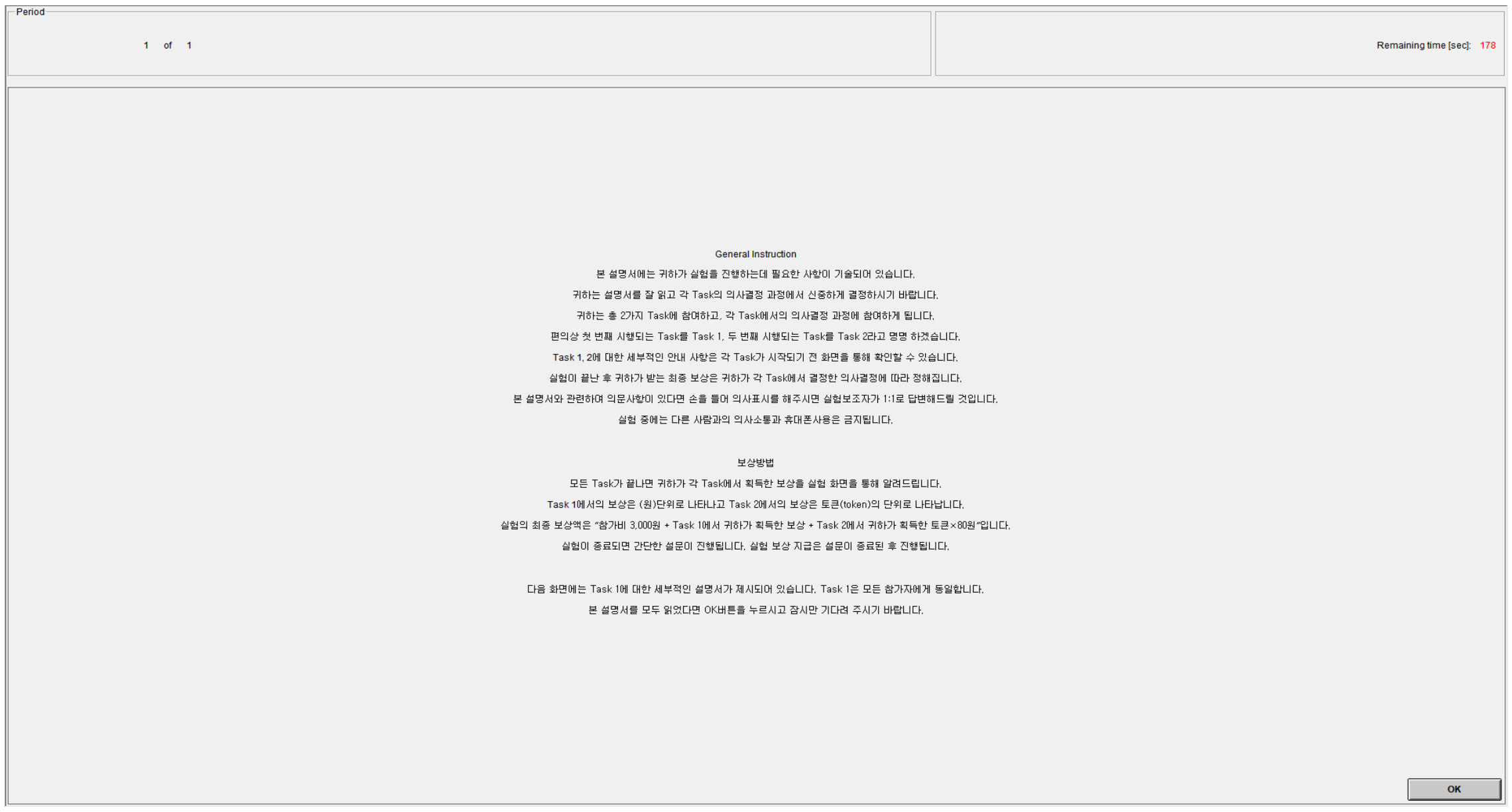

### A.5.2. Decision-making Screen for Task 1

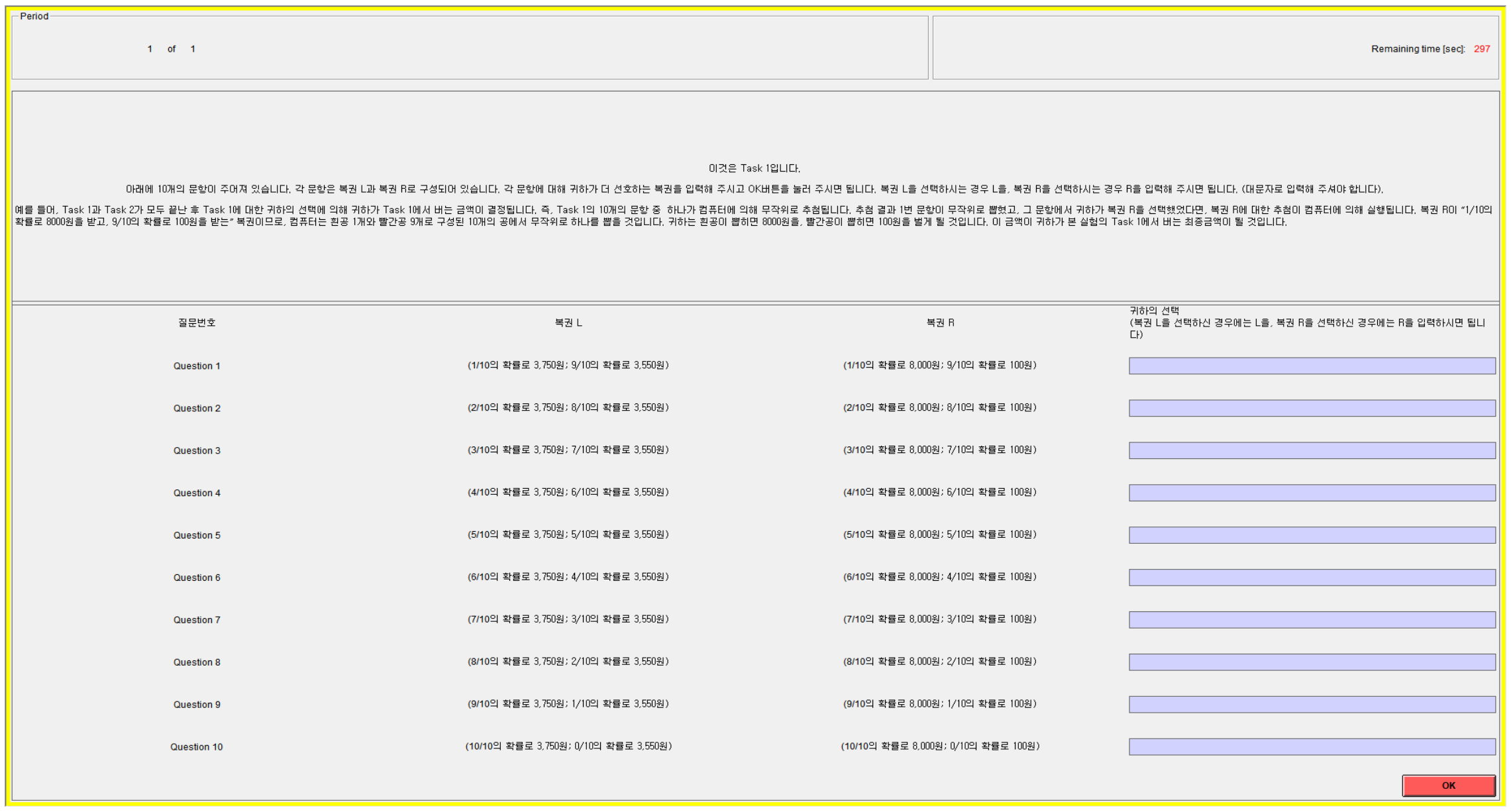

## A.5.3. Task 2

i) Specific instruction for Player C in P&I0 treatment

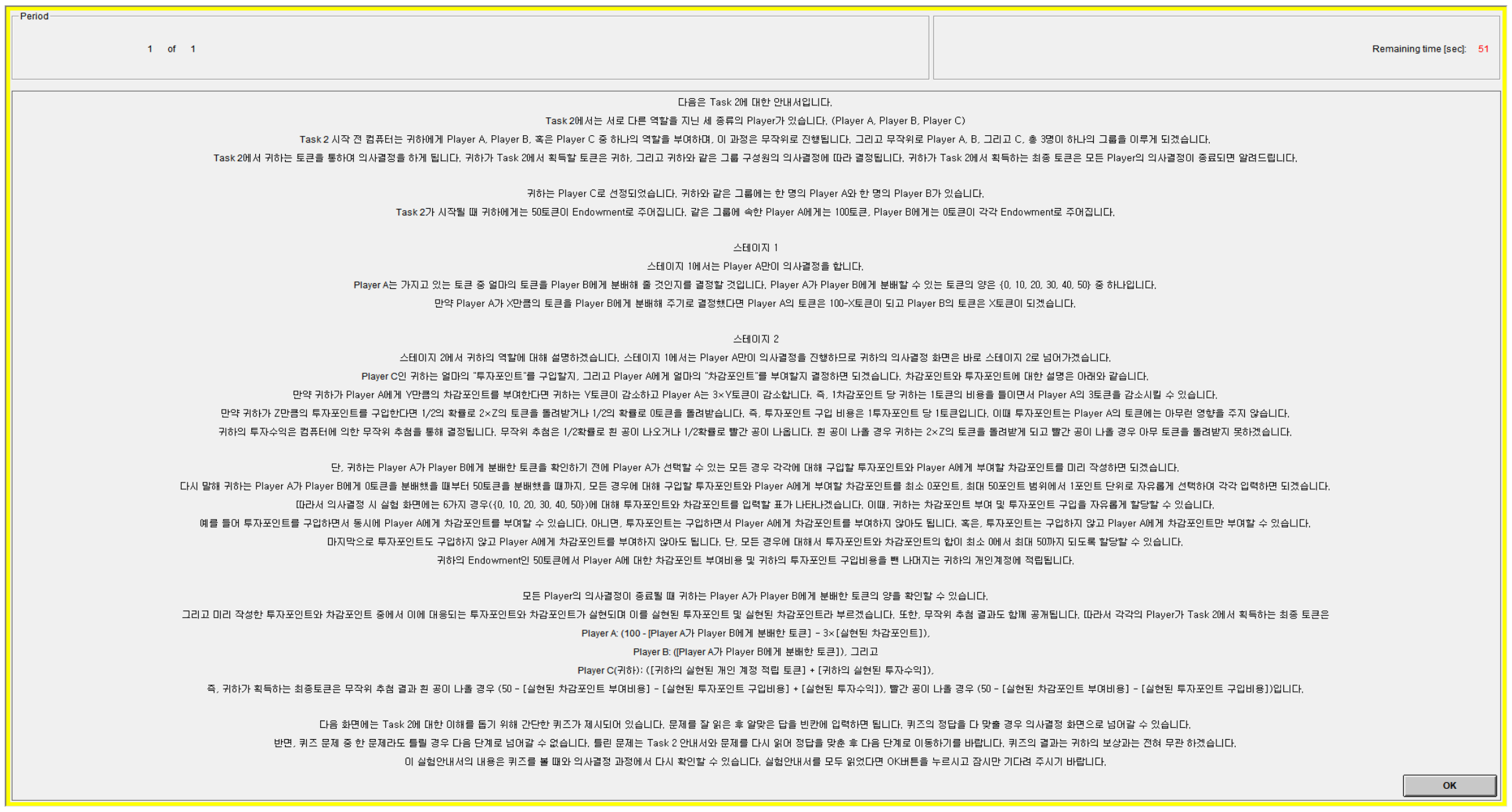

ii) Control questions and checking answers for Player C in P&I0 treatment

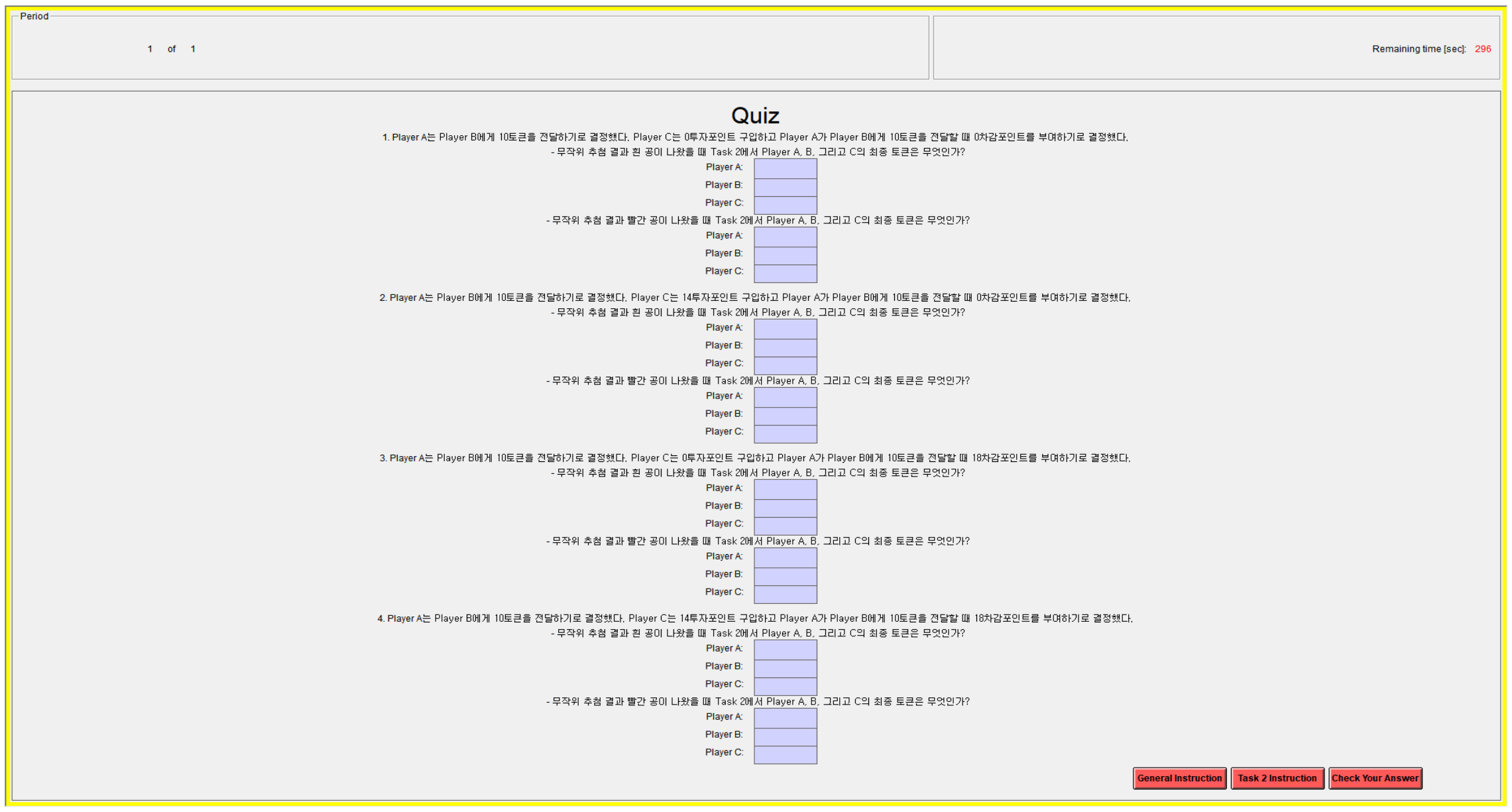

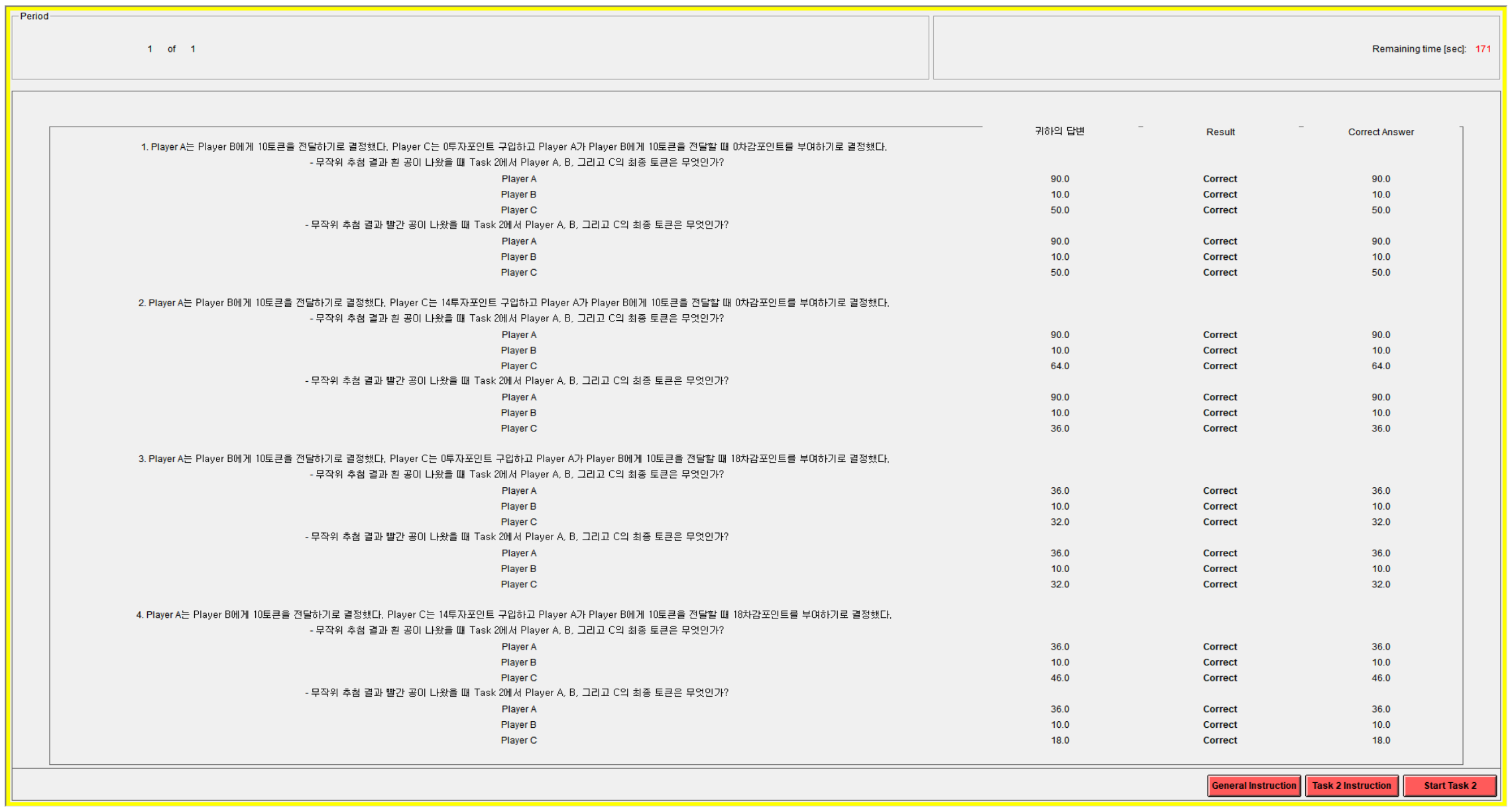

iii) Decision-making procedure for Player C in P&I0 treatment

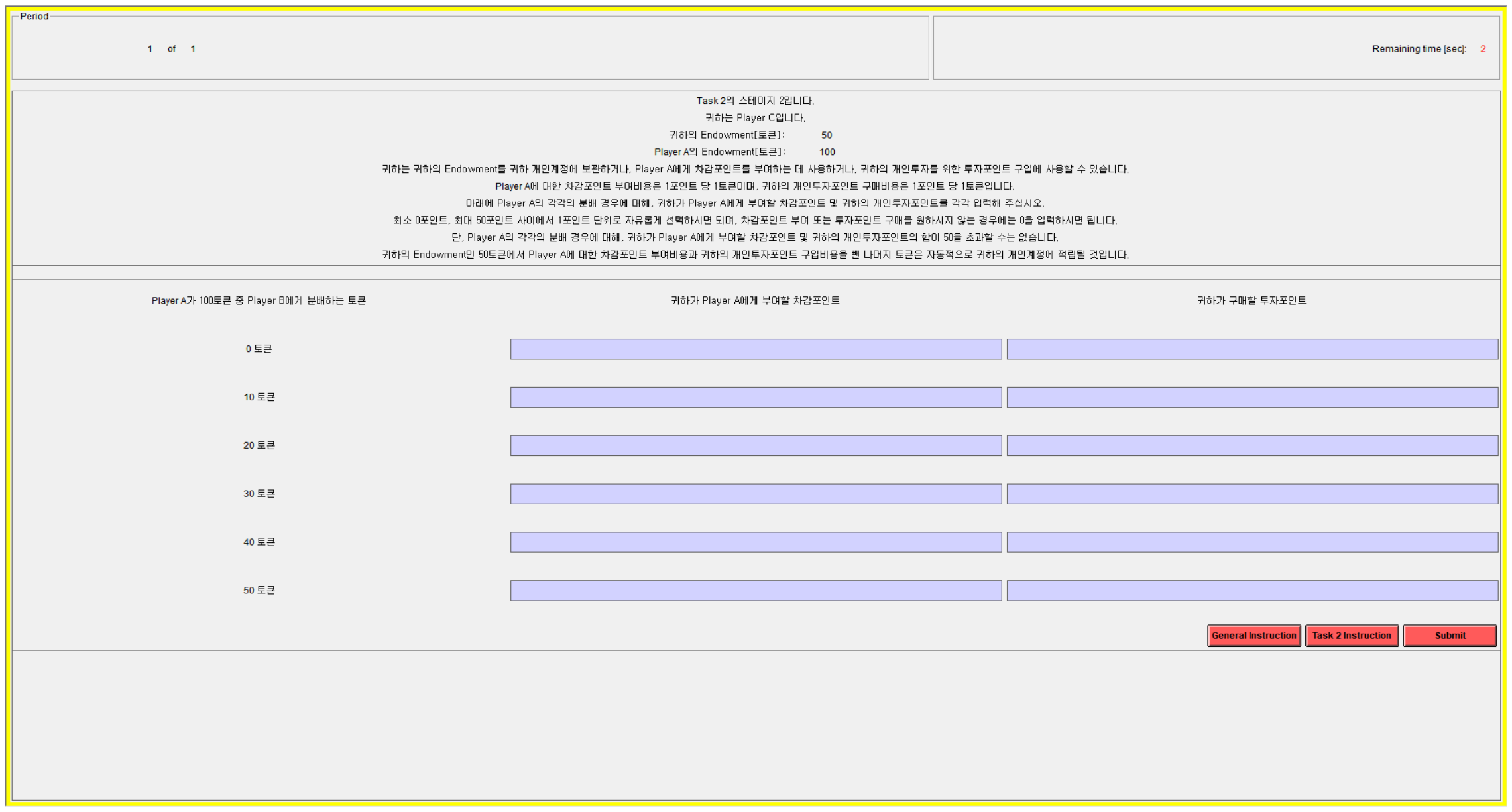

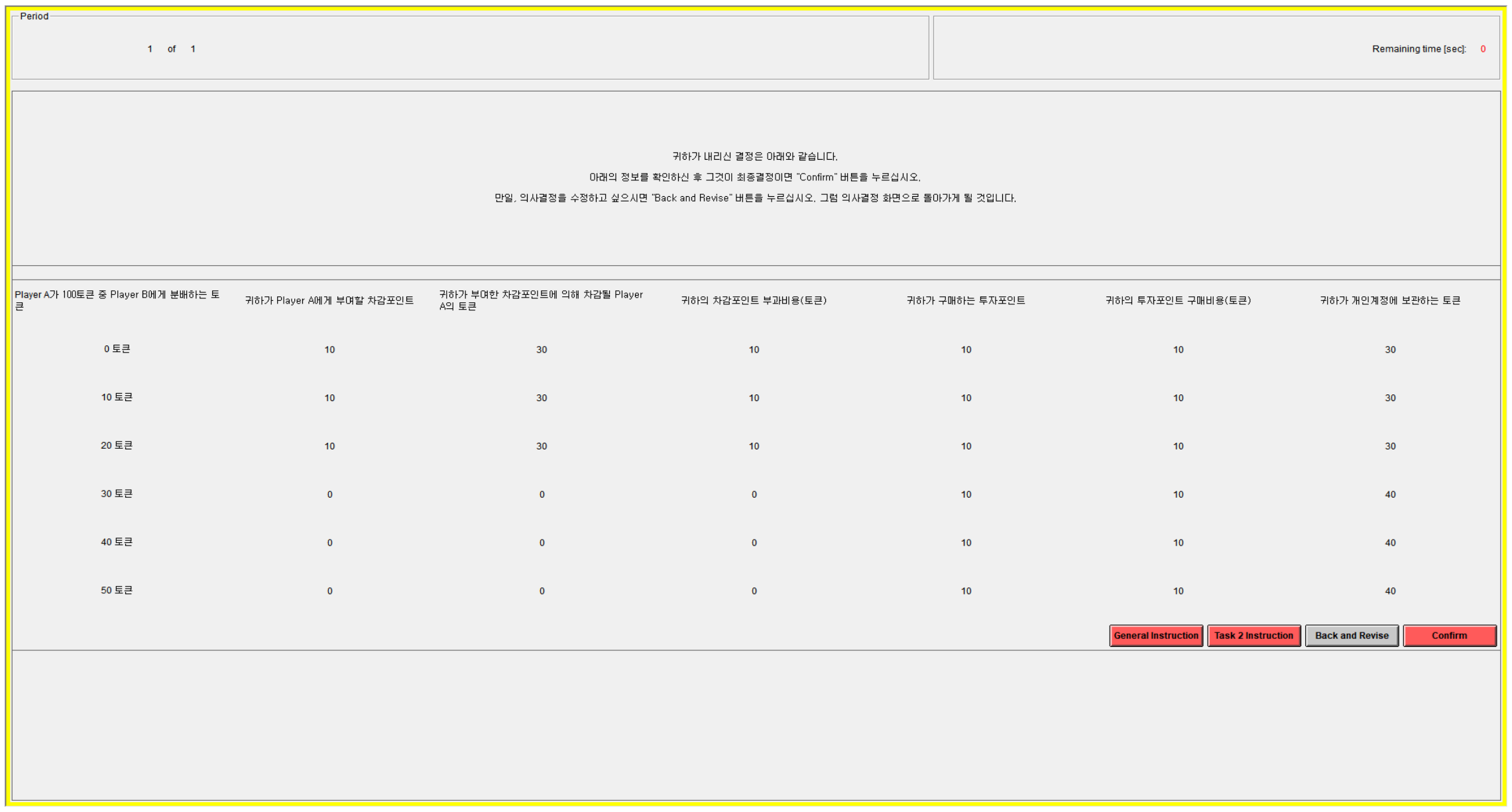

iv) Task 2 payoff display for Player C in P&I0 treatment

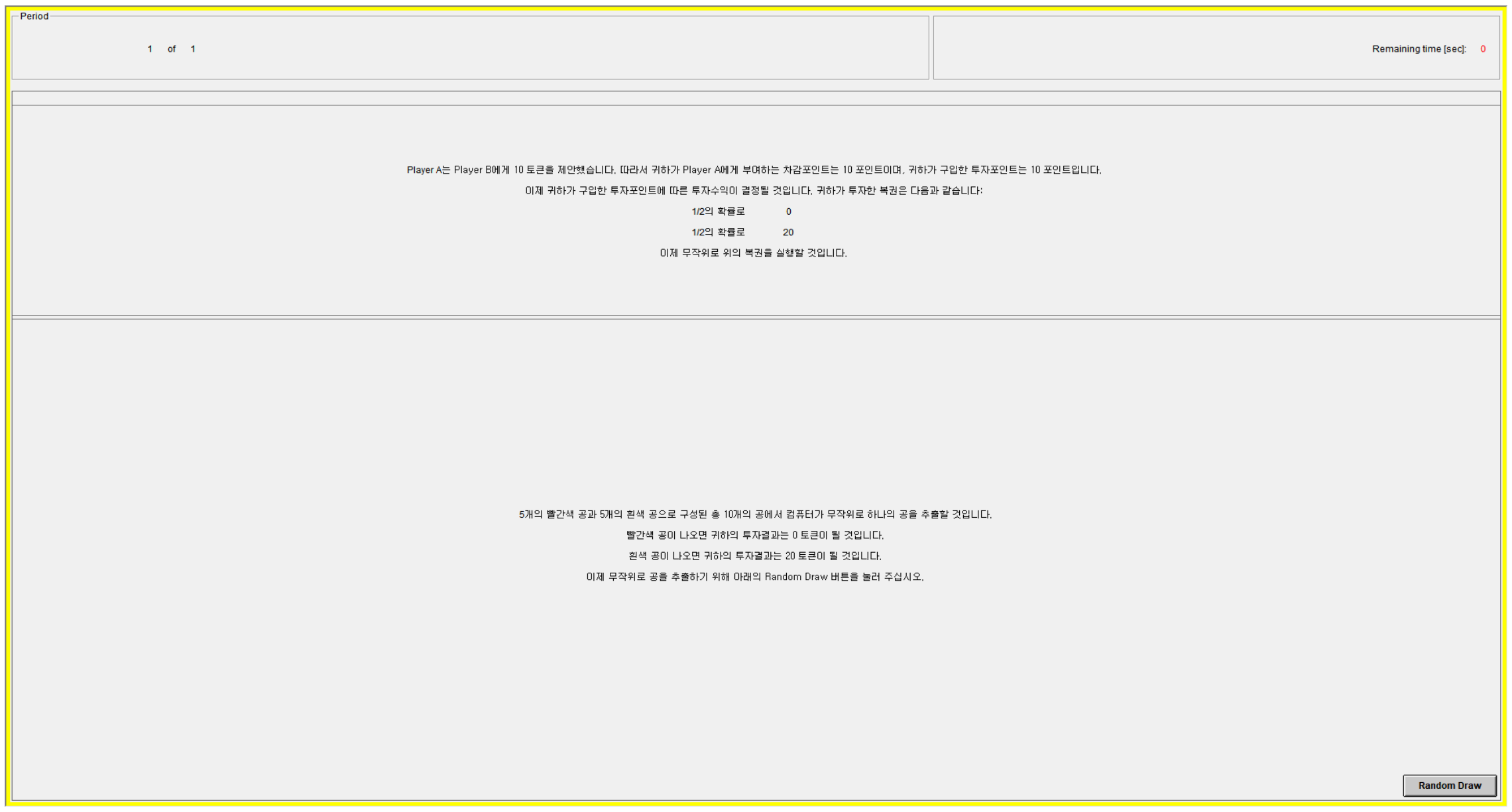

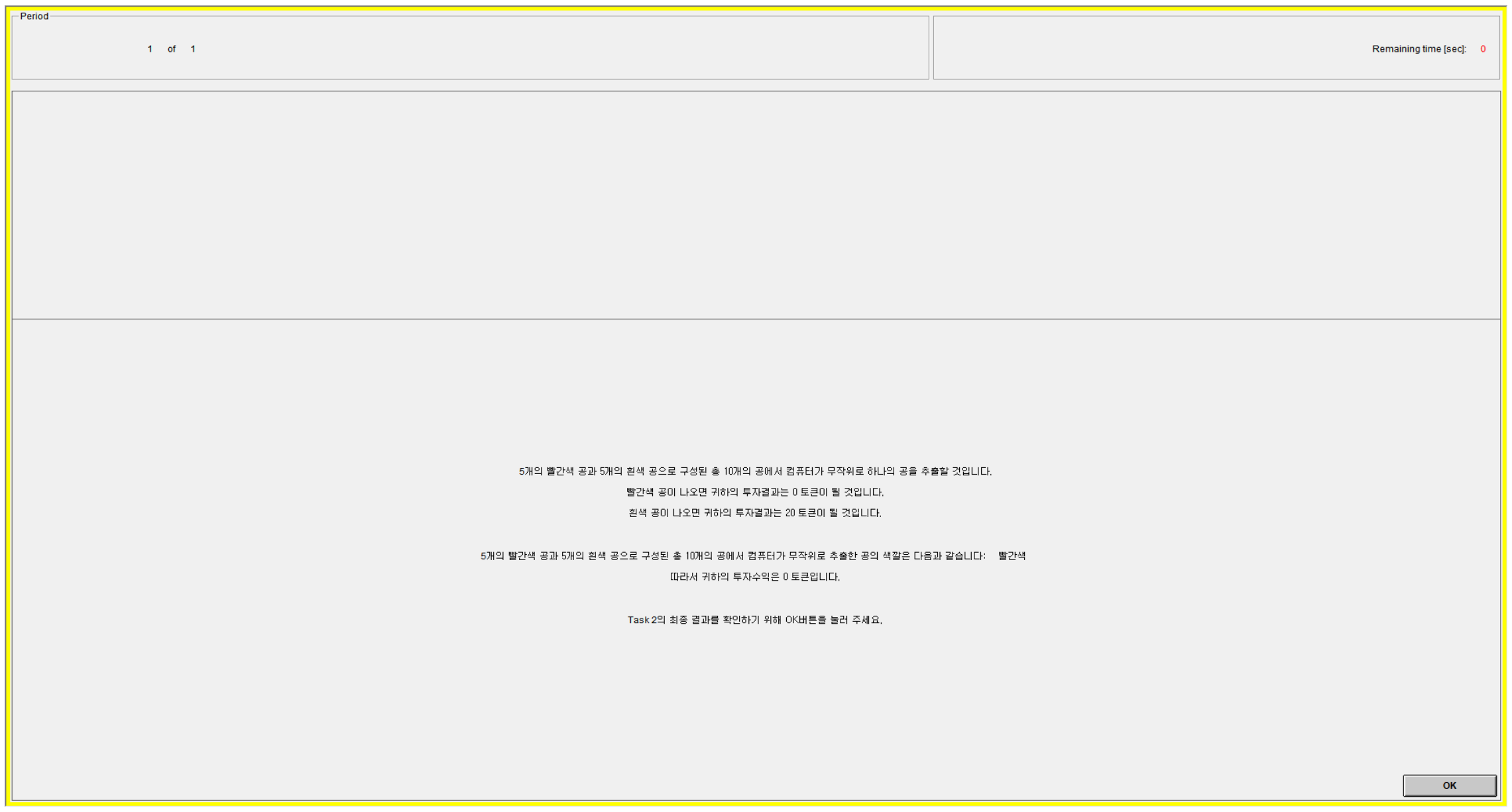
Period
1 of 1
Remaining time [sec]: 0
5개의 빨간색 공과 5개의 흰색 공으로 구성된 총 10개의 공에서 컴퓨터가 무작위로 하나의 공을 추출할 것입니다.
빨간색 공이 나오면 귀하의 투자결과는 0 토큰이 될 것입니다.
흰색 공이 나오면 귀하의 투자결과는 20 토큰이 될 것입니다.
5개의 빨간색 공과 5개의 흰색 공으로 구성된 총 10개의 공에서 컴퓨터가 무작위로 추출한 공의 색깔은 다음과 같습니다: 빨간색
따라서 귀하의 투자수익은 0 토큰입니다.
Task 2의 최종 결과를 확인하기 위해 OK버튼을 눌러 주세요.
OK

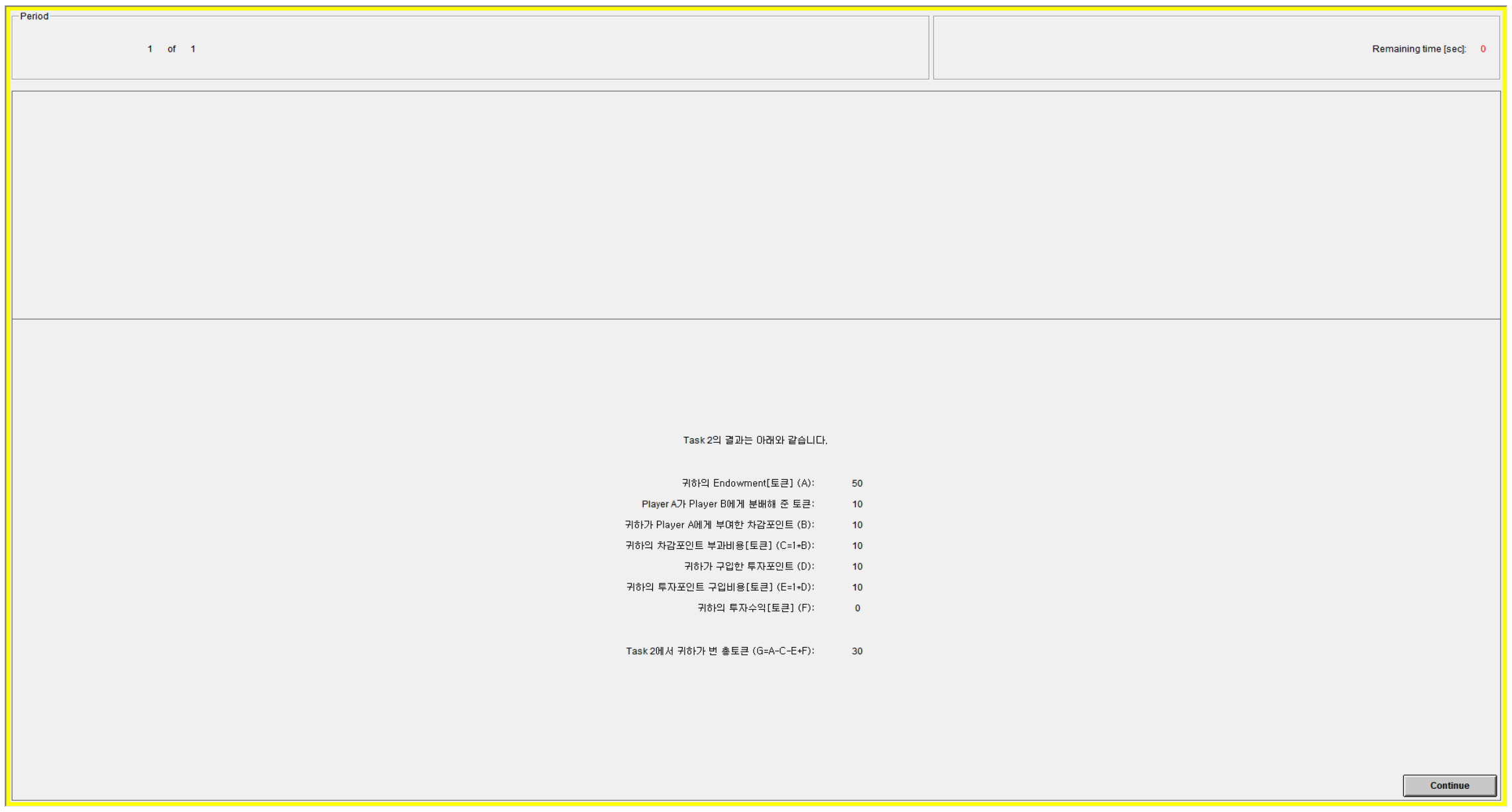
Period
1 of 1
Remaining time [sec]: 0
Task 2의 결과는 아래와 같습니다.
귀하의 Endowment[토큰] (A): 50
Player A가 Player B에게 분배해 준 토큰: 10
귀하가 Player A에게 부여한 차감포인트 (B): 10
귀하의 차감포인트 부과비용[토큰] (C=1*B): 10
귀하가 구입한 투자포인트 (D): 10
귀하의 투자포인트 구입비용[토큰] (E=1*D): 10
귀하의 투자수익[토큰] (F): 0
Task 2에서 귀하가 번 총토큰 (G=A-C-E+F): 30
Continue

## A.5.4. Payoff Decision Procedure for Task 1

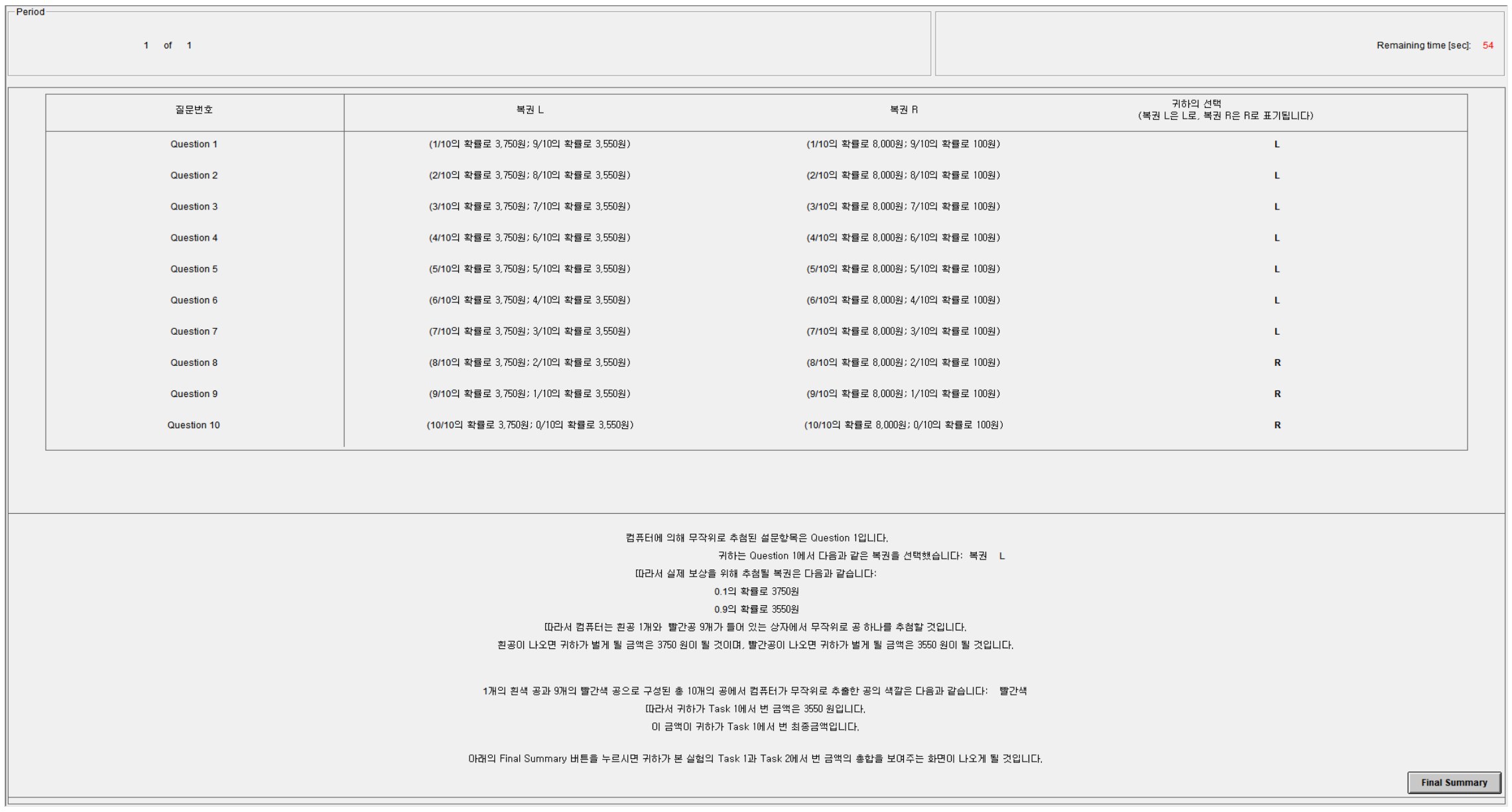

Period

1 of 1

Remaining time [sec]: 54

| 질문번호 | 복권 L | 복권 R | 귀하의 선택 (복권 L은 L로, 복권 R은 R로 표기됩니다) |
|---|---|---|---|
| Question 1 | (1/10의 확률로 3,750원; 9/10의 확률로 3,550원) | (1/10의 확률로 8,000원; 9/10의 확률로 100원) | L |
| Question 2 | (2/10의 확률로 3,750원; 8/10의 확률로 3,550원) | (2/10의 확률로 8,000원; 8/10의 확률로 100원) | L |
| Question 3 | (3/10의 확률로 3,750원; 7/10의 확률로 3,550원) | (3/10의 확률로 8,000원; 7/10의 확률로 100원) | L |
| Question 4 | (4/10의 확률로 3,750원; 6/10의 확률로 3,550원) | (4/10의 확률로 8,000원; 6/10의 확률로 100원) | L |
| Question 5 | (5/10의 확률로 3,750원; 5/10의 확률로 3,550원) | (5/10의 확률로 8,000원; 5/10의 확률로 100원) | L |
| Question 6 | (6/10의 확률로 3,750원; 4/10의 확률로 3,550원) | (6/10의 확률로 8,000원; 4/10의 확률로 100원) | L |
| Question 7 | (7/10의 확률로 3,750원; 3/10의 확률로 3,550원) | (7/10의 확률로 8,000원; 3/10의 확률로 100원) | L |
| Question 8 | (8/10의 확률로 3,750원; 2/10의 확률로 3,550원) | (8/10의 확률로 8,000원; 2/10의 확률로 100원) | R |
| Question 9 | (9/10의 확률로 3,750원; 1/10의 확률로 3,550원) | (9/10의 확률로 8,000원; 1/10의 확률로 100원) | R |
| Question 10 | (10/10의 확률로 3,750원; 0/10의 확률로 3,550원) | (10/10의 확률로 8,000원; 0/10의 확률로 100원) | R |

컴퓨터에 의해 무작위로 추첨된 설문항목은 Question 1입니다.
귀하는 Question 1에서 다음과 같은 복권을 선택했습니다: 복권 L
따라서 실제 보상을 위해 추첨될 복권은 다음과 같습니다:
0.1의 확률로 3750원
0.9의 확률로 3550원
따라서 컴퓨터는 흰공 1개와 빨간공 9개가 들어 있는 상자에서 무작위로 공 하나를 추첨할 것입니다.
흰공이 나오면 귀하가 벌게 될 금액은 3750 원이 될 것이며, 빨간공이 나오면 귀하가 벌게 될 금액은 3550 원이 될 것입니다.

1개의 흰색 공과 9개의 빨간색 공으로 구성된 총 10개의 공에서 컴퓨터가 무작위로 추출한 공의 색깔은 다음과 같습니다: 빨간색
따라서 귀하가 Task 1에서 번 금액은 3550 원입니다.
이 금액이 귀하가 Task 1에서 번 최종금액입니다.

아래의 Final Summary 버튼을 누르시면 귀하가 본 실험의 Task 1과 Task 2에서 번 금액의 총합을 보여주는 화면이 나오게 될 것입니다.

Final Summary

## A.5.5. Final Total Payoff Display

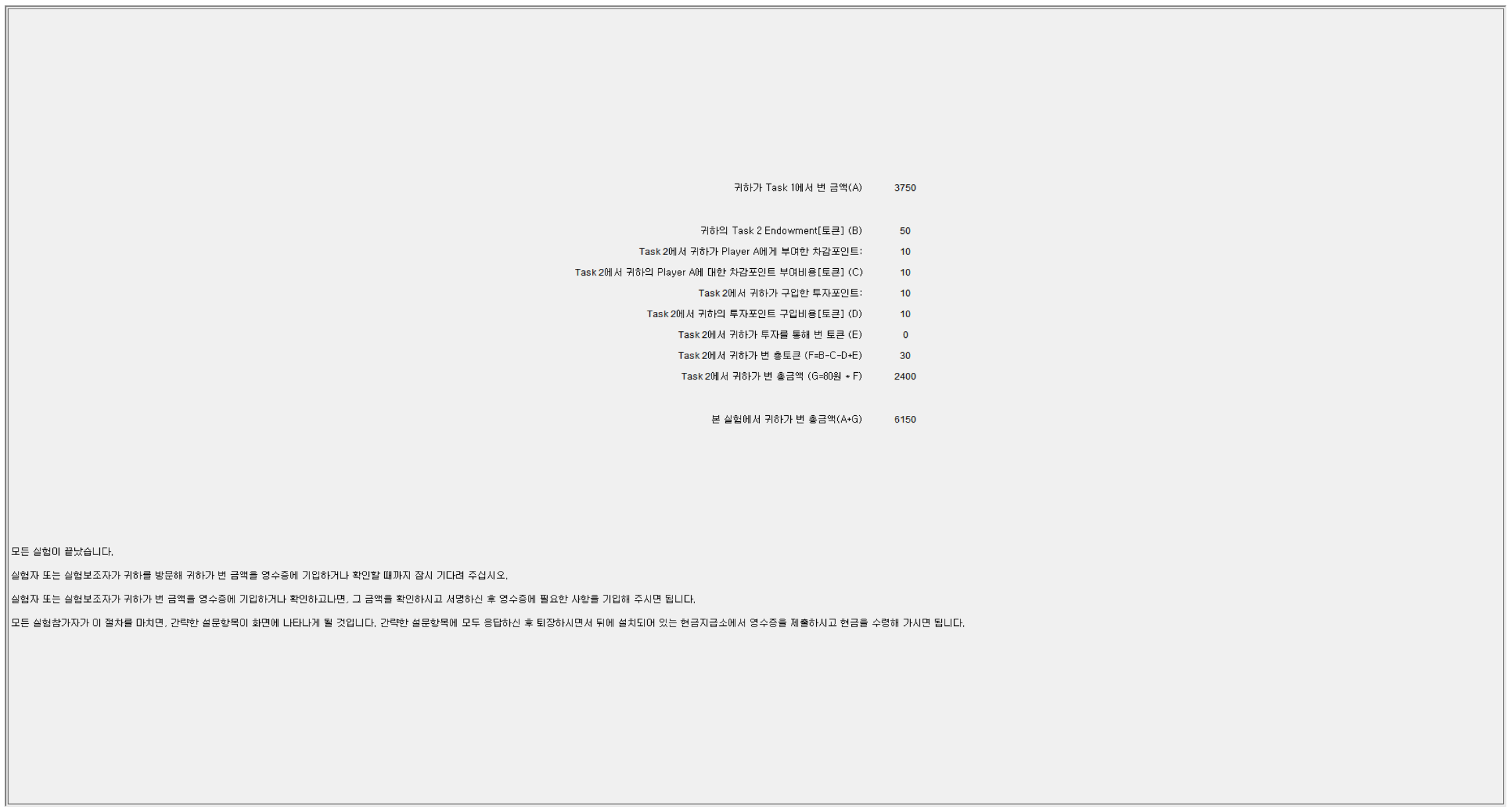

| | |
|---|---|
| 귀하가 Task 1에서 번 금액(A) | 3750 |
| 귀하의 Task 2 Endowment[토큰] (B) | 50 |
| Task 2에서 귀하가 Player A에게 부여한 차감포인트: | 10 |
| Task 2에서 귀하의 Player A에 대한 차감포인트 부여비용[토큰] (C) | 10 |
| Task 2에서 귀하가 구입한 투자포인트: | 10 |
| Task 2에서 귀하의 투자포인트 구입비용[토큰] (D) | 10 |
| Task 2에서 귀하가 투자를 통해 번 토큰 (E) | 0 |
| Task 2에서 귀하가 번 총토큰 (F=B-C-D+E) | 30 |
| Task 2에서 귀하가 번 총금액 (G=80원 * F) | 2400 |
| 본 실험에서 귀하가 번 총금액(A+G) | 6150 |

모든 실험이 끝났습니다.

실험자 또는 실험보조자가 귀하를 방문해 귀하가 번 금액을 영수증에 기입하거나 확인할 때까지 잠시 기다려 주십시오.

실험자 또는 실험보조자가 귀하가 번 금액을 영수증에 기입하거나 확인하고나면, 그 금액을 확인하시고 서명하신 후 영수증에 필요한 사항을 기입해 주시면 됩니다.

모든 실험참가자가 이 절차를 마치면, 간략한 설문항목이 화면에 나타나게 될 것입니다. 간략한 설문항목에 모두 응답하신 후 퇴장하시면서 뒤에 설치되어 있는 현금지급소에서 영수증을 제출하시고 현금을 수령해 가시면 됩니다.

## Appendix B. Formal Descriptions of Hypotheses

In this section, we formally describe how the investment option can work as a compromise or decoy in our experiment. As we mentioned in the main text, we can consider the psychological-payoff attribute in two points of view: the "emotion-venting" viewpoint or the "inequity-averse" viewpoint. Appendix B.1 focuses on the emotion-venting viewpoint and shows that the investment option works as a compromise. Appendix B.2 focuses on the inequity-averse viewpoint and shows that the investment option becomes a decoy for the safe option.

### Appendix B.1. Hypotheses Based on a Compromise Effect Generated by the Emotion-venting Viewpoint

Suppose that a third party considers the material-payoff attribute and the psychological-payoff attribute when evaluating an alternative. Denote $A_M$ as the material-payoff attribute and $A_P$ as the psychological-payoff attribute. In this subsection, we focus on the "emotion-venting" viewpoint for the psychological-payoff attribute since venting negative emotions such as anger is known as one of the motivations of TPP (Jordan et al., 2016; Tan and Xiao, 2018; Nelissen and Zeelenberg, 2009).

Let $X$ be the set of all alternatives, that is, $X = \{TP, I, S\}$, where $TP$ is the punishment option, $I$ is the investment option, and $S$ is the safe option. The investment option yields the zero expected net return in P&I0 and I0 treatments and a strictly negative net return in P&Ineg and Ineg treatments. Let $X^r \subset X$ be the set of available alternatives under treatment $r \in \{P, P\&I0, I0, P\&Ineg, Ineg\}$. For instance, $X^P = \{TP, S\}$, $X^{I0} = \{I, S\}$, and $X^{P\&I0} = \{TP, I, S\}$. Denote $t$ as a transfer level from Player A to Player B, where $t \in \{0, 10, 20, 30, 40, 50\}$. In this setting, the *context* where the third party faces will vary by the available options and by the transfer level. Each treatment offers different options; hence, a context is represented by $(r, t)$.

The normalized contextual concavity model (NCCM) (Kivetz et al., 2004) is able to capture how a compromise effect can affect a third party's choice in our experiment. According to the model, individual $i$'s deterministic component of the utility of alternative $j$ in context $(r, t)$ is equal to the sum of attribute-specific terms that are *concave* functions of the normalized partworth gains in context $(r, t)$:

$$M_{i,j}^{(r,t)} = \sum_k \left(W_{i,max,k}^{(r,t)} - W_{i,min,k}^{(r,t)}\right)\left(\frac{W_{i,j,k}^{(r,t)} - W_{i,min,k}^{(r,t)}}{W_{i,max,k}^{(r,t)} - W_{i,min,k}^{(r,t)}}\right)^{c_k} \quad (1)$$

$$= \sum_k \left(W_{i,max,k}^{(r,t)} - W_{i,min,k}^{(r,t)}\right)^{1-c_k}\left(W_{i,j,k}^{(r,t)} - W_{i,min,k}^{(r,t)}\right)^{c_k} \quad (2)$$

where $M_{i,j}^{(r,t)}$ is $i$'s deterministic component of the utility of alternative $j$ in context $(r,t)$, $W_{i,j,k}^{(r,t)}$ is $i$'s partworth of alternative $j$ with respect to attribute $k$ in context $(r,t)$, $W_{i,max,k}^{(r,t)}$ and $W_{i,min,k}^{(r,t)}$ are the maximum and the minimum partworths, respectively, among available alternatives with respect to attribute $k$ in context $(r,t)$, and $c_k \in (0,1)$ is the concavity parameter of attribute $k$.[1] By observing Equation (2), we see that the first term, $\left(W_{i,max,k}^{(r,t)} - W_{i,min,k}^{(r,t)}\right)^{1-c_k}$, can substantially affect the deterministic component, $M_{i,j}^{(r,t)}$, if the concavity parameter, $c_k$, is close to $0$. This implies that the deterministic component of the utility of the *same* alternative can differ depending on a given context[2] and the impact would be greater as the function is *more* concave.

The overall utility of alternative $j$ in context $(r,t)$ consists of its deterministic component $M_{i,j}^{(r,t)}$ and an error term $\varepsilon_{ij}$. Assuming that $\varepsilon_{ij}$ is i.i.d., the probability of $i$ choosing alternative $j$ in context $(r,t)$ follows the multinomial logit model (McFadden, 1973):

$$\Pr_i\left(j|(r,t)\right) = \frac{\exp\left(bM_{i,j}^{(r,t)}\right)}{\sum_h \exp\left(bM_{i,h}^{(r,t)}\right)} \quad (3)$$

where $b$ is the logit scale parameter.

Considering risk-neutral and risk-averse third parties, we assume the followings:

---

[1] For example, $W_{i,TP,A_M}^{(P,10)}$ is $i$'s partworth of the punishment option ($TP$) with respect to the material-payoff attribute ($A_M$) when Player A transfers 10 tokens to Player B in P treatment $(P,10)$.

[2] For instance, fixing attribute $k$ and transfer level $t$, the difference between the maximum and minimum partworths, $W_{i,max,k}^{(r,t)} - W_{i,min,k}^{(r,t)}$, can be varied by treatment ($r$) since each treatment has a different set of available alternatives. Fixing attribute $k$ and treatment $r$, the difference between the maximum and minimum partworths can also be varied if the partworths fluctuate by transfer level ($t$).

**Assumption 1.** *In terms of the material-payoff attribute, we assume that* $W_{i,S,A_M}^{(r,t)} > W_{i,I,A_M}^{(r,t)} > W_{i,TP,A_M}^{(r,t)}$ *for all* $(r,t)$. *Moreover, the partworths are constant for all* $(r,t)$.

**Assumption 2.** *In terms of the psychological-payoff attribute focusing on the emotion-venting viewpoint, we assume that* $W_{i,TP,A_P}^{(r,t)} > W_{i,I,A_P}^{(r,t)} > W_{i,S,A_P}^{(r,t)}$ *for all* $(r,t)$ *where* $t = 0, 10, \dots, 40$. *Moreover, when t is fixed, the partworths are constant for all* $r$. *When* $r$ *is fixed,* $W_{i,TP,A_P}^{(r,t)} - W_{i,S,A_P}^{(r,t)}$ *and* $W_{i,I,A_P}^{(r,t)} - W_{i,S,A_P}^{(r,t)}$ *decrease in* $t$ *where* $t = 0, 10, \dots, 40$.

Assumption 1 is the partworth assumption for the material-payoff attribute. As discussed in Section 3 of the main text, such partworth inequalities exist because the risk-neutral and risk-averse evaluate the safe option the best, the investment option the second-best, and the costly punishment option the worst in terms of the material-payoff attribute. We also assume that the partworths are constant for all treatments and transfer levels.

Assumption 2 is the partworth assumption for the psychological-payoff attribute focusing on the emotion-venting viewpoint. As we discussed in the main text, such partworth inequalities exist because a third party can directly express one's negative emotion using the punishment option while the safe option means doing nothing. We additionally assume that the size of partworths is constant across all treatments given any unequal transfer. However, the size of partworths fluctuates depending on unequal transfer levels given any treatment. This assumption comes from the characteristics of the emotion-venting viewpoint. In particular, decreasing $W_{i,TP,A_P}^{(r,t)} - W_{i,S,A_P}^{(r,t)}$ and $W_{i,I,A_P}^{(r,t)} - W_{i,S,A_P}^{(r,t)}$ in $t$ implies that negative emotions such as anger are alleviated as Player C observes more fair allocation determined by Player A.

Given Assumptions 1 and 2, we can now attain $M_{i,j}^{(r,t)}$ for each alternative in each context. Throughout the following analysis, we focus on P&I0, P, and I0 treatments[3], and $t \in \{0, \dots, 40\}$. In P&I0 treatment, we have $X^{P\&I0} = \{TP, I, S\}$. Thus, we obtain

$$M_{i,TP}^{(P\&I0,t)} = W_{i,TP,A_P}^{(P\&I0,t)} - W_{i,S,A_P}^{(P\&I0,t)}, \tag{4.1}$$

[3] We can obtain an analogous analysis using P&Ineg and Ineg treatments.

$$M_{i,I}^{(P\&I0,t)} = \left(W_{i,S,A_M}^{(P\&I0,t)} - W_{i,TP,A_M}^{(P\&I0,t)}\right)^{1-c_M}\left(W_{i,I,A_M}^{(P\&I0,t)} - W_{i,TP,A_M}^{(P\&I0,t)}\right)^{c_M}$$
$$+\left(W_{i,TP,A_P}^{(P\&I0,t)} - W_{i,S,A_P}^{(P\&I0,t)}\right)^{1-c_P}\left(W_{i,I,A_P}^{(P\&I0,t)} - W_{i,S,A_P}^{(P\&I0,t)}\right)^{c_P}, \quad (4.2)$$

$$M_{i,S}^{(P\&I0,t)} = W_{i,S,A_M}^{(P\&I0,t)} - W_{i,TP,A_M}^{(P\&I0,t)}. \quad (4.3)$$

In $P$ treatment, we have $X^P = \{TP, S\}$. Thus, we obtain

$$M_{i,TP}^{(P,t)} = W_{i,TP,A_P}^{(P,t)} - W_{i,S,A_P}^{(P,t)}, \quad (5.1)$$

$$M_{i,S}^{(P,t)} = W_{i,S,A_M}^{(P,t)} - W_{i,TP,A_M}^{(P,t)}. \quad (5.2)$$

In $I0$ treatment, we have $X^{I0} = \{I, S\}$. Thus, we obtain

$$M_{i,I}^{(I0,t)} = W_{i,I,A_P}^{(I0,t)} - W_{i,S,A_P}^{(I0,t)}, \quad (6.1)$$

$$M_{i,S}^{(I0,t)} = W_{i,S,A_M}^{(I0,t)} - W_{i,I,A_M}^{(I0,t)}. \quad (6.2)$$

We now derive the first proposition implying the decrease in the probability of choosing the punishment option when the investment option is additionally available.

**Proposition 1.** *The probability of choosing option $TP$ in P&I0 [P&Ineg] treatment is lower than the probability of choosing option $TP$ in P treatment.*

[Proof of Proposition 1] For simplicity, we suppress the subscript $i$. Note that for any $b \in (0,1)$ and $t \in \{0, \dots, 40\}$, the probability of choosing alternative $TP$ in P&I0 treatment is $\Pr\left(TP|(P\&I0,t)\right) = \frac{\exp\left(bM_{TP}^{(P\&I0,t)}\right)}{\exp\left(bM_{TP}^{(P\&I0,t)}\right)+\exp\left(bM_{I}^{(P\&I0,t)}\right)+\exp\left(bM_{S}^{(P\&I0,t)}\right)}$, and the probability of choosing alternative $TP$ in P treatment is $\Pr\left(TP|(P,t)\right) = \frac{\exp\left(bM_{TP}^{(P,t)}\right)}{\exp\left(bM_{TP}^{(P,t)}\right)+\exp\left(bM_{S}^{(P,t)}\right)}$. By Assumptions 1 and 2, we have $W_{TP,A_P}^{(P\&I0,t)} - W_{S,A_P}^{(P\&I0,t)} = W_{TP,A_P}^{(P,t)} - W_{S,A_P}^{(P,t)} > 0$ and $W_{S,A_M}^{(P\&I0,t)} - W_{TP,A_M}^{(P\&I0,t)} = W_{S,A_M}^{(P,t)} - W_{TP,A_M}^{(P,t)} > 0$, respectively. Then, we obtain $M_{TP}^{(P\&I0,t)} = M_{TP}^{(P,t)}$ by Equation (4.1) and (5.1). We also obtain $M_{S}^{(P\&I0,t)} = M_{S}^{(P,t)}$ by Equation (4.3) and (5.2). Since $bM_{I}^{(P\&I0,t)}$ is strictly positive for all $t = 0, \dots, 40$, it follows that $\Pr\left(TP|(P\&I0,t)\right) < \Pr\left(TP|(P,t)\right)$ for all unequal allocations.

The crucial part of the proof of Proposition 1 is an additional term, $\exp\left(bM_I^{(P\&I0,t)}\right) > 0$, in the denominator of $\Pr\left(TP|(P\&I0,t)\right)$. The term is due to the different set of available alternatives between the treatments.

Similar to the proof of Proposition 1, the probabilities of choosing the investment option when the punishment option is available and when it is not available can also be compared by considering P&I0 and I0 treatment. However, in this case, we have an ambiguous result. In particular, the probabilities of choosing the investment option are given by $\Pr\left(I|(P\&I0,t)\right) = \frac{\exp\left(bM_I^{(P\&I0,t)}\right)}{\exp\left(bM_{TP}^{(P\&I0,t)}\right)+\exp\left(bM_I^{(P\&I0,t)}\right)+\exp\left(bM_S^{(P\&I0,t)}\right)}$ and $\Pr\left(I|(I0,t)\right) = \frac{\exp\left(bM_I^{(I0,t)}\right)}{\exp\left(bM_I^{(I0,t)}\right)+\exp\left(bM_S^{(I0,t)}\right)}$ in P&I0 and I0 treatments, respectively. First, from Equation (4.3) and (6.2), we obtain $M_S^{(P\&I0,t)} > M_S^{(I0,t)}$ since $W_{S,A_M}^{(P\&I0,t)} = W_{S,A_M}^{(I0,t)}$ and $W_{TP,A_M}^{(P\&I0,t)} < W_{I,A_M}^{(P\&I0,t)} = W_{I,A_M}^{(I0,t)}$ by Assumption 1. Second, we have $W_{I,A_P}^{(I0,t)} - W_{S,A_P}^{(I0,t)} = \left(W_{I,A_P}^{(I0,t)} - W_{S,A_P}^{(I0,t)}\right)^{1-c_P}\left(W_{I,A_P}^{(I0,t)} - W_{S,A_P}^{(I0,t)}\right)^{c_P} < \left(W_{TP,A_P}^{(P\&I0,t)} - W_{S,A_P}^{(P\&I0,t)}\right)^{1-c_P}\left(W_{I,A_P}^{(P\&I0,t)} - W_{S,A_P}^{(P\&I0,t)}\right)^{c_P}$ since $W_{I,A_P}^{(I0,t)} = W_{I,A_P}^{(P\&I0,t)} < W_{TP,A_P}^{(P\&I0,t)}$ and $W_{S,A_P}^{(I0,t)} = W_{S,A_P}^{(P\&I0,t)}$ by Assumption 2. Thus, from Equation (4.2) and (6.1), we obtain $M_I^{(P\&I0,t)} > M_I^{(I0,t)}$. Third, $bM_{TP}^{(P\&I0,t)}$ is strictly positive. Hence, we observe that both numerator and denominator of $\Pr\left(I|(P\&I0,t)\right)$ are larger than those of $\Pr\left(I|(I0,t)\right)$, respectively. In other words, we cannot conclude whether the probability of choosing the investment option is larger if the punishment option is available as well. However, by postulating the following additional assumption, we can proceed with the analysis of a third party's investment behavior.

**Assumption 3.** *For any* $b \in (0,1)$ *and for any* $t = 0, 10, \ldots, 40$, *we assume that* $\exp\left(bM_{i,I}^{(P\&I0,t)} + bM_{i,S}^{(I0,t)}\right) > \exp\left(bM_{i,I}^{(I0,t)} + bM_{i,TP}^{(P\&I0,t)}\right) + \exp\left(bM_{i,I}^{(I0,t)} + bM_{i,S}^{(P\&I0,t)}\right)$.

Assumption 3 can be satisfied if we have a sufficiently large deterministic component of the utility of option $I$ in P&I0 treatment, $M_{i,I}^{(P\&I0,t)}$, and we can achieve large $M_{i,I}^{(P\&I0,t)}$ when the concavity parameters, $c_M$ and $c_P$, are sufficiently small enough. Recall that the concavity parameters $c_M$ and $c_P$ are only involved with $M_{i,I}^{(P\&I0,t)}$ which is an *intermediate* option (see Equation (4.2)). This is crucial to capture the compromise effect because the value of the attribute-specific deterministic component for the intermediate investment option gets higher as its function gets more concave.[4]

One might ask how reasonable Assumption 3 is, and we can partially answer it by the following simulation result. In the simulation, we assume that $b = 0.05$. Moreover, for simplicity, we assume that $c_M = c_P$, and $W_{i,TP,A_M}^{(r,t)} = 0$, $W_{i,I,A_M}^{(r,t)} = 25$, $W_{i,S,A_M}^{(r,t)} = 50$ and $W_{i,TP,A_P}^{(r,t)} = 50$, $W_{i,I,A_P}^{(r,t)} = 25$, $W_{i,S,A_P}^{(r,t)} = 0$ for each $t$.[5] Table B1 reports the left-hand side value (LHS) and the right-hand side value (RHS) of Assumption 3 based on the simulation. We observe that LHS is higher than RHS for all $c_M = c_P \leq 0.5$ and the difference in values gets larger as the value of the concavity parameters becomes smaller. In fact, two values are equal when $c_M = c_P = 0.6469$. Therefore, we see that Assumption 3 is satisfied when the concavity parameters are small enough. Given that estimated concavity parameters from the datasets in Kivetz et al. (2004) are 0.160, 0.315, and 0.364, we conclude that Assumption 3 is not a pathological assumption.

---

[4] According to Equation (4.2), as the size of the concavity parameters gets smaller, it puts more weight on the term $W_{i,S,A_M}^{(P\&I0,t)} - W_{i,TP,A_M}^{(P\&I0,t)}$ for the material payoff attribute and/or the term $W_{i,TP,A_P}^{(P\&I0,t)} - W_{i,S,A_P}^{(P\&I0,t)}$ for the psychological payoff attribute, which are larger than the other multiplied terms. Of course, the context, what types of options are available, also plays a crucial role to capture the compromise effect since it determines the minimum and the maximum partworths.

[5] The values that we used in the simulation are based on the estimation results in Kivetz et al. (2004). The estimations of $b$ are 0.051, 0.064, and 0.075 from the datasets in Kivetz et al. (2004). When eliciting the partworths in Kivetz et al. (2004), the paper let the sum of ranges of partworths for all attributes equals 100. Following this in our setting, we need to have $W_{i,TP,A_P}^{(r,t)} - W_{i,S,A_P}^{(r,t)} + W_{i,S,A_M}^{(r,t)} - W_{i,TP,A_M}^{(r,t)} = 50 + 50 = 100$. Given this condition, we give the most moderate values to partworths, i.e., $W_{i,TP,A_M}^{(r,t)} = 0$, $W_{i,I,A_M}^{(r,t)} = 25$, $W_{i,S,A_M}^{(r,t)} = 50$ and $W_{i,TP,A_P}^{(r,t)} = 50$, $W_{i,I,A_P}^{(r,t)} = 25$, $W_{i,S,A_P}^{(r,t)} = 0$. The assumption $c_M = c_P$ is consistent with the result of Kivetz et al. (2004), and estimated values of concavity parameters from the datasets in Kivetz et al. (2004) are 0.160, 0.315, and 0.364.

**Table B1. Simulation results**

| $c_M = c_P$ | 0.5 | 0.4 | 0.3 | 0.2 | 0.1 |
|---|---|---|---|---|---|
| LHS | 119.77 | 154.36 | 202.6058 | 271.17 | 370.62 |
| RHS | 85.04 | 85.04 | 85.04 | 85.04 | 85.04 |

Note. LHS indicates the left-hand side value of $\exp\left(bM_{i,I}^{(P\&I0,t)} + bM_{i,S}^{(I0,t)}\right)$ and RHS indicates the right-had side value of $\exp\left(bM_{i,I}^{(I0,t)} + bM_{i,TP}^{(P\&I0,t)}\right) + \exp\left(bM_{i,I}^{(I0,t)} + bM_{i,S}^{(P\&I0,t)}\right)$ in Assumption 3. The values are calculated by assuming $b = 0.05$, $c_M = c_P$, and $W_{i,TP,A_M}^{(r,t)} = 0$, $W_{i,I,A_M}^{(r,t)} = 25$, $W_{i,S,A_M}^{(r,t)} = 50$ and $W_{i,TP,A_P}^{(r,t)} = 50$, $W_{i,I,A_P}^{(r,t)} = 25$, $W_{i,S,A_P}^{(r,t)} = 0$ for each $t$.

With Assumption 3, we derive the second proposition implying the increase in the probability of choosing the investment option when the punishment option is additionally available.

**Proposition 2.** *The probability of choosing option I in P&I0 [P&Ineg] treatment is higher than the probability of choosing option I in I0 [Ineg] treatment.*

[Proof of Proposition 2] For any $b \in (0, 1)$ and $t \in \{0, \dots, 40\}$, the probabilities of choosing option $I$ are given by $\Pr\left(I|(P\&I0, t)\right) = \frac{\exp\left(bM_I^{(P\&I0,t)}\right)}{\exp\left(bM_{TP}^{(P\&I0,t)}\right)+\exp\left(bM_I^{(P\&I0,t)}\right)+\exp\left(bM_S^{(P\&I0,t)}\right)}$ and $\Pr\left(I|(I0, t)\right) = \frac{\exp\left(bM_I^{(I0,t)}\right)}{\exp\left(bM_I^{(I0,t)}\right)+\exp\left(bM_S^{(I0,t)}\right)}$ in P&I0 and I0 treatments, respectively. Then, by Assumption 3, we have $\Pr\left(I|(P\&I0, t)\right) > \Pr\left(I|(I0, t)\right)$ for all unequal allocations.

To sum up, when focusing on the emotion-venting viewpoint in the psychological-payoff attribute, Propositions 1 and 2 show that the probability of investing increases and the probability of punishing decreases if all options are available. This indicates that the investment option works as a compromise, and we construct hypotheses for our experiment based on the compromise effect as below:

**Hypothesis 1.** *For the risk-neutral and the risk-averse, the choice frequency and the expenditure of punishment in P&I0 [P&Ineg] treatment are lesser than those in P treatment.*

**Hypothesis 2.** *For the risk-neutral and the risk-averse, the choice frequency and the expenditure of investment in P&I0 [P&Ineg] treatment are greater than those in I0 [Ineg] treatment.*

**Appendix B.2. Hypotheses Based on a Decoy Effect Generated by the Inequity-averse Viewpoint**

In this subsection, we consider the psychological-payoff attribute in the "inequity-averse" point of view instead of the "emotion-venting" point of view. Since inequity aversion is also known as one of the main motivations for third-party punishment (Leibbrandt and López-Pérez, 2011, 2012; Kamei, 2020), it is worth noting to build hypotheses based on the inequity aversion viewpoint and compare them with the previously derived hypotheses based on the emotion-venting viewpoint.

When risk-neutral and risk-averse third parties rank alternatives in terms of the material-payoff attribute, we admit Assumption 1, i.e., $W_{i,S,A_M}^{(r,t)} > W_{i,I,A_M}^{(r,t)} > W_{i,TP,A_M}^{(r,t)}$. We follow the model of Saito (2013) when ranking alternatives in terms of the inequity-averse viewpoint since the model suggests how to apply the notion of inequity aversion under uncertainty. The Saito (2013) model basically follows the Fehr and Schmidt (1999) model when capturing inequity aversion, i.e., if the opponent's payoff is higher, then the disutility parameter of *envy*, $\alpha$, is involved, and if the opponent's payoff is lower, then the disutility parameter of *guilt*, $\beta$, is involved. Here, we assume that $\alpha + \beta > 1$ and $\alpha > \beta > 0$. The novelty of the Saito (2013) model is that it considers inequity aversion both from an *ex-ante* point of view and an *ex-post* point of view, i.e., inequity aversion is associated with both *ex-ante expected allocations* and *ex-post allocations*.

To describe formally, let $p$ be a punishment level where $0 \leq p \leq 50$ and $z$ be an investment level where $0 \leq z \leq 50$. Remind that $t$ is a transfer level where $t \in \{0, 10, 20, 30, 40, 50\}$. Let $[A, B, C]$ be a vector of allocations where $A$ is the allocation for Player A, $B$ is the allocation for Player B, and $C$ is the allocation for Player C. Since the investment option in P&I0 treatment returns double or nothing with equal probability, a vector of *ex-ante expected allocations* is given by $[100 - t - 3p, t, 50 - p]$, and vectors of *ex-post allocations* are given by $[100 - t - 3p, t, 50 - p + z]$ with probability half and $[100 - t - 3p, t, 50 - p - z]$ with probability half.

To compute the partworth of the investment option, we assume that $p = 0$ and $z > 0$. If $z \leq 50 - t$, the *ex-ante expected allocations* gives an utility of $50 - (100 - t - 50)\alpha - (50 - t)\beta$, and the *ex-post allocations* gives an utility of $\frac{1}{2}\{50 + z - (100 - t - 50 - z)\alpha - (50 + z - t)\beta\} + \frac{1}{2}\{50 - z - (100 - t - 50 + z)\alpha - (50 - z - t)\beta\}$. The linear combination, $\delta[50 -$ $(100 - t - 50)\alpha - (50 - t)\beta] + (1 - \delta)\left[\frac{1}{2}\{50 + z - (100 - t - 50 - z)\alpha - (50 + z - t)\beta\} + \frac{1}{2}\{50 - z - (100 - t - 50 + z)\alpha - (50 - z - t)\beta\}\right]$, where $\delta$ is the weighting parameter, is the final partworth of the investment option. Similarly, if $z > 50 - t$, then the final partworth of the investment option is computed by $\delta[50 - (100 - t - 50)\alpha - (50 - t)\beta] + (1 - \delta)\left[\frac{1}{2}\{50 + z - (50 + z - 100 + t + 50 + z - t)\beta\} + \frac{1}{2}\{50 - z - (100 - t - 50 + z + t - 50 + z)\alpha\}\right]$. Thus, we obtain

$$W_{i,I,A_P}^{(r,t)} = \begin{cases} 50 - (50 - t)(\alpha + \beta) & if\ z \leq 50 - t \\ 50 - (50 - t)(\alpha + \beta) + (1 - \delta)(50 - t - z)(\alpha + \beta) & if\ z > 50 - t \end{cases}. \quad (7)$$

For the safe option, we assume that $p = 0$ and $z = 0$. We obtain

$$W_{i,S,A_P}^{(r,t)} = 50 - (50 - t)(\alpha + \beta) \quad (8)$$

since the *ex-ante* and *ex-post* allocations are the same. If $z \leq 50 - t$, then $W_{i,S,A_P}^{(r,t)} = W_{i,I,A_P}^{(r,t)}$. If $z > 50 - t$, then $W_{i,S,A_P}^{(r,t)} > W_{i,I,A_P}^{(r,t)}$ since $(1 - \delta)(50 - t - z)(\alpha + \beta) > 0$. Hence, we verify that $W_{i,S,A_P}^{(r,t)} \geq W_{i,I,A_P}^{(r,t)}$ for all $z$.

Next, we show that $W_{i,TP,A_P}^{(r,t)} \geq W_{i,S,A_P}^{(r,t)}$ for all $p$. Assume $p > 0$ and $z = 0$ to get the partworth of the punishment option. Then we obtain

$$W_{i,TP,A_P}^{(r,t)} = \begin{cases} 50 - (50 - t)(\alpha + \beta) - p(1 - 2\alpha - \beta) & if\ p < \frac{50 - t}{2} \\ 50 - p\beta & if\ \frac{50 - t}{2} \leq p < 50 - t \\ 50 - (t - 50)(\alpha + \beta) - p(1 - \alpha - 2\beta) & if\ p \geq 50 - t \end{cases}. \quad (9)$$

If $p \leq \frac{50-t}{2}$, then $-p(1 - 2\alpha - \beta) > 0$. If $\frac{50-t}{2} \leq p < 50 - t$, then $50 - p\beta > 50 - (50 - t)\beta \geq 50 - (50 - t)\alpha - (50 - t)\beta$. If $p \geq 50 - t$, then $50 - (t - 50)(\alpha + \beta) - p(1 - \alpha - 2\beta) \geq 50 + (50 - t)(\alpha + \beta) - (50 - t)(1 - \alpha - 2\beta) = 50 + (50 - t)(2\alpha + 3\beta - 1) \geq 50$. Thus, we verify that $W_{i,TP,A_P}^{(r,t)} \geq W_{i,S,A_P}^{(r,t)}$ for all $p$.

To summarize, we have $W_{i,TP,A_P}^{(r,t)} \geq W_{i,S,A_P}^{(r,t)} \geq W_{i,I,A_P}^{(r,t)}$ in the inequity-averse point of view.

Remind that we have $W^{(r,t)}_{i,S,A_M} > W^{(r,t)}_{i,I,A_M} > W^{(r,t)}_{i,TP,A_M}$ in terms of the material-payoff attribute. Combining these, we find that the investment option works as a *decoy* for the safe option instead of being a compromise. The decoy effect implies that the safe option would be more attractive in P&I0 treatment where three options are available compared to P treatment where only the punishment option and the safe option are available. Then, due to the increased attractiveness of the safe option, the choice probability of the punishment option would decrease if the investment option is additionally available.

When the investment option yields a zero expected net return, we have a weak inequality between the partworths of the investment and the safe options in the psychological-payoff attribute. In the supplementary treatments where the investment option has a negative expected net return, the investment option works as a decoy for the safe option with strict inequalities in both attributes. Specifically, when the investment option yields a negative expected net return, a vector of *ex-ante expected allocations* is given by $\left[100 - t - 3p, t, 50 - p - \frac{1}{4}z\right]$, and vectors of *ex-post allocations* are given by $\left[100 - t - 3p, t, 50 - p + \frac{1}{2}z\right]$ with probability half and $[100 - t - 3p, t, 50 - p - z]$ with probability half. Then the partworth of the investment option is given by

$$W^{(r,t)}_{i,I,A_P} = \begin{cases} F & if\ z < 50 - t \\ F + (1-\delta)\left(25 - \frac{t}{2} - \frac{z}{2}\right)(\alpha+\beta) & if\ 50 - t \le z < 2(50-t) \\ F + (1-\delta)\left(50 - t - \frac{3z}{4}\right)(\alpha+\beta) & if\ 2(50-t) \le z < 4(50-t) \\ 50 - \frac{z}{4} - \frac{z}{2}\alpha - (1-\delta)\frac{z}{2}(\alpha+\beta) & if\ z \ge 4(50-t) \end{cases} \quad (10)$$

where $F = 50 - (50-t)(\alpha+\beta) - \frac{z}{4}(1+\alpha-\beta)$. If $z < 50 - t$, then $F < 50 - (50 - t)(\alpha+\beta)$ since $1+\alpha-\beta > 0$. If $50 - t \le z < 2(50-t)$, then $25 - \frac{t}{2} - \frac{z}{2} < 0$. If $2(50 - t) \le z < 4(50-t)$, then $50 - t - \frac{3z}{4} < 0$. If $z \ge 4(50-t)$, then $50 - \frac{z}{4} - \frac{z}{2}\alpha - (1 - \delta)\frac{z}{2}(\alpha+\beta) \le 50 - (50-t) - 2(50-t)\alpha - (1-\delta)2(50-t)(\alpha+\beta) = t - 2(50-t)\alpha - 2(1-\delta)(50-t)(\alpha+\beta) < 50 - (50-t)(\alpha+\beta)$ since $t \le 50$ and $\alpha > \beta$. Therefore, we verify that $W^{(r,t)}_{i,I,A_P} > W^{(r,t)}_{i,S,A_P}$ for all $z$.

To summarize, when the investment option yields a negative expected net return, we have $W^{(r,t)}_{i,TP,A_P} \geq W^{(r,t)}_{i,S,A_P} > W^{(r,t)}_{i,I,A_P}$ in the inequity-averse viewpoint and $W^{(r,t)}_{i,S,A_M} > W^{(r,t)}_{i,I,A_M} > W^{(r,t)}_{i,TP,A_M}$ in terms of the material-payoff attribute. In this case, the investment option is a decoy for the safe option with strict inequalities in both attributes. Hence, the argument of the decoy effect is valid as well.

Two hypotheses for our experiment based on the decoy effect are summarized below:

**Hypothesis 3.** *For the risk-neutral and the risk-averse, the choice frequency and the expenditure of punishment in P&I0 [P&Ineg] treatment are lesser than those in P treatment.*

**Hypothesis 4.** *For the risk-neutral and the risk-averse, the choice frequency and the expenditure of the safe option in P&I0 [P&Ineg] treatment are larger than those in P treatment.*

Hypothesis 3 is identical to Hypothesis 1, which means that both the compromise effect and the decoy effect agree that the demand for punishment would decrease when the investment option is available. However, the two effects provide different predictions for other behaviors (Hypothesis 2 and Hypothesis 4).

**References for Appendix B**

## Appendix C. Robustness Check for the Third Party's Behavior

In Section 4.1 of the main text, we have shown that TPP is fragile in the presence of the investment option where the expected net return equals zero. In addition, the findings from Section 4.2 show that Player C is likely to, or at least not decrease the amount of, investment in the unattractive lottery if TPP is available simultaneously. In this section, we check the robustness of our findings.

### C.1. All Third Parties in P, P&I0, and I0 Treatments

In this subsection, we check whether our findings are robust even if we consider all third parties regardless of their choices in Task 1. Note that there are 27, 28, and 23 third parties in P, P&I0, and I0 treatments, respectively (see Table 4 in the main text).

Figures C1a and C1b show that Player C is less willing to punish Player A at unequal transfer levels when the investment option was additionally available compared with a circumstance when the punishment and the safe options were only available. The percentages of punishers are significantly smaller in P treatment than in P&I0 treatment at the transfer levels of 10, 20, and 30 at the significance level of 5% (p=[0.106, 0.031, 0.015, 0.029, 0.227, 1.000], two-sided Fisher's exact test)[6]. Likewise, the expenditures for deduction points are significantly lower in P treatment than in P&I0 treatment at the transfer levels of 0 to 30 (p=[0.023, 0.010, 0.005, 0.012, 0.153, 1.000], two-sided Wilcoxon rank-sum test). We further examine a third party's punishing behavior using the individual-level data as presented in Table C1. The averages of mean [median] punishment are 7.54 [7.28] in P treatment and 3.10 [2.71] in P&I0 treatment, and the difference is statistically significant (p=0.017 [p=0.004], two-sided Wilcoxon rank-sum test).

Figures C1c and C1d show that Player C is more willing to invest in the lottery when the punishment option is additionally available. The difference in investment proportions is statistically significant at all transfer levels (p=[0.023, 0.030, 0.003, 0.007, 0.007, 0.016], two-sided Fisher's exact test). Likewise, the difference in investment expenditures is significantly different at most of the transfer levels (p=[0.035, 0.077, 0.023, 0.057, 0.022, 0.041], two-sided Wilcoxon rank-sum test). Using the individual-level data as described in Table C1, we obtain 14.74

[6] Consistent with the main text, p is a vector collecting p-values at transfer levels of 0, 10, 20, 30, 40, and 50.

[14.52] and 23.10 [23.36] as the averages of mean [median] investment in I0 treatment and P&I0 treatment, respectively, and the difference is marginally insignificant [significant] (p=0.051 [p=0.031], two-sided Wilcoxon rank-sum test).

Hence, we conclude that our main findings are robust even if we consider all third parties.

**Table C1. Punishment and investment using the individual-level data (with all third parties)**

| | Punishment All third parties | | | | Investment All third parties | | | |
|---|---|---|---|---|---|---|---|---|
| | Mean | p-value | Median | p-value | Mean | p-value | Median | p-value |
| P [I0] | 7.54 | | 7.28 | | 14.74 | | 14.52 | |
| | (7.13) | 0.017 | (7.01) | 0.004 | (15.81) | 0.051 | (16.11) | 0.031 |
| P&I0 | 3.10 | | 2.71 | | 23.10 | | 23.36 | |
| | (5.79) | | (5.88) | | (16.54) | | (16.80) | |

Note: Mean and median refers to an average using mean and median individual-level data, respectively. Standard deviations are in parentheses. p-values are from the two-sided Wilcoxon rank-sum test. There are 27, 28, and 23 third parties in P, P&I0, and I0 treatments, respectively.

**Figure C1. Punishment and investment in P, I0, and P&I0 treatments (with all third parties)**

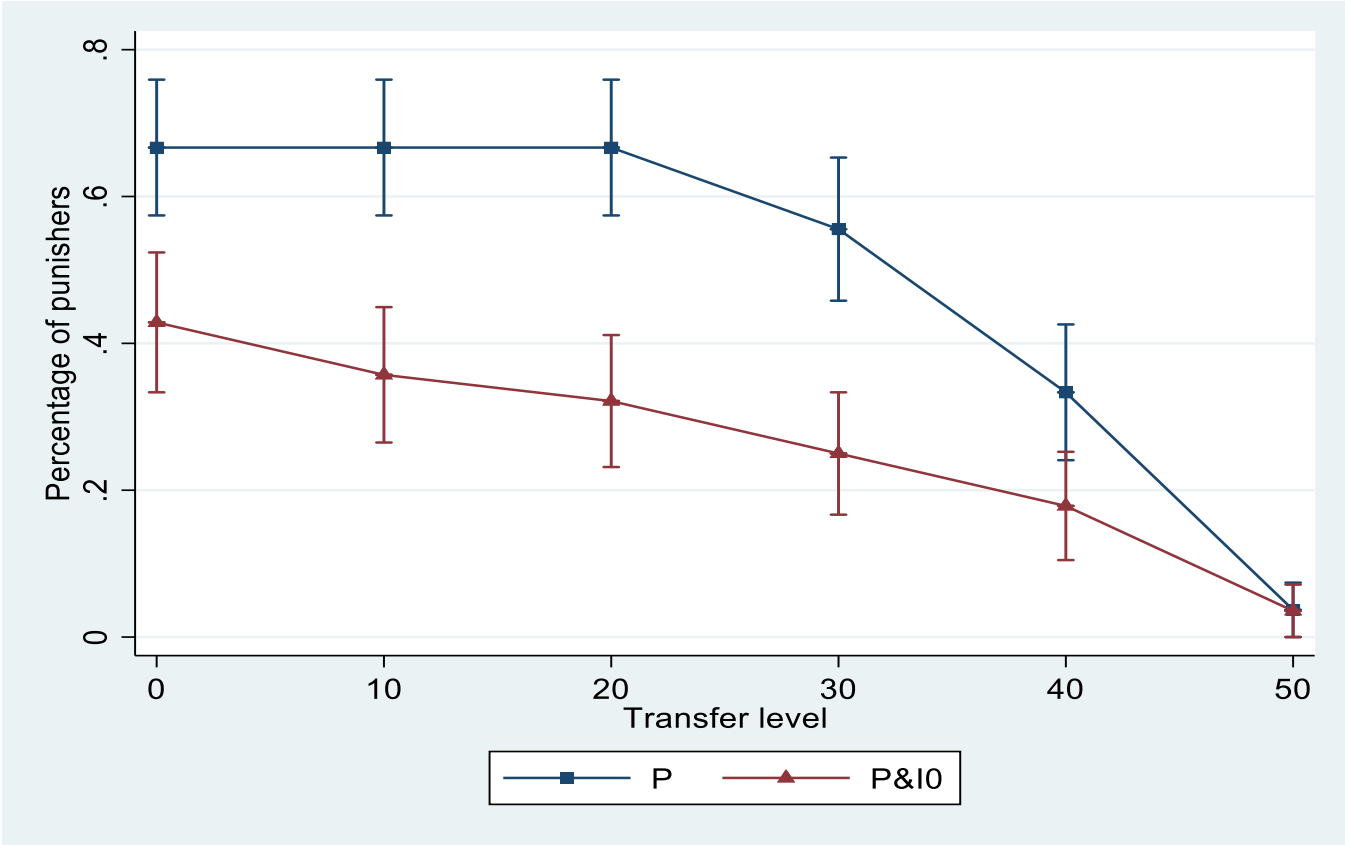

**a.** Percentage of punishers

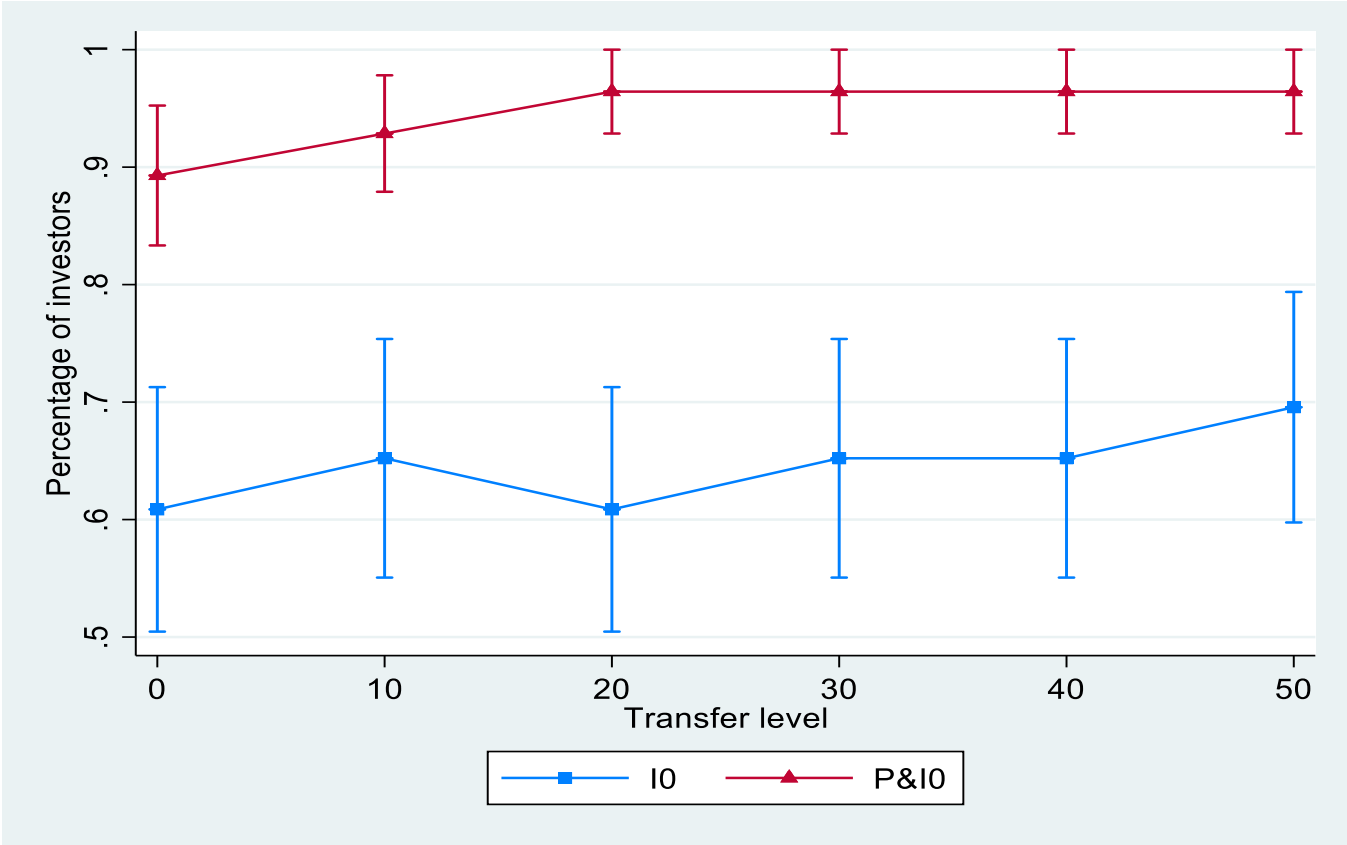

**c.** Percentage of investors

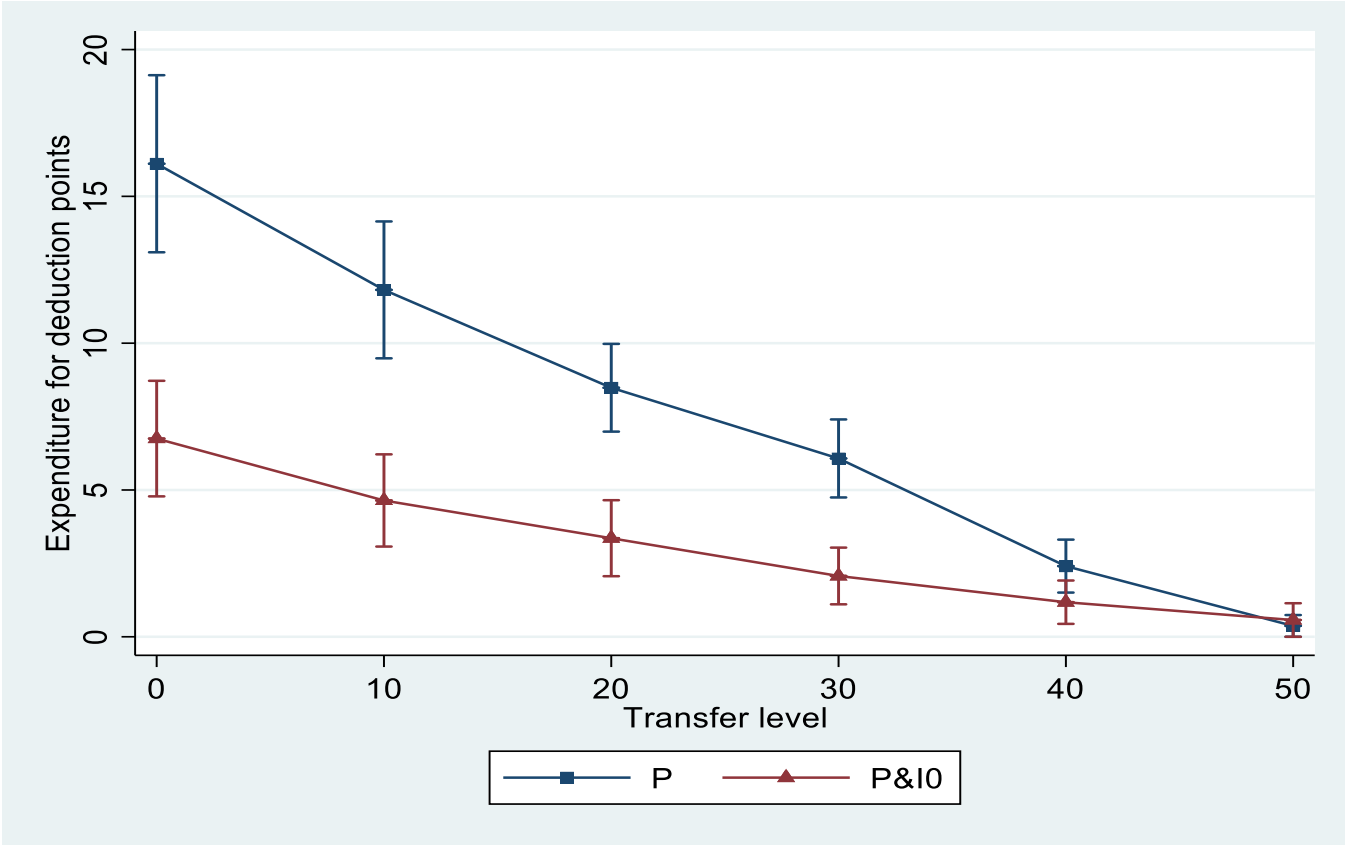

**b.** Average expenditure for deduction points

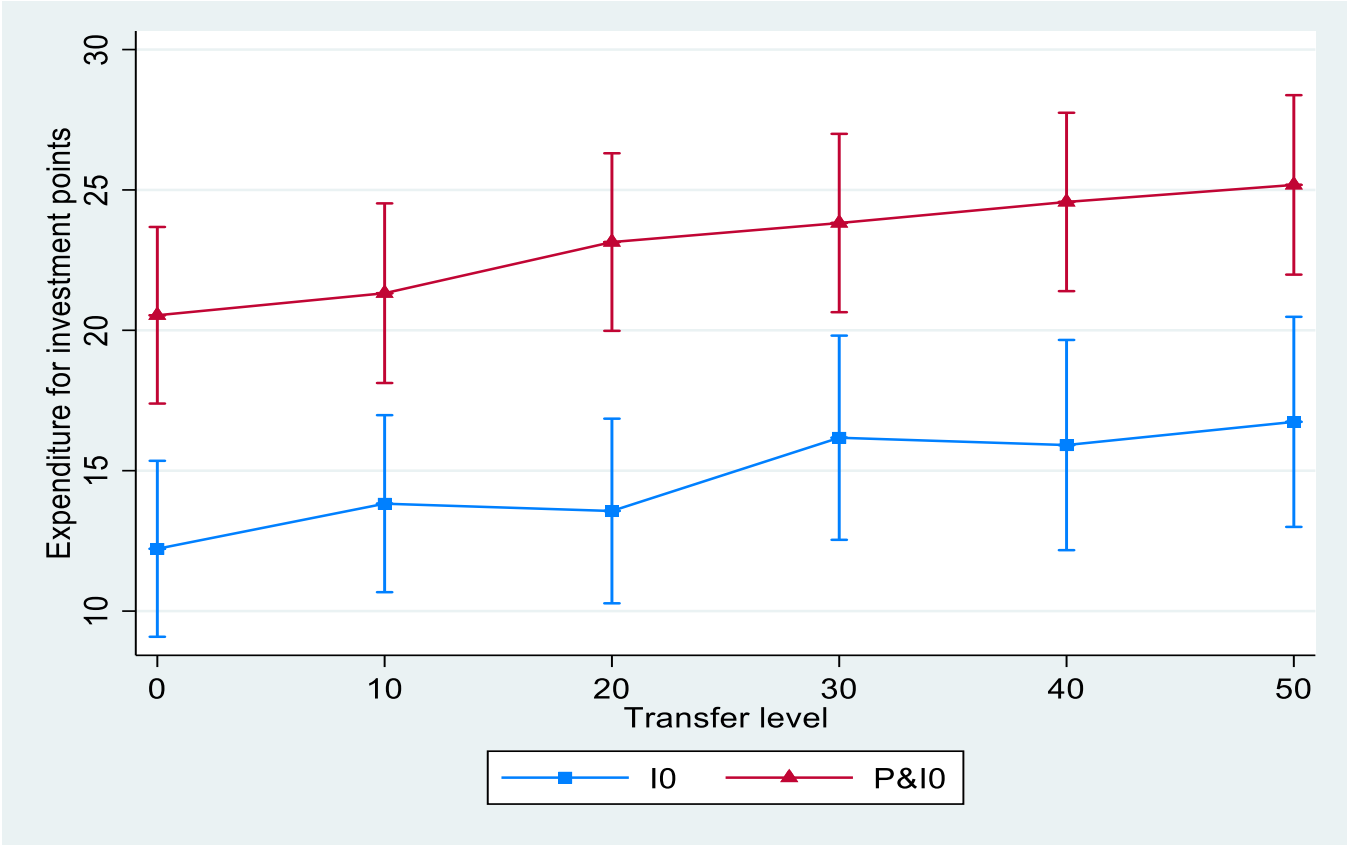

**d.** Average expenditure for investment points

Note: A punisher [investor] is defined as Player C who spent at least one token to deduction [investment] points. Expenditure for deduction [investment] points is the number of tokens that Player C spent for deduction [investment] points among 50 tokens. Transfer level is the number of tokens that Player A transferred to Player B. A vertical bar indicates one standard error.

### C.2. Risk-consistent Third Parties in P, P&I0, and I0 Treatments

In this subsection, we check the robustness of our findings by considering risk-consistent third parties. Note that there are 27, 26, and 22 risk-consistent third parties in P, P&I0, and I0 treatments, respectively (see Table 4 in the main text).

Consistent with our main findings, Figures C2a and C2b show that risk-consistent Player C is less likely to punish Player A at unequal transfer levels when the investment option was also available. We find that the percentages of punishers are significantly smaller in P treatment than in P&I0 treatment at the transfer levels of 10, 20, and 30 (p=[0.101, 0.029, 0.013, 0.024, 0.202, 1.000], two-sided Fisher's exact test) and the expenditures for deduction points are significantly lower in P treatment than in P&I0 treatment at the transfer levels of 0 to 30 (p=[0.032, 0.013, 0.007, 0.014, 0.123, 0.957], two-sided Wilcoxon rank-sum test). We have consistent results using the individual-level data as well. According to Table C2, the averages of mean [median] punishment are 7.54 [7.28] and 3.23 [2.83] in P treatment and P&I0 treatments, respectively, and the difference is statistically significant (p=0.023 [p=0.005], two-sided Wilcoxon rank-sum test).

One may argue that the diminished demand for TPP can be attributed to risk-loving third parties when considering all the risk-consistent parties. Specifically, in P&I0 treatment, a risk-loving punisher's problem is deciding how many tokens one will allocate to deduction and investment points within one's budget constraint. Due to the constraint, the risk-loving may decrease the punishment expenditure and increase the investment expenditure. Hence, if risk-loving third parties are greater in P&I0 treatment than in P treatment, then the diminished demand for TPP could occur without the treatment effect. However, we find no strong evidence that the distribution of Player C's risk attitude is heterogeneous between the treatments.[7]

Figures C2c and C2d show that risk-consistent Player C is more likely to invest in P&I0 treatment than in I0 treatment. We find that the difference in investment proportions is statistically significant at all transfer levels (p=[0.029, 0.017, 0.001, 0.002, 0.002, 0.006], two-sided Fisher's exact test) and the difference in investment expenditures is significantly different at most of the transfer levels (p=[0.038, 0.070, 0.021, 0.060, 0.021, 0.039], two-sided Wilcoxon rank-sum test).

---

[7] The averages of Player C's *Switching point* are 6.00 and 5.81 in P and P&I0 treatments, respectively, and the difference is insignificant (p>0.742, two-sided Wilcoxon rank-sum test). This implies the non-heterogeneity of distribution of Player C's risk attitude between the treatments.

We also have a similar result using the individual-level data. According to Table C2, the averages of mean [median] investment are 15.41 [15.18] and 24.15 [24.38] in I0 treatment and P&I0 treatment, respectively, and the difference is statistically significant (p=0.046 [p=0.032], two-sided Wilcoxon rank-sum test).

Likewise, the increased demand for investment can be caused by the biased selection of Player C in terms of risk attitude. For instance, when risk-loving third parties are greater in P&I0 treatment than in I0 treatment, then the increased demand we observed above could happen regardless of the treatment effect. However, the distributions of Player C's risk attitude between P&I0 and I0 treatments do not seem to be heterogeneous.[8]

Therefore, we check the similar punishment and investment behaviors even when we consider the risk-consistent third parties.

**Table C2. Punishment and investment using the individual-level data (with risk-consistent)**

| | Punishment Risk-consistent | | | | Investment Risk-consistent | | | |
|---|---|---|---|---|---|---|---|---|
| | Mean | p-value | Median | p-value | Mean | p-value | Median | p-value |
| P [I0] | 7.54 | | 7.28 | | 15.41 | | 15.18 | |
| | (7.13) | 0.023 | (7.01) | 0.005 | (15.84) | 0.046 | (16.17) | 0.032 |
| P&I0 | 3.23 | | 2.83 | | 24.15 | | 24.38 | |
| | (5.99) | | (6.09) | | (16.50) | | (16.78) | |

Note: Mean and median refers to an average using mean and median individual-level data, respectively. Standard deviations are in parentheses. p-values are from the two-sided Wilcoxon rank-sum test. There are 27, 26, and 22 risk-consistent third parties in P, P&I0, and I0 treatments, respectively.

[8] The averages of Player C's *Switching point* are 6.23 and 5.81 in I0 and P&I0 treatments, respectively, and the difference is not statistically significant (p>0.404, two-sided Wilcoxon rank-sum test). This implies the non-heterogeneity of distribution of Player C's risk attitude between the treatments. In addition, for risk-loving Player Cs, the averages of investment are 24.44 (n=3) in I0 treatment and 24.75 (n=4) in P&I0 treatment. Although we cannot conduct any statistical test due to the small sample size, the difference seems not large.

**Figure C2. Punishment and investment in P, I0, and P&I0 treatments (with risk-consistent)**

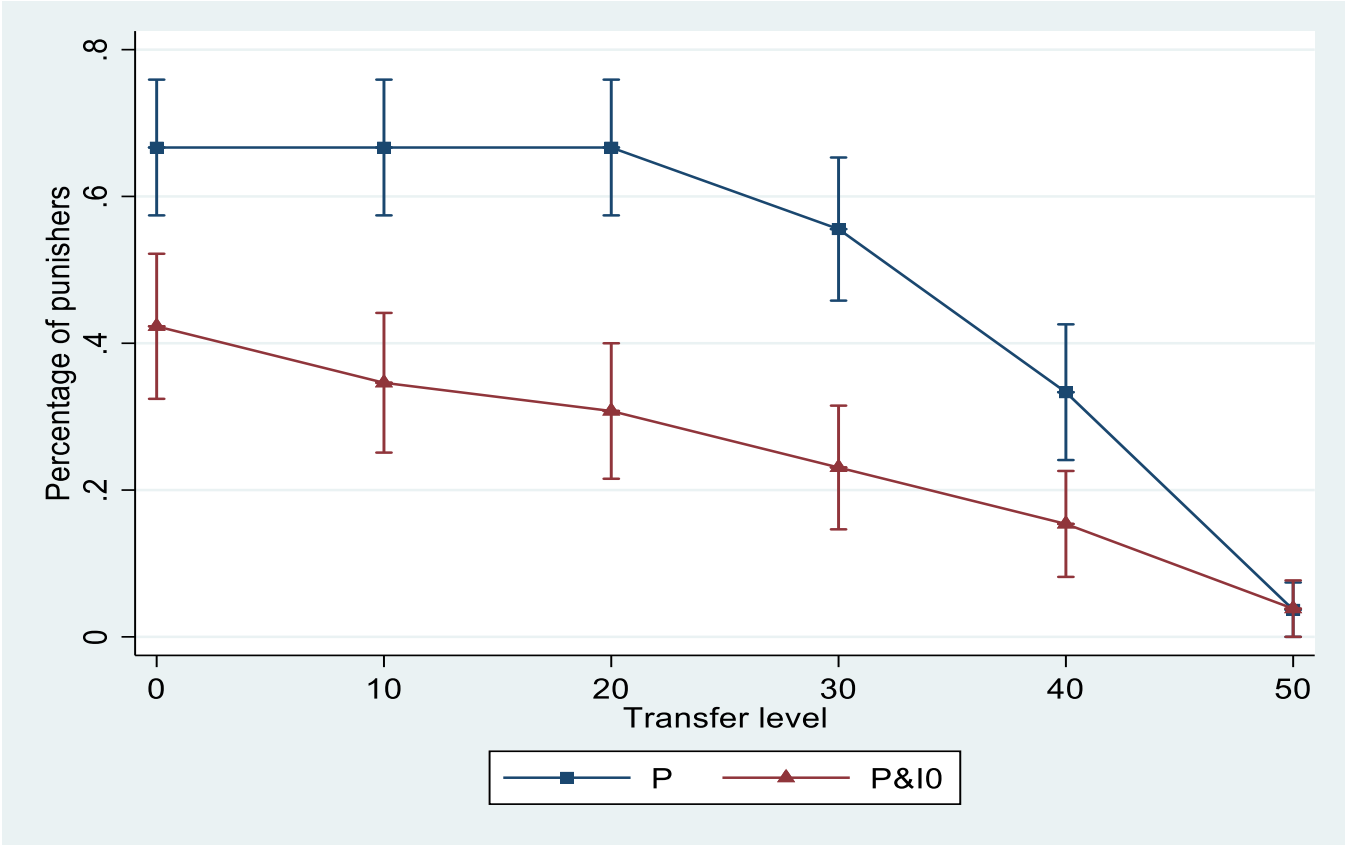

**a.** Percentage of punishers

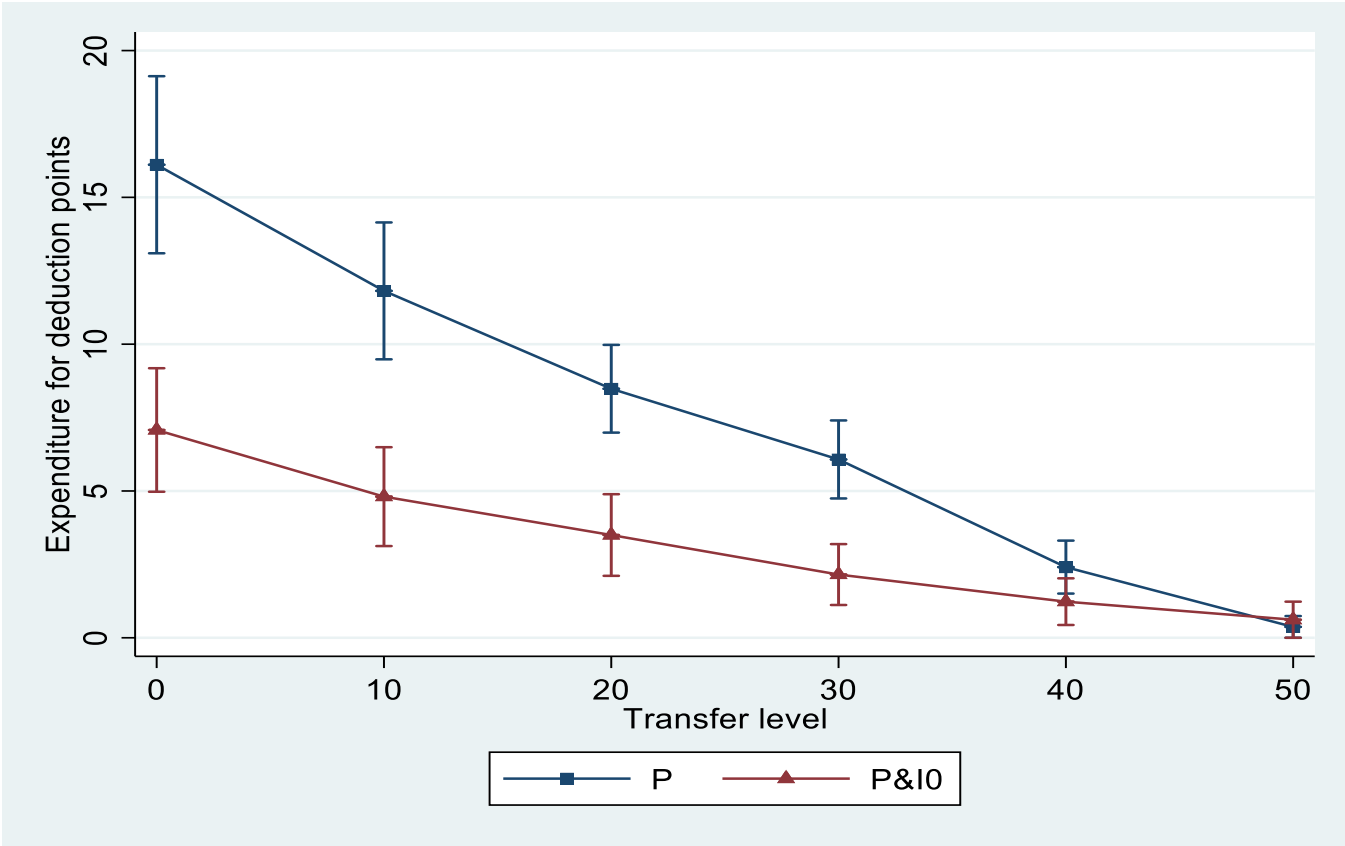

**b.** Average expenditure for deduction points

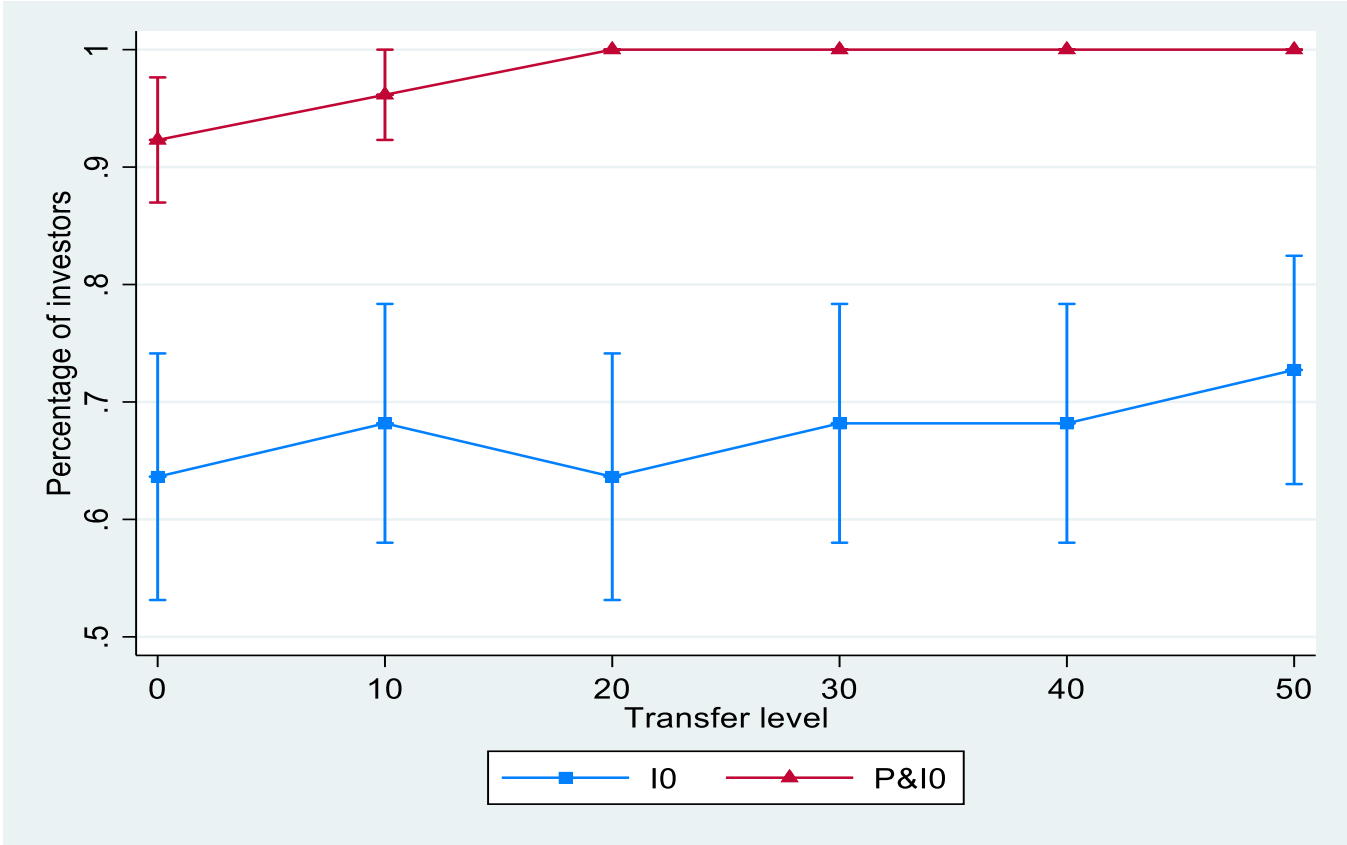

**c.** Percentage of investors

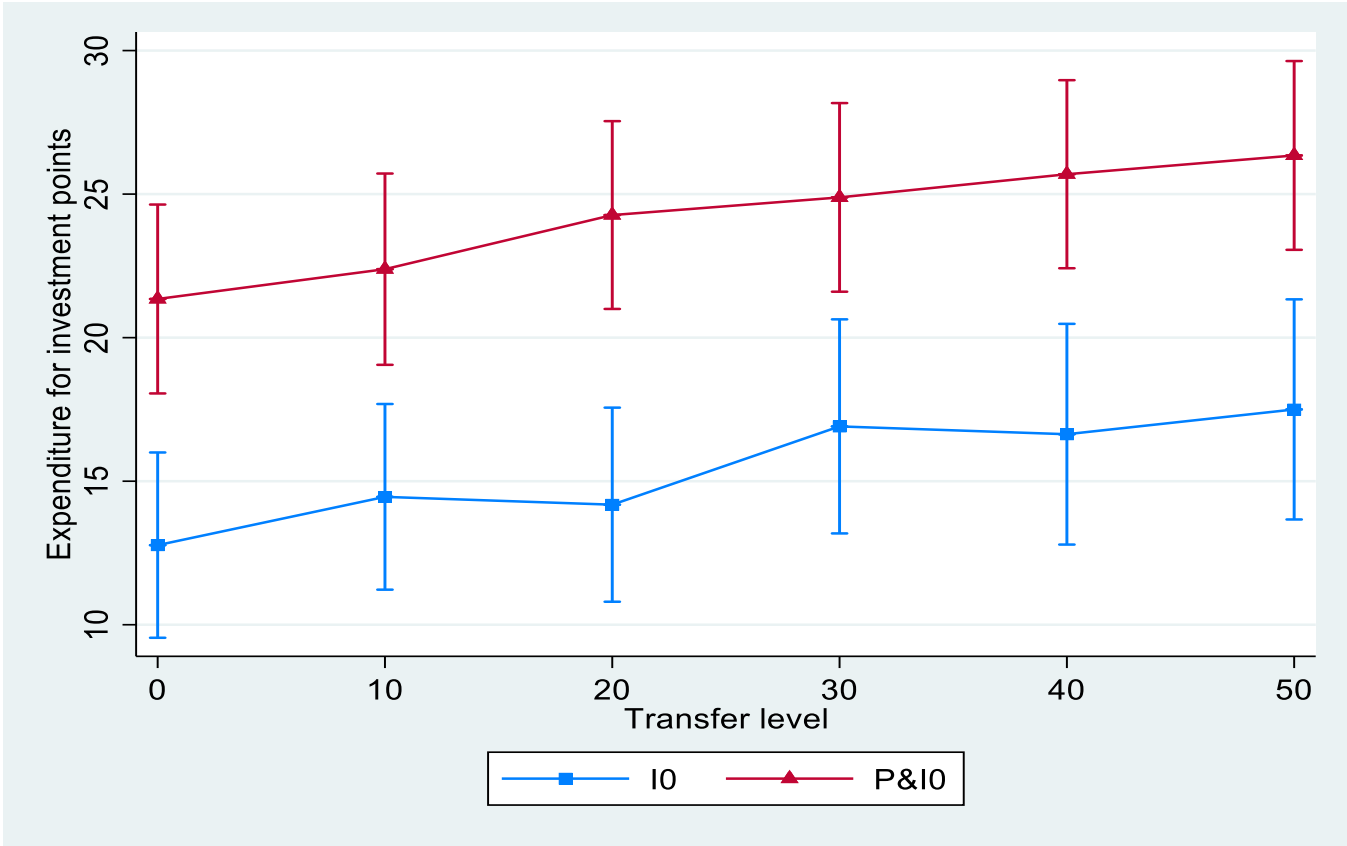

**d.** Average expenditure for investment points

Note: A punisher [investor] is defined as Player C who spent at least one token to deduction [investment] points. Expenditure for deduction [investment] points is the number of tokens that Player C spent for deduction [investment] points among 50 tokens. Transfer level is the number of tokens that Player A transferred to Player B. A vertical bar indicates plus and minus one standard error.

### C.3. Risk-neutral and Risk-averse Third Parties in P, P&I0, and I0 Treatments

In the main text, we examine a third party's behavior using regression methods. In this subsection, we show that these results are robust even if we consider the *Interaction* term, which is the product of *Transfer* and *Treatment*, as an independent variable in the models.

Analysis of punishing behavior considering the risk-neutral and the risk-averse is illustrated in Section 4.1, and related regression results are reported in Table 5 in the main text. Following the models in the text, we regress *Punisher* or *Punishment* on *Transfer*, *Treatment*, *Interaction*, and other controlling variables using the LPM, Probit, OLS, and Tobit methods. The regression results are presented in columns 1–4 in Table C3. All *Treatment* variables indicate negative coefficients, and they are significantly different from zero.[9] Thus, the *Interaction* term does not qualitatively affect the regression results implying the diminished demand for punishment.

Meanwhile, the analysis of investment behavior considering the risk-neutral and the risk-averse is illustrated in Section 4.2, and related regression results are provided in Table 7. Likewise, for further analysis, we regress *Investor* or *Investment* on *Transfer*, *Treatment*, *Interaction*, and other controlling variables using the LPM, Probit, OLS, and Tobit methods. The results are reported in columns 5–8 in Table C3, and they show that the coefficients of *Treatment* are all positive and mostly significantly different from zero.[10] Hence, we confirm that the regression results implying the increased demand for investment are robust even if we consider the *Interaction* term in the models as well.

---

[9] Note that in column 3 in Table C3, the coefficient of *Interaction* is positive and significant. This indicates that when the OLS but not the Tobit method is used, the marginal effect of *Transfer* on *Punishment* is smaller in P&I0 treatment than in P treatment. The different results between the two methods may be due to truncated data. Indeed, the Tobit model in column 4 considers 176 left-censored and 1 right-censored data.

[10] Note that in column 6 in Table C3, the *Interaction* term is omitted because all Player Cs chose to invest in the lottery at the transfer levels 10–50 in P&I0 treatment.

**Table C3. Regression on punishment and investment (with risk-neutral and risk-averse)**

| Treatment: | P, P&I0 | | | | I0, P&I0 | | | |
|---|---|---|---|---|---|---|---|---|
| Level of data: | Transfer level | | | | Transfer level | | | |
| Model: | LPM | Probit | OLS | Tobit | LPM | Probit | OLS | Tobit |
| Dependent variable: | Punisher | | Punishment | | Investor | | Investment | |
| | (1) | (2) | (3) | (4) | (5) | (6) | (7) | (8) |
| Transfer | −0.012*** | −0.038*** | −0.264*** | −0.521*** | 0.002 | 0.007 | 0.064 | 0.119 |
| | (0.002) | (0.007) | (0.054) | (0.061) | (0.001) | (0.006) | (0.044) | (0.075) |
| Treatment: P&I0 | −0.387** | −1.254** | −8.702*** | −15.286*** | 0.368*** | 1.270* | 7.018* | 14.675** |
| | (0.153) | (0.493) | (2.886) | (4.897) | (0.101) | (0.704) | (3.835) | (5.839) |
| Interaction: Transfer × Treatment | 0.004 | 0.002 | 0.165** | 0.039 | −0.001 | — | 0.015 | −0.004 |
| | (0.003) | (0.010) | (0.063) | (0.104) | (0.002) | — | (0.054) | (0.087) |
| Switching Point | −0.052 | −0.234 | −0.117 | −1.750 | 0.024 | 0.130 | −0.908 | −0.506 |
| | (0.045) | (0.155) | (0.674) | (1.528) | (0.038) | (0.223) | (1.641) | (2.568) |
| Constant | 0.225 | −1.520 | 4.044 | −12.440 | 0.056 | −0.963 | −37.669** | −74.533** |
| | (0.600) | (2.113) | (10.980) | (22.502) | (0.430) | (1.593) | (18.235) | (29.373) |
| Socio-demographic | Yes | Yes | Yes | Yes | Yes | Yes | Yes | Yes |
| Personality traits | Yes | Yes | Yes | Yes | Yes | Yes | Yes | Yes |
| Observations | 270 | 270 | 270 | 270 | 246 | 136 | 246 | 246 |
| Number of individuals | 45 | 45 | 45 | 45 | 41 | 41 | 41 | 41 |
| Left-censored | — | — | — | 176 | — | — | — | 44 |
| Right-censored | — | — | — | 1 | — | — | — | 44 |
| (Pseudo) R-square | 0.331 | 0.317 | 0.366 | 0.132 | 0.405 | 0.349 | 0.512 | 0.114 |

Note: The dependent variable in Tobit models is truncated at 0 (left-censored) and 50 (right-censored). Socio-demographic includes gender, income, and Economics major. Clustered standard errors at the individual level are in parentheses. ***, **, and * denote significance at 1%, 5%, and 10% level, respectively.

### C.4. Risk-averse Third Parties in P, P&I0, and I0 Treatments

As discussed in Section 3 of the main text, punishing and investing behaviors only considering risk-averse third parties are worthy to examine since some risk-neutral third parties may not evaluate the safe option better than the investment option in terms of the material-payoff attribute. Note that there are 16, 13, and 12 risk-averse third parties in P, P&I0, and I0 treatments, respectively (see Table 4 in the main text)..

Figures C3a and C3b show that risk-averse Player Cs are still reluctant to punish Player A in P&I0 treatment than in P treatment. The differences in the proportion of punishers as well as the differences punishment expenditures are significant or marginally insignificant at most of the unfair transfer levels (p=[0.264, 0.130, 0.052, 0.020, 0.020, 1.000], two-sided Fisher's exact test; p=[0.092, 0.030, 0.018, 0.012, 0.016, 0.367], two-sided Wilcoxon rank-sum tests). Moreover, as presented in Table C4, the averages of mean [median] punishment are 6.47 [6.16] and 1.35 [0.77] in P and P&I0 treatments, respectively, and the difference is marginally insignificant [significant] (p=0.058 [p=0.015], two-sided Wilcoxon rank-sum tests). All columns in Table C5 and columns 1–4 in Table C7 report related regression results. Note that all coefficients of *Treatment* have negative values, and all except one are significantly different from zero. Thus, the diminished demand for TPP holds even when we only consider risk-averse third parties.

Next, we examine the investing behavior of risk-averse third parties. Figure C3c shows that the proportion of investors is higher in P&I0 treatment than in I0 treatment, and the differences in proportions between the treatments at the transfer levels of 0, 20, and 50 are statistically significant (p=[0.039, 0.096, 0.039, 0.096, 0.096, 0.039], two-sided Fisher's exact test). This implies the apparent effect of the punishment opportunity in P&I0 treatment on the investment behavior of risk-averse Player Cs. Figure C3d shows that although the differences are insignificant, the investment expenditure is higher in P&I0 treatment than in I0 treatment at all transfer levels (p=[0.202, 0.411, 0.144, 0.426, 0.258, 0.162], two-sided Wilcoxon rank-sum tests). Similarly, as shown in Table C4, the averages of mean [median] investments are 14.06 [13.88] and 20.97 [21.23] in I0 and P&I0 treatments, respectively. Although the differences are statistically insignificant, we can still confirm that the demand for investment in P&I0 treatment is at least not smaller than the demand for investment in I0 treatment. The related regression results are presented in all columns in Table C6 and columns 5–8 in Table C7. The coefficients of *Treatment* indicate positive values.[11] Hence, the results considering only the risk-averse weakly support the increased demand for investment when both the punishment and the investment options are available.

**Table C4. Punishment and investment using the individual-level data (with risk-averse)**

| | Punishment Risk-aversion | | | | Investment Risk-aversion | | | |
|---|---|---|---|---|---|---|---|---|
| | Mean | p-value | Median | p-value | Mean | p-value | Median | p-value |
| P [I0] | 6.47 | | 6.16 | | 14.06 | | 13.88 | |
| | (7.20) | 0.058 | (6.96) | 0.015 | (13.77) | 0.299 | (13.89) | 0.247 |
| P&I0 | 1.35 | | 0.77 | | 20.97 | | 21.23 | |
| | (2.58) | | (1.88) | | (16.97) | | (17.16) | |

Note: Mean and median refer to an average using mean and median individual-level data, respectively. Standard deviations are in parentheses. p-values are from the two-sided Wilcoxon rank-sum test. There are 16, 13, and 12 risk-averse third parties in P, P&I0, and I0 treatments, respectively.

[11] We omit the regression results of Probit models in columns 3 and 4 in Table C6 since the convergence is not achieved. The failure of the convergence is due to the failure of the identification for Probit in P&I0 treatment, where all subjects choose to invest for all transfer levels.

## Figure C3. Punishment and Investment in P, I0 and P&I0 treatments (with risk-averse)

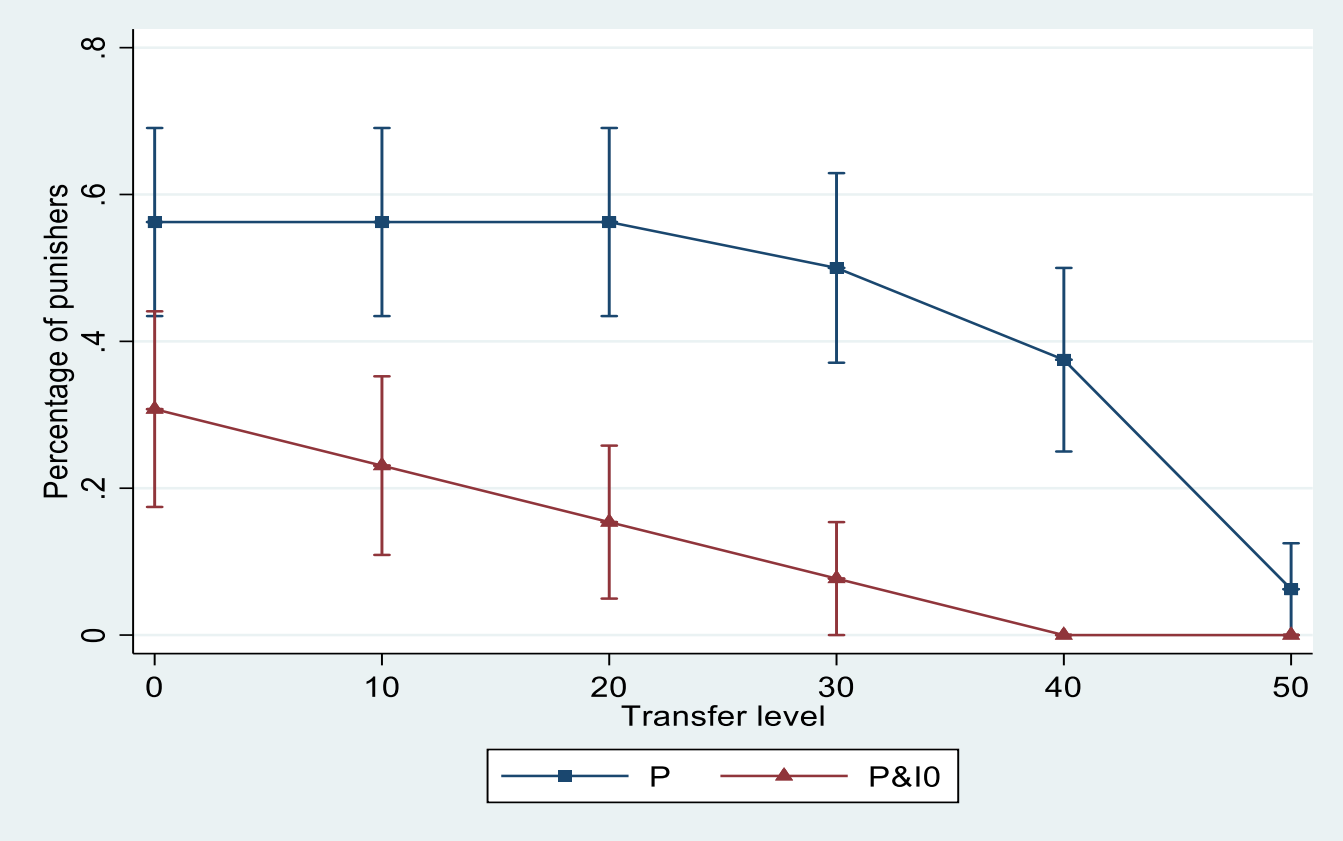

**a.** Percentage of punishers

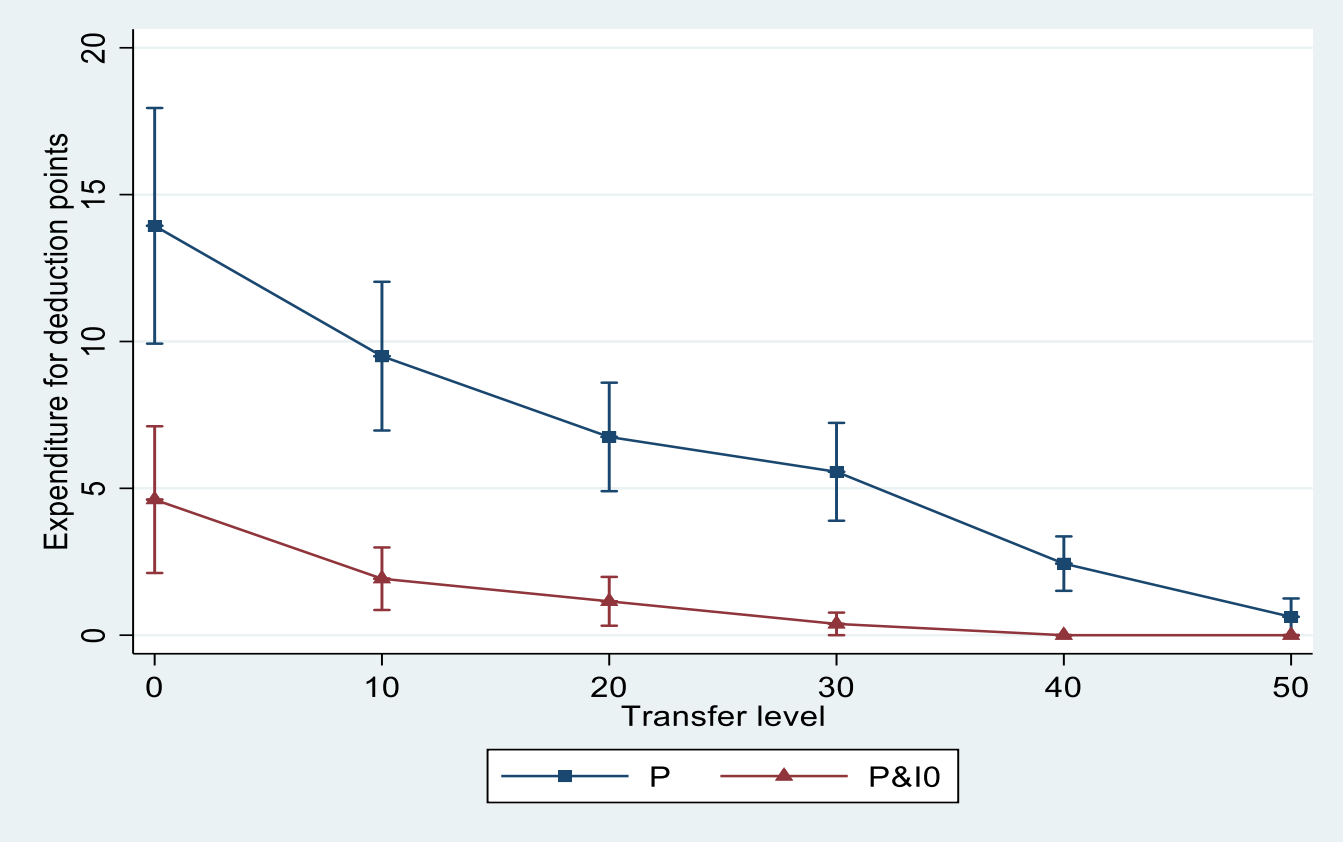

**b.** Average expenditure for deduction points

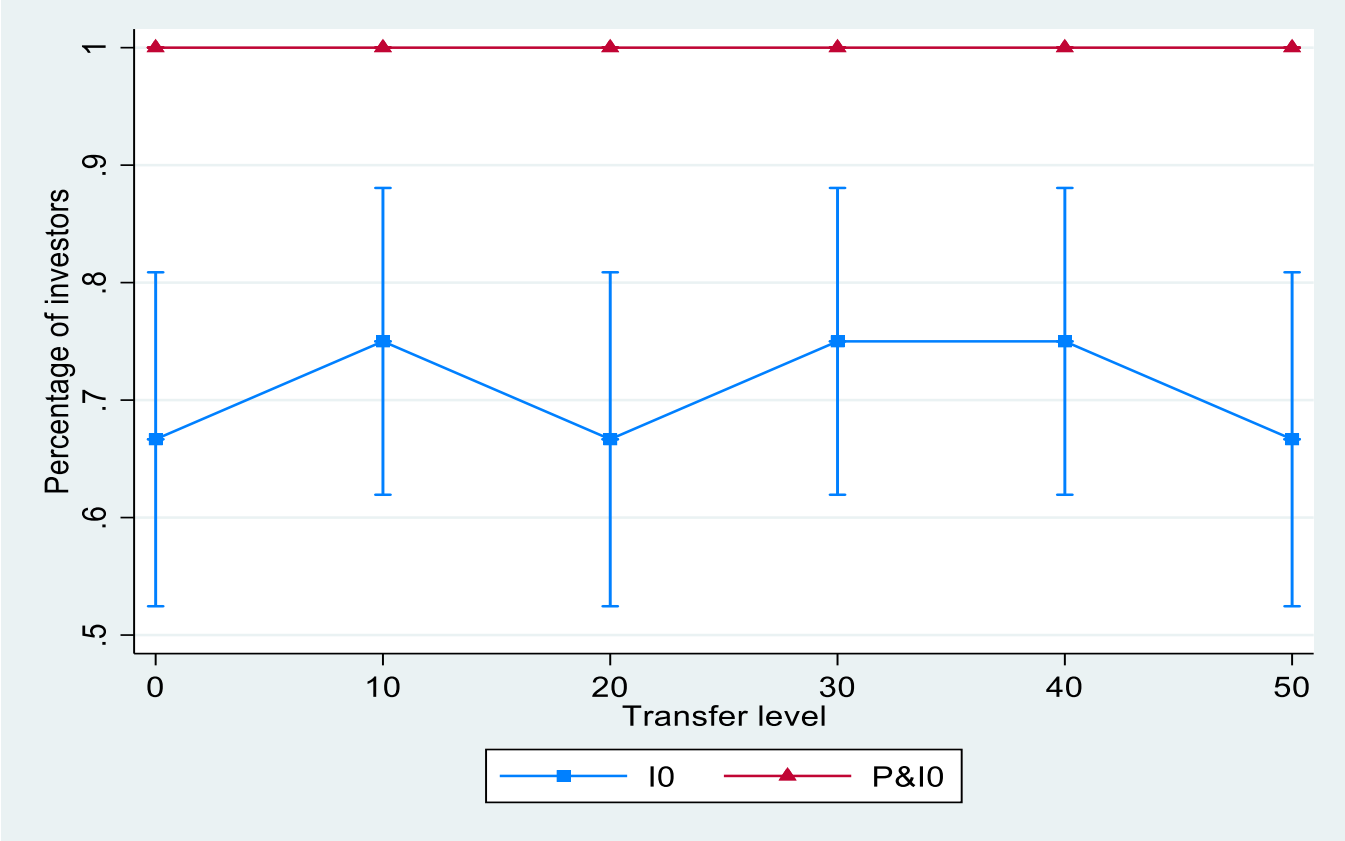

**c.** Percentage of investors

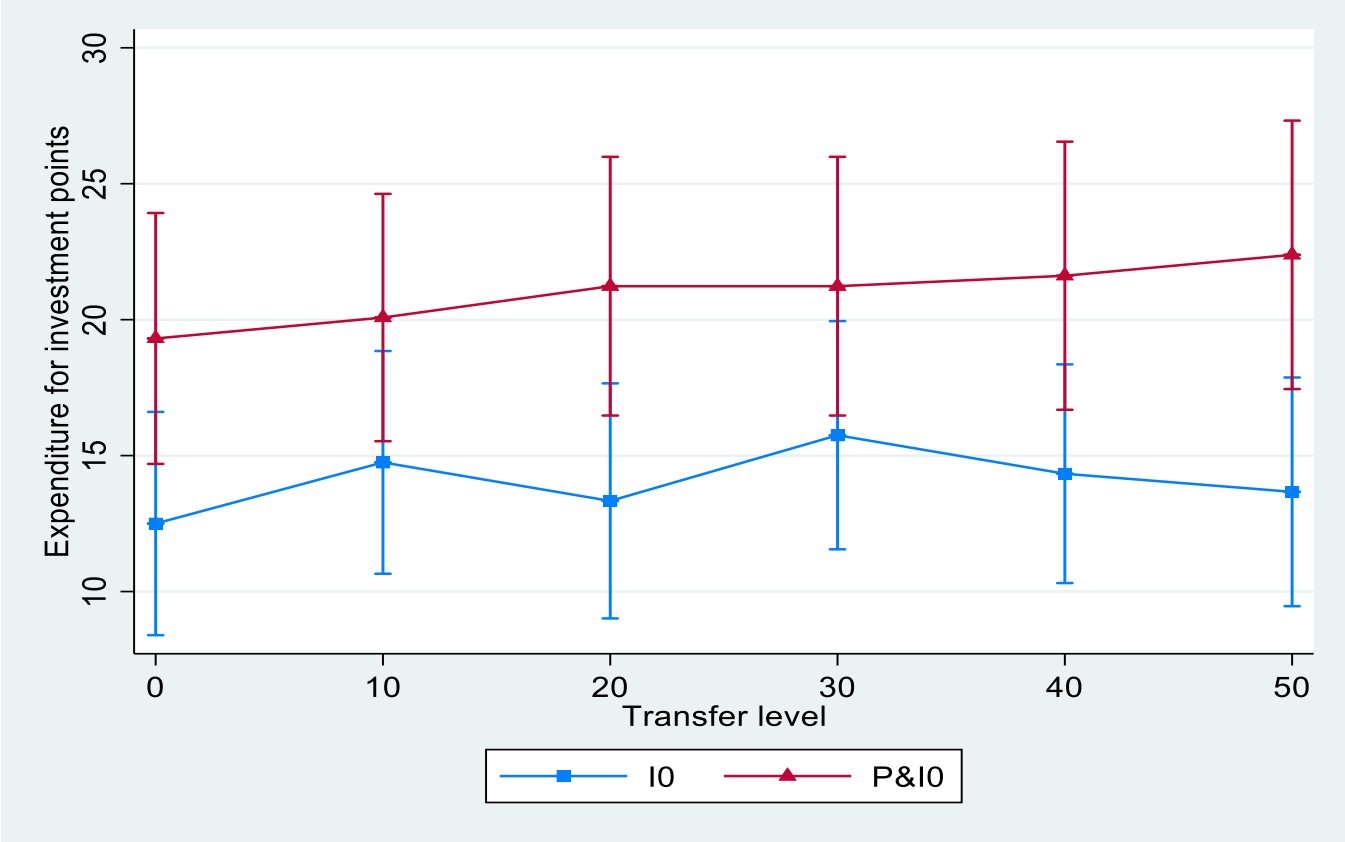

**d.** Average expenditure for investment points

Note: A punisher [investor] is defined as Player C who spent at least one token to deduction [investment] points. Expenditure for deduction [investment] points is the number of tokens that Player C spent for deduction [investment] points among 50 tokens. Transfer level is the number of tokens that Player A transferred to Player B. A vertical bar indicates plus and minus one standard error.

**Table C5. Regression on punishment (with risk-averse)**

| Treatment: | P, P&I0 | | | | | | | |
|---|---|---|---|---|---|---|---|---|
| Level of data: | Transfer level | | | | Transfer level | | | |
| Model: | LPM | | Probit | | OLS | | Tobit | |
| Dependent variable: | Punisher | | Punisher | | Punishment | | Punishment | |
| | (1) | (2) | (3) | (4) | (5) | (6) | (7) | (8) |
| Transfer | −0.008*** | −0.009*** | −0.044*** | −0.039*** | −0.178*** | −0.254*** | −0.578*** | −0.538*** |
| | (0.002) | (0.003) | (0.008) | (0.008) | (0.047) | (0.071) | (0.080) | (0.076) |
| Treatment: P&I0 | −0.376*** | −0.434** | −2.439*** | −2.076*** | −5.563*** | −9.800** | −25.904*** | −22.716*** |
| | (0.117) | (0.177) | (0.524) | (0.661) | (2.010) | (3.716) | (6.826) | (7.518) |
| Interaction: Transfer × Treatment | — | 0.002 | — | −0.020 | — | 0.169* | — | −0.224 |
| | — | (0.004) | — | (0.018) | — | (0.085) | — | (0.220) |
| Switching point | 0.031 | 0.031 | −0.010 | 0.008 | 0.135 | 0.135 | −0.605 | −0.583 |
| | (0.067) | (0.067) | (0.290) | (0.284) | (1.442) | (1.447) | (3.013) | (2.993) |
| Constant | −0.469 | −0.443 | −5.116 | −5.336 | −1.761 | 0.139 | −42.815 | −44.327 |
| | (0.749) | (0.760) | (3.479) | (3.405) | (16.765) | (17.110) | (33.108) | (32.679) |
| Socio-demographic | Yes | Yes | Yes | Yes | Yes | Yes | Yes | Yes |
| Personality traits | Yes | Yes | Yes | Yes | Yes | Yes | Yes | Yes |
| Observations | 174 | 174 | 174 | 174 | 174 | 174 | 174 | 174 |
| Left-censored | — | — | — | — | — | — | 122 | 122 |
| Right-censored | — | — | — | — | — | — | 1 | 1 |
| Number of individuals | 29 | 29 | 29 | 29 | 29 | 29 | 29 | 29 |
| (Pseudo) R-square | 0.398 | 0.400 | 0.470 | 0.475 | 0.335 | 0.366 | 0.175 | 0.176 |

Note: The dependent variable in Tobit models is truncated at 0 (left-censored) and 50 (right-censored). Socio-demographic includes gender, income, and Economics major. Clustered standard errors at the individual level are in parentheses. ***, **, and * denote significance at 1%, 5%, and 10% levels, respectively.

**Table C6. Regression on investment (with risk-averse)**

| Treatment: | I0, P&I0 | | | | | | | |
|---|---|---|---|---|---|---|---|---|
| Level of data: | Transfer level | | | | Transfer level | | | |
| Model: | LPM | | Probit | | OLS | | Tobit | |
| Dependent variable: | Investor | | Investor | | Investment | | Investment | |
| | (1) | (2) | (3) | (4) | (5) | (6) | (7) | (8) |
| Transfer | 0.000 | 0.000 | — | — | 0.039 | 0.020 | 0.053 | 0.031 |
| | (0.001) | (0.002) | — | — | (0.031) | (0.049) | (0.040) | (0.068) |
| Treatment: P&I0 | 0.274* | 0.280* | — | — | 5.595 | 4.666 | 11.017 | 10.016 |
| | (0.134) | (0.141) | — | — | (4.877) | (4.953) | (6.786) | (6.724) |
| Interaction: Transfer × Treatment | — | 0.000 | — | — | — | 0.037 | — | 0.040 |
| | — | (0.002) | — | — | — | (0.062) | — | (0.082) |
| Switching point | −0.076 | −0.076 | — | — | −1.094 | −1.094 | −2.249 | −2.248 |
| | (0.047) | (0.047) | — | — | (2.418) | (2.427) | (3.369) | (3.368) |
| Constant | 0.915* | 0.912* | — | — | −43.643* | −43.160* | −64.798* | −64.239* |
| | (0.465) | (0.469) | — | — | (24.867) | (24.904) | (35.525) | (35.429) |
| Socio-demographic | Yes | Yes | — | — | Yes | Yes | Yes | Yes |
| Personality traits | Yes | Yes | — | — | Yes | Yes | Yes | Yes |
| Observations | 150 | 150 | — | — | 150 | 150 | 150 | 150 |
| Left-censored | — | — | — | — | — | — | 21 | 21 |
| Right-censored | — | — | — | — | — | — | 20 | 20 |
| Number of individuals | 25 | 25 | — | — | 25 | 25 | 25 | 25 |
| (Pseudo) R-square | 0.493 | 0.493 | — | — | 0.505 | 0.505 | 0.107 | 0.107 |

Note: The dependent variable in Tobit models is truncated at 0 (left-censored) and 50 (right-censored). Socio-demographic includes gender, income, and Economics major. Clustered standard errors at the individual level are in parentheses. Columns 3 and 4 are omitted because the identification for Probit failed in P&I0 treatment in which every subject invested in all transfer levels. ***, **, and * denote significance at 1%, 5%, and 10% levels, respectively.

**Table C7. Regression on punishment and investment (with risk-averse)**

| Treatment: | P, I0, P&I0 | | | | | | | |
|---|---|---|---|---|---|---|---|---|
| Level of data: | Individual level | | Individual level | | Individual level | | Individual level | |
| Model: | OLS | Tobit | OLS | Tobit | OLS | Tobit | OLS | Tobit |
| Dependent variable: | Mean punishment | | Median punishment | | Mean investment | | Median investment | |
| | (1) | (2) | (3) | (4) | (5) | (6) | (7) | (8) |
| Treatment: P&I0 | −5.563* | −13.855** | −5.655** | −18.676*** | 5.595 | 10.273 | 5.826 | 10.571 |
| | (2.678) | (5.393) | (2.551) | (6.478) | (7.311) | (7.234) | (7.359) | (7.273) |
| Switching point | 0.135 | −0.205 | −0.256 | −1.545 | −1.094 | −2.209 | −1.156 | −2.290 |
| | (1.586) | (2.456) | (1.511) | (2.549) | (3.736) | (3.598) | (3.761) | (3.618) |
| Constant | −6.214 | −26.122 | −4.959 | −32.364 | −42.660 | −60.670 | −43.043 | −61.293 |
| | (15.849) | (24.578) | (15.099) | (26.374) | (36.803) | (36.616) | (37.046) | (36.824) |
| Socio-demographic | Yes | Yes | Yes | Yes | Yes | Yes | Yes | Yes |
| Personality traits | Yes | Yes | Yes | Yes | Yes | Yes | Yes | Yes |
| Observations | 29 | 29 | 29 | 29 | 25 | 25 | 25 | 25 |
| Left-censored | — | 16 | — | 18 | — | 3 | — | 3 |
| Right-censored | — | 0 | — | 0 | — | 3 | — | 3 |
| Number of individuals | 29 | 29 | 29 | 29 | 25 | 25 | 25 | 25 |
| (Pseudo) R-square | 0.366 | 0.143 | 0.384 | 0.176 | 0.5285 | 0.115 | 0.534 | 0.117 |

Note: The dependent variable in Tobit models is truncated at 0 (left-censored) and 50 (right-censored). Socio-demographic includes gender, income, and Economics major. Standard errors are in parentheses. ***, **, and * denote significance at 1%, 5%, and 10% levels, respectively.

### C.5. Risk-neutral and Risk-averse Third Parties in P, P&Ineg, and Ineg Treatments

For the extended analysis in the main text, we examine a third party's behavior using the supplementary treatments, namely, P&Ineg and Ineg treatments. In this subsection, we provide related regression results of the supplementary treatments and confirm the consistency of these results with the main findings. Note that there are 23, 10, and 11 third parties in P, P&Ineg, and Ineg treatments, respectively (see Table 4 in the main text).

Consider risk-neutral and risk-averse third parties. First, we regress *Punisher* or *Punishment* on *Transfer, Treatment*, *Interaction*, and other controlling variables. Here, *Treatment* is defined as 0 if P treatment and 1 if P&Ineg treatment. The regression results of *Punisher* are reported in columns 1–4 in Table C8, whereas the results of *Punishment* are reported in columns 5–8 in Table C8. In accordance with the findings in the main text, all *Treatment* variables indicate negative coefficients and are significantly different from zero.[12] In addition, we regress *Punishment* on *Treatment* and other controlling variables using the individual-level data. The results are presented in columns 1–4 in Table C10. Consistently, *Treatment* variables indicate negative values and are significantly different from zero.

Next, we regress *Investor* or *Investment* on *Transfer, Treatment*, *Interaction*, and other controlling variables. Here, *Treatment* is defined as 0 if Ineg treatment and 1 if P&Ineg treatment. The results of *Investor* are presented in columns 1–4 in Table C9, whereas the results of *Investment* are presented in columns 5–8 in Table C9. We find that all *Treatment* variables indicate positive coefficients and are significantly different from zero. Moreover, we regress *Investment* on *Treatment* and other controlling variables using the individual-level data. The results are presented in columns 5–8 in Table C10. Consistent with the findings from the transfer level data, all *Treatment* variables are positive and significantly different from zero.

Therefore, the results in P&Ineg, Ineg, and P treatments confirm the robustness of the findings in P&I0, I0, and P treatments in the main text.

[12] Note that the significance of coefficients of *Interaction* is different between the OLS method and the Tobit method as displayed in columns 6 and 8, respectively, in Table C8. This may be due to truncated data where theTobit method, in column 8, considers 125 left-censored and 1 right-censored data.

**Table C8. Regression on punishment (with risk-neutral and risk-averse)**

| Treatment: | P, P&Ineg | | | | | | | |
|---|---|---|---|---|---|---|---|---|
| Level of data: | Transfer level | | | | Transfer level | | | |
| Model: | LPM | | Probit | | OLS | | Tobit | |
| Dependent variable: | Punisher | | Punisher | | Punishment | | Punishment | |
| | (1) | (2) | (3) | (4) | (5) | (6) | (7) | (8) |
| Transfer | −0.010*** | −0.012*** | −0.037*** | −0.036*** | −0.206*** | −0.264*** | −0.539*** | −0.531*** |
| | (0.002) | (0.002) | (0.006) | (0.006) | (0.043) | (0.054) | (0.063) | (0.065) |
| Treatment: P&Ineg | −0.508*** | −0.668*** | −2.104*** | −1.988*** | −7.449*** | −12.233*** | −25.618*** | −24.442*** |
| | (0.128) | (0.184) | (0.566) | (0.669) | (2.507) | (3.621) | (7.115) | (8.412) |
| Interaction: Transfer × Treatment | — | 0.006 | — | −0.008 | — | 0.191** | — | −0.089 |
| | — | (0.004) | — | (0.016) | — | (0.071) | — | (0.233) |
| Switching Point | 0.006 | 0.006 | −0.011 | −0.012 | 0.545 | 0.545 | 0.640 | 0.630 |
| | (0.045) | (0.045) | (0.173) | (0.174) | (0.568) | (0.570) | (1.552) | (1.564) |
| Constant | 0.460 | 0.509 | 0.016 | −0.005 | 10.127 | 11.577 | 10.249 | 10.037 |
| | (0.622) | (0.629) | (2.163) | (2.171) | (10.342) | (10.576) | (23.324) | (23.479) |
| Socio-demographic | Yes | Yes | Yes | Yes | Yes | Yes | Yes | Yes |
| Personality traits | Yes | Yes | Yes | Yes | Yes | Yes | Yes | Yes |
| Observations | 198 | 198 | 198 | 198 | 198 | 198 | 198 | 198 |
| Left-censored | — | — | — | — | — | — | 125 | 125 |
| Right-censored | — | — | — | — | — | — | 1 | 1 |
| Number of individuals | 33 | 33 | 33 | 33 | 33 | 33 | 33 | 33 |
| (Pseudo) R-square | 0.320 | 0.331 | 0.301 | 0.302 | 0.377 | 0.407 | 0.133 | 0.134 |

Note: The dependent variable in Tobit models is truncated at 0 (left-censored) and 50 (right-censored). Socio-demographic includes gender, income, and Economics major. Clustered standard errors at individual level are in parentheses. ***, **, and * denote significance at 1%, 5%, and 10% levels, respectively.

**Table C9. Regression on investment (with risk-neutral and risk-averse)**

| Treatment: | Ineg, P&Ineg | | | | | | | |
|---|---|---|---|---|---|---|---|---|
| Level of data: | Transfer level | | | | Transfer level | | | |
| Model: | LPM | | Probit | | OLS | | Tobit | |
| Dependent variable: | Investor | | Investor | | Investment | | Investment | |
| | (1) | (2) | (3) | (4) | (5) | (6) | (7) | (8) |
| Transfer | 0.000 | −0.001 | 0.001 | −0.005 | 0.018 | 0.034 | 0.049 | 0.090 |
| | (0.002) | (0.003) | (0.009) | (0.013) | (0.054) | (0.101) | (0.134) | (0.295) |
| Treatment: P&Ineg | 0.676*** | 0.621*** | 7.433** | 7.232** | 18.257*** | 19.066*** | 54.155*** | 55.929*** |
| | (0.140) | (0.152) | (3.450) | (3.653) | (5.584) | (6.365) | (17.571) | (16.825) |
| Interaction: Transfer × Treatment | — | 0.002 | — | 0.015 | — | −0.032 | — | −0.071 |
| | — | (0.004) | — | (0.019) | — | (0.103) | — | (0.310) |
| Switching Point | 0.067 | 0.067 | −0.132 | −0.141 | 2.257 | 2.257 | 4.903 | 4.902 |
| | (0.043) | (0.043) | (0.432) | (0.432) | (2.449) | (2.460) | (4.335) | (4.326) |
| Constant | −0.010 | 0.016 | −1.244 | −1.104 | −10.698 | −11.083 | −70.651 | −71.388 |
| | (0.751) | (0.765) | (3.397) | (3.433) | (35.306) | (35.299) | (82.972) | (82.544) |
| Socio-demographic | Yes | Yes | Yes | Yes | Yes | Yes | Yes | Yes |
| Personality traits | Yes | Yes | Yes | Yes | Yes | Yes | Yes | Yes |
| Observations | 126 | 126 | 126 | 126 | 126 | 126 | 126 | 126 |
| Left-censored | — | — | — | — | — | — | 66 | 66 |
| Right-censored | — | — | — | — | — | — | 14 | 14 |
| Number of individuals | 21 | 21 | 21 | 21 | 21 | 21 | 21 | 21 |
| (Pseudo) R-square | 0.538 | 0.539 | 0.544 | 0.547 | 0.525 | 0.526 | 0.196 | 0.196 |

Note: The dependent variable in Tobit models is truncated at 0 (left-censored) and 50 (right-censored). Socio-demographic includes gender, income, and Economics major. Clustered standard errors at the individual level are in parentheses. ***, **, and * denote significance at 1%, 5%, and 10% levels, respectively.

**Table C10. Regression on punishment and investment (with risk-neutral and risk-averse)**

| Treatment: | P, P&Ineg | | | | Ineg, P&Ineg | | | |
|---|---|---|---|---|---|---|---|---|
| Level of data: | Individual level | | | | Individual level | | | |
| Model: | OLS | Tobit | OLS | Tobit | OLS | Tobit | OLS | Tobit |
| Dependent variable: | Mean punishment | | Median punishment | | Mean investment | | Median investment | |
| | (1) | (2) | (3) | (4) | (5) | (6) | (7) | (8) |
| Treatment: P&Ineg | −7.449** | −13.993*** | −7.734** | −22.895*** | 18.257** | 34.143*** | 19.556** | 144.677*** |
| | (3.079) | (4.966) | (3.133) | (7.759) | (7.914) | (9.778) | (8.106) | (38.792) |
| Switching Point | 0.545 | 0.540 | 0.280 | −1.017 | 2.257 | 3.526 | 2.099 | −3.177 |
| | (0.882) | (1.238) | (0.897) | (1.549) | (2.832) | (2.996) | (2.901) | (3.478) |
| Constant | 4.971 | 5.992 | 7.020 | 6.737 | −10.238 | −30.055 | −2.598 | −309.780** |
| | (11.764) | (16.479) | (11.968) | (18.332) | (45.664) | (53.023) | (46.773) | (102.421) |
| Socio-demographic | Yes | Yes | Yes | Yes | Yes | Yes | Yes | Yes |
| Personality traits | Yes | Yes | Yes | Yes | Yes | Yes | Yes | Yes |
| Observations | 33 | 33 | 33 | 33 | 21 | 21 | 21 | 21 |
| Left-censored | — | 15 | — | 17 | — | 7 | — | 11 |
| Right-censored | — | 0 | — | 0 | — | 2 | — | 2 |
| Number of individuals | 33 | 33 | 33 | 33 | 21 | 21 | 21 | 21 |
| (Pseudo) R-square | 0.396 | 0.095 | 0.450 | 0.155 | 0.598 | 0.200 | 0.589 | 0.366 |

Note: The dependent variable in Tobit models is truncated at 0 (left-censored) and 50 (right-censored). Socio-demographic includes gender, income, and Economics major. Standard errors are in parentheses. ***, **, and * denote significance at 1%, 5%, and 10% levels, respectively..

### C.6. Substitution Effect Between Punishment and Investment

In the main text, we have suggested that the investment option may work as a compromise rather than a decoy by showing the possible substitution effect between the punishment option and the investment option. One way to verify the substitution effect is to compare the extent of investment conditional on a non-punisher in P&I0 treatment to the extent of investment in I0 treatment. In this subsection, we verify that the substitution effect we claim is indeed robust even if the *Interaction* term is included as an independent variable in the regression models. Furthermore, the claim is also supported by the regression results from the supplementary treatments with other controlled variables.

In Table C11, the additional regression results from I0 and P&I0 treatments are reported in columns 1 and 2, whereas those from Ineg and P&Ineg treatments are reported in columns 3–6. Note that all *Treatment* variables have positive values, and most of them are significantly different from zero. This supports our claim about the substitution effect and the hypothesis about the compromise effect.

**Table C11. The substitution effect (with risk-neutral and risk-averse)**

| Treatments | I0, P&I0 | | Ineg, P&Ineg | | | |
|---|---|---|---|---|---|---|
| Level of data: | Transfer level | | Transfer level | | | |
| Model: | OLS | Tobit | OLS | | Tobit | |
| Dependent variable: | Investment | Investment | Investment | | Investment | |
| | (conditional on a non-punisher) | | (conditional on a non-punisher) | | | |
| | (1) | (2) | (3) | (4) | (5) | (6) |
| Transfer | 0.064 | 0.123 | −0.018 | 0.034 | −0.048 | 0.084 |
| | (0.044) | (0.077) | (0.057) | (0.101) | (0.149) | (0.286) |
| Treatment: P&I0 or P&Ineg | 7.556* | 16.033** | 19.105*** | 22.084*** | 56.047*** | 62.496*** |
| | (3.905) | (6.577) | (5.259) | (6.165) | (15.989) | (14.851) |
| Interaction: Transfer × Treatment | −0.009 | −0.042 | — | −0.116 | — | −0.256 |
| | (0.068) | (0.102) | — | (0.114) | — | (0.294) |
| Switching Point | −1.504 | −1.786 | 1.868 | 1.854 | 4.058 | 4.028 |
| | (1.841) | (3.071) | (2.358) | (2.362) | (3.860) | (3.819) |
| Constant | −38.140** | −77.428** | −2.418 | −3.140 | −42.947 | −43.729 |
| | (18.445) | (30.203) | (33.081) | (32.961) | (80.014) | (78.137) |
| Socio-demographic | Yes | Yes | Yes | Yes | Yes | Yes |
| Personality traits | Yes | Yes | Yes | Yes | Yes | Yes |
| Observations | 219 | 219 | 120 | 120 | 120 | 120 |
| Left-censored | — | 43 | — | — | 63 | 63 |
| Right-censored | — | 44 | — | — | 14 | 14 |
| Number of individuals | 41 | 41 | 21 | 21 | 21 | 21 |
| (Pseudo) R-square | 0.556 | 0.128 | 0.562 | 0.566 | 0.219 | 0.221 |

Note: The dependent variable in Tobit models is truncated at 0 (left-censored) and 50 (right-censored). Socio-demographic includes gender, income, and Economics major. Clustered standard errors at the individual level are in parentheses. ***, **, and * denote significance at 1%, 5%, and 10% levels, respectively.

## C.7. Non-increasing Safe Option Under the Supplementary Treatment

In this section, we present analyses on the non-increasing safe option under the supplementary treatment where the investment option with a negative expected net return is available.

Remind that (almost) all risk-neutral and risk-averse third parties spent at least one token for the safe option at all transfer levels in P treatment. In P&Ineg treatment, 80% of those third parties spent at least one token for the safe option at all transfer levels. The proportion in P&Ineg treatment is smaller than the proportion in P treatment at all transfer levels, while the difference is statistically insignificant at all transfer levels (p=[0.212, 0.085, 0.085, 0.085, 0.085, 0.085], two-sided Fisher's exact test). Also, in P treatment, remind that risk-neutral and risk-averse third parties spent 35 to 47 tokens on average for the safe option for all transfer levels. In P&Ineg treatment, those third parties spent 32 to 36 tokens on average for the safe option for all transfer levels. The expenditure in P&Ineg treatment is smaller than the expenditure in P treatment, and the difference is statistically significant at some transfer levels (p=[0.704, 0.387, 0.328, 0.291, 0.027, 0.001], two-sided Wilcoxon rank-sum test). According to the individual-level data, the averages of mean [median] safe expenditures are 43.12 [42.98] in P treatment and 33.67 [34.25] in P&Ineg treatment, while the difference between the treatments is statistically insignificant (p=0.164 [p=0.310], two-sided Wilcoxon rank-sum test).

Although most of the differences are statistically insignificant, we still observe that risk-neutral and risk-averse third parties are directionally less willing to choose the safe option if the investment option with a negative expected net return is additionally available. Thus, from our experimental data, we find no evidence to support Hypothesis 4 that the investment option works as a decoy for the safe option.

## Appendix D. Dictator's Actual Transfer

We examine Player A's transfer behavior under P, P&I0, and I0 treatments. Throughout this section, we consider risk-consistent Player As. However, the overall results are not qualitatively affected even though the entire data set is considered.

By comparing P and P&I0 treatments, we can investigate the impact of the investment option on Player A's transfer behavior. We find that the difference in percentages of Player As who transferred a positive number of tokens to Player B (58% [n=26] in P treatment vs. 67% [n=24] in P&I0 treatment, p=0.570, two-sided Fisher's exact test) and the difference in Player A's transfer level (20 tokens in P treatment vs. 18 tokens in P&I0 treatment, p=0.849, two-sided Wilcoxon rank-sum test) between the two treatments are not substantial and are statistically insignificant at the significance level of 5%. This implies that the presence of the investment option might not affect Player A's belief in Player C's punishment behavior and hence does not alter Player A's transfer behavior.

By comparing I0 and P&I0 treatments, we can examine the impact of the punishment option on Player A's transfer behavior. Despite the insignificant statistical results, we observe that the difference in the percentages (39% [n=23] in I0 treatment vs. 67% [n=24] in P&I0 treatment, p=0.082, two-sided Fisher's exact test) and the difference in the transfer levels (10 tokens in I0 treatment vs. 18 tokens in P&I0 treatment, p=0.086, two-sided Wilcoxon rank-sum test) between the two treatments are substantial. In other words, our experimental data show that Player A is directionally more willing to transfer tokens to Player B when the punishment option is available to Player C. This implies that Player A might be threatened by the sanctioning mechanism and increases the amount of transfer to avoid Player C's punishment.

Furthermore, we regress *Transfer Dummy* and *Transfer Level* on *Treatment* and other controlling variables. *Transfer Dummy* is defined as 0 if Player A transferred nothing, and 1 if Player A transferred any positive token amount to Player B. *Transfer Level* is defined as the token amount Player A transferred to Player B. The results comparing P and P&I0 treatments are reported in columns 1–4 in Table D1 and show that the coefficients of *Treatment* indicate inconsistent results. The results comparing I0 and P&I0 treatments are reported in columns 5–8 in Table D1 and show that the *Treatment* indicate consistently positive coefficients although they are insignificant. Thus, our data weakly suggest that the sanctioning mechanism influences Player A's behavior to transfer more tokens to Player B, whereas the investment mechanism does not.

**Table D1. Player A's actual transfer (with risk-consistent)**

| Treatments | P, P&I0 | | | | I0, P&I0 | | | |
|---|---|---|---|---|---|---|---|---|
| Model: | LPM | Probit | OLS | Tobit | LPM | Probit | OLS | Tobit |
| Dependent variable: | Transfer Dummy | | Transfer Level | | Transfer Dummy | | Transfer Level | |
| | (1) | (2) | (3) | (4) | (5) | (6) | (7) | (8) |
| Treatment: P&I0 | 0.020 | 0.170 | −3.557 | −4.103 | 0.251 | 0.893* | 7.026 | 15.063 |
| | (0.140) | (0.450) | (5.066) | (9.312) | (0.163) | (0.481) | (4.785) | (8.983) |
| Switching Point | −0.007 | −0.031 | −0.506 | −1.583 | −0.017 | −0.056 | −1.207 | −1.929 |
| | (0.036) | (0.117) | (1.306) | (2.526) | (0.044) | (0.130) | (1.303) | (2.419) |
| Constant | 0.220 | −1.273 | −22.955 | −73.269 | −0.595 | −4.749* | −20.052 | −88.393* |
| | (0.727) | (2.330) | (26.340) | (49.908) | (0.843) | (2.862) | (24.813) | (50.763) |
| Socio-demographic | Yes | Yes | Yes | Yes | Yes | Yes | Yes | Yes |
| Personality traits | Yes | Yes | Yes | Yes | Yes | Yes | Yes | Yes |
| Observations | 50 | 50 | 50 | 50 | 47 | 47 | 47 | 47 |
| Number of individuals | 50 | 50 | 50 | 50 | 47 | 47 | 47 | 47 |
| Left-censored | — | — | — | 19 | — | — | — | 22 |
| Right-censored | — | — | — | 8 | — | — | — | 3 |
| (Pseudo) R-square | 0.293 | 0.252 | 0.362 | 0.082 | 0.226 | 0.205 | 0.330 | 0.074 |

Note: The dependent variable in Tobit models is truncated at 0 (left-censored) and 50 (right-censored). Socio-demographic includes gender, income, and Economics major. Standard errors are in parentheses. ***, **, and * denote significance at 1%, 5%, and 10% levels, respectively.